\documentclass[aps,prc,twocolumn,floatfix,superscriptaddress]{revtex4}

\usepackage{graphicx,graphics}
\usepackage{epsfig}
\usepackage{subfigure}
\usepackage{amsmath}
\usepackage{color} % for color comments
\usepackage{ulem}  % for strikeout
\usepackage{multirow}

\newcommand{\be}{\begin{equation}}
\newcommand{\ee}{\end{equation}}
\newcommand{\ba}{\begin{eqnarray}}
\newcommand{\ea}{\end{eqnarray}}
\newcommand{\ban}{\begin{eqnarray*}}
\newcommand{\ean}{\end{eqnarray*}}
\newcommand \nn {\nonumber}

\newcommand{\wv}[1]{\mathbf{#1}}

\newcommand{\eq}[1]{Eq.~(\ref{#1})}
\newcommand{\eqtwo}[2]{Eqs.~(\ref{#1}) and (\ref{#2})}
\newcommand{\fig}[1]{Fig.~\ref{#1}}

\newcommand{\wk}{\mathbf{k}}
\newcommand{\wq}{\mathbf{q}}
\newcommand{\tab}[1]{Table \ref{#1}}

\begin{document}

\title{Comparison of Jet Quenching Formalisms for a Quark-Gluon Plasma ``Brick''}

\affiliation{Departamento de F\'{\i}sica de Part\'{\i}culas and IGFAE, Universidade de Santiago de Compostela, E-15706 Santiago de Compostela, Spain}
\affiliation{Columbia University, Physics Department and Nevis Laboratories, PO Box 137, Irvington NY, 10533}
\affiliation{Department of Physics, McGill University, Montreal QC H3A-2T8, Canada}
\affiliation{Department of Physics, The Ohio State University, Columbus, OH 43210, USA}
\affiliation{Department of Physics, Wayne State University,  Detroit, MI 48201}
\affiliation{Department of Physics, University of Cape Town, Private Bag X3, Rondebosch 7701, South Africa}
\affiliation{Nuclear Science Division, Lawrence Berkeley National Laboratory, Berkeley, CA 94720, USA}
\affiliation{Nikhef, National Institute for Subatomic Physics and Institute for Subatomic Physics of Utrecht University, Utrecht, Netherlands}
\affiliation{Department of Physics \& CTMS, Duke University, Durham, NC 27708, USA}
\affiliation{Physics Department, Bldg. 510A, Brookhaven National Laboratory, Upton, NY 11973, USA}
\affiliation{Institute of Particle Physics, Central China Normal University, Wuhan 430079, China}%

\author{Nestor~Armesto}\affiliation{Departamento de F\'{\i}sica de Part\'{\i}culas and IGFAE, Universidade de Santiago de Compostela, E-15706 Santiago de Compostela, Spain}
\author{Brian~Cole}\affiliation{Columbia University, Physics Department and Nevis Laboratories, PO Box 137, Irvington NY, 10533}
\author{Charles~Gale}\affiliation{Department of Physics, McGill University, Montreal QC H3A-2T8, Canada}
\author{William A.~Horowitz}\affiliation{Department of Physics, The Ohio State University, Columbus, OH 43210, USA}\affiliation{Department of Physics, University of Cape Town, Private Bag X3, Rondebosch 7701, South Africa}
\author{Peter~Jacobs}\affiliation{Nuclear Science Division, Lawrence Berkeley National Laboratory, Berkeley, CA 94720, USA}
\author{Sangyong~Jeon}\affiliation{Department of Physics, McGill University, Montreal QC H3A-2T8, Canada}
\author{Marco~van Leeuwen}\affiliation{Nikhef, National Institute for Subatomic Physics and Institute for Subatomic Physics of Utrecht University, Utrecht, Netherlands}
\author{Abhijit~Majumder}%
\affiliation{Department of Physics, The Ohio State University, Columbus, OH 43210, USA}%
\affiliation{Department of Physics, Wayne State University,  Detroit, MI 48201}
\author{Berndt~M\"uller}\affiliation{Department of Physics \& CTMS, Duke University, Durham, NC 27708, USA}
\author{Guang-You~Qin}\affiliation{Department of Physics \& CTMS, Duke University, Durham, NC 27708, USA}
\author{Carlos A.~Salgado}\affiliation{Departamento de F\'{\i}sica de Part\'{\i}culas and IGFAE, Universidade de Santiago de Compostela, E-15706 Santiago de Compostela, Spain}
\author{Bj\"orn Schenke}\affiliation{Department of Physics, McGill University, Montreal QC H3A-2T8, Canada}\affiliation{Physics Department, Bldg. 510A, Brookhaven National Laboratory, Upton, NY 11973, USA}
\author{Marta~Verweij}\affiliation{Nikhef, National Institute for Subatomic Physics and Institute for Subatomic Physics of Utrecht University, Utrecht, Netherlands}
\author{Xin-Nian Wang}%
\affiliation{Institute of Particle Physics, Central China Normal University, Wuhan 430079, China}%
\affiliation{Nuclear Science Division, Lawrence Berkeley National Laboratory, Berkeley, CA 94720, USA}
\author{Urs Achim Wiedemann}\affiliation{Department of Physics, CERN, Theory Unit, CH-1211 Geneva 23, Switzerland}

\date{\today}

\begin{abstract}
We review the currently available formalisms for radiative energy loss
of a high-momentum parton in a dense strongly interacting medium. The
underlying theoretical framework of the four commonly used formalisms
is discussed and the differences and commonalities between the
formalisms are highlighted. A quantitative comparison of the single
gluon emission spectra as well as the energy loss distributions is
given for a model system consisting of a uniform medium  
with a fixed
length of $L=2$ fm and $L=5$ fm (the `Brick'). Sizable quantitative differences are
found. The largest differences can be attributed to specific approximations
that are made in the calculation of the radiation spectrum.
\end{abstract}

\maketitle

%
%    Intro and executive summary
%

\section{Introduction}

QCD jets are produced in hadronic interactions of all kinds, by the
hard scattering of the constituent partons (quarks and gluons) of the
colliding projectiles. One of the most significant experimental
discoveries in the collision of heavy nuclei at $\sqrt{s_{NN}}=200$
GeV at the Relativistic Heavy Ion Collider (RHIC) was the observation
of a strong suppression of the inclusive yield of high momentum (high $p_T$)
hadrons \cite{Adcox:2001jp,Adler:2002xw} and the semi-inclusive rate
of azimuthally back-to-back high $p_T$ hadron pairs relative to
expectations from p+p and d+Au collisions
\cite{Adams:2003im,Adams:2006yt,Adare:2010ry}, as expected from jet
quenching by the hot and dense medium formed in these
collisions. Evidence for jet quenching has also been observed recently
in Pb+Pb collisions at the LHC
\cite{Aad:2010bu,Aamodt:2010jd,Chatrchyan:2011sx}.

Most generally, the term `jet quenching' refers to the modification of
the evolution of an energetic parton induced by its interactions with
a colored medium. The expected manifestations of modified evolution of
parton showers include the suppression of hadronic spectra at high
$p_T$ and their back-to-back azimuthal correlations, as well as the
enhancement of hadron spectra at low $p_T$ and possibly the angular
broadening of internal jet structure and di-jet acoplanarity.

Data
from the RHIC heavy ion program strongly support the picture that jet
quenching is caused by the loss of energy of the primary parton,
either by collisions with constituents of the medium (collisional or
elastic energy loss \cite{Thomas:1991ea,Mustafa:2003vh}) or by gluon
bremsstrahlung (radiative or inelastic energy loss
\cite{Gyulassy:1993hr,Baier:1996kr,Baier:2000mf}), prior to
hadronization in the vacuum.  The loss of energy
suffered by an energetic quark or gluon penetrating a QCD medium
probes dynamical properties of the medium.  Jet quenching
provides unique and powerful tools for studying the properties of the hot and dense matter
produced in heavy ion collisions.

In recent years, a number of different dynamical models of jet
quenching has been formulated and compared to the jet quenching
signatures measured at RHIC (see
\cite{Gyulassy:2003mc,dEnterria:2009am,Wiedemann:2009sh,Majumder:2010qh}
for reviews). These models are based on the common assumption that
interactions of the energetic parton and the radiated gluon with the
medium can be calculated via perturbative QCD (pQCD), regardless
whether the properties of the medium itself can be treated
perturbatively. Within such perturbative approaches, however,
different approximations
exist in the calculation of radiative energy loss of the leading parton. Direct comparisons of the various pQCD-based
models show large quantitative discrepancies in  the medium density
that is needed to describe the same suppression of the inclusive production of light leading
mesons \cite{Bass:2008rv,Adare:2008qa,Adare:2008cg}.  The full modeling of
jet quenching in the complex, dynamic process of heavy ion collisions contains a large number of
components, including the initial production spectrum, a time-dependent
medium density profile, and fragmentation. The complexity of the modeling, with each model at present employing a different dynamical framework, makes it difficult to
isolate the specific differences between the energy loss formalisms, and thereby to constrain the underlying mechanisms of jet quenching through full model comparisons.

In this paper we review the current state of the art of pQCD-based
radiative energy loss models, including a discussion of their
theoretical foundations and limitations, their kinematic region of
applicability, as well as various model-dependent sources of
uncertainty. In order to make a direct comparison of the models, we
provide a systematic quantitative comparison of energy loss formalisms
using a highly simplified model problem: the energy loss of light
quarks in a static medium of constant density and a fixed length,
known as the QGP brick.We will restrict the discussion to leading
parton energy loss, for which the largest body of theoretical work
exists.

The paper is organized as follows. The first section provides an
executive summary of our main findings, with particular emphasis on
common features, technical and conceptual differences, and
uncertainties of different jet quenching models.  This section is a
self-contained narrative with a short introduction to the different
classes of jet quenching models. It is intended to be accessible to a non-expert reader. Many of the technical details will be omitted from the executive
summary, which refers to following sections for more detailed
explanations and numerical studies supporting the main statements.
For more extensive
descriptions of the various jet quenching formalisms and of the available
phenomenology including model--data comparisons, we refer to the
existing reviews 
\cite{Gyulassy:2003mc,dEnterria:2009am,Wiedemann:2009sh,Majumder:2010qh}

%%%%%%%%%%%%%%%%%%%%%%%%%%%%%%%%%%%%%%%%%
\section{Executive Summary}

\subsection{Jet quenching is caused by partonic interactions in dense matter}
We first discuss the factorized perturbative QCD framework for calculating jet and hadron production in the absence
of medium effects.  In elementary collisions ($e^+e^-$, pp, and
$\bar{\mathrm{p}}$p), our understanding of hadron production at high
transverse momentum is relatively mature. High-$p_T$ hadron production results
from partonic processes with large momentum transfer that can be described
with controlled uncertainty as the convolution of incoming
parton distribution functions (PDFs), a hard partonic collision process, and
the fragmentation of the partonic final state \cite{Collins:1985ue}.
The PDFs and Fragmentation Functions (FFs) are universal,
non-perturbative, scale-dependent distributions that obey DGLAP
evolution and are determined by global fits to data from elementary
collisions \cite{Pumplin:2002vw,Martin:2003sk,Hirai:2007sx,deFlorian:2007hc,Albino:2008fy}.

In ultra-relativistic
heavy ion collisions, there is currently no firm theoretical
argument that production cross sections factorize. Rather,
factorization is a working assumption that is consistent with
phenomenological analyses made so far and that underlies all models
discussed in this report. General considerations indicate that if
factorization is assumed, the hard partonic interaction itself cannot
be modified by the medium, since it occurs on temporal and spatial
scales too short to be resolved by the medium. However, it is
also expected on general grounds that the parton distribution functions in nuclei differ from
those in nucleons, and it is possible that the fragmentation of
outgoing partons is modified by the presence of a hot QCD medium.

We know with certainty that jet quenching receives its main
contribution on the level of the outgoing fragmenting parton. This
follows from several lines of argument.  In particular, it is
found that the production of direct photons, which have no strong
interaction with medium, does not show the large suppressions characteristic
for jet quenching \cite{Adler:2005ig}.  Moreover, there are by now
several global PDF fits that parametrize the nuclear dependence of
incoming parton distribution functions (nPDFs).  For the typical
momentum fractions $x$ and virtualities $Q$ relevant for hard
processes, these nPDF-fits show nuclear modifications of the order of
10-30 \% that cannot account for the observed factor 4-5 suppression
of single inclusive hadron spectra \cite{Eskola:2009uj,Arleo:2006xb}. Therefore, the energy
degradation observed in nucleus-nucleus collision for all hadronic
high transverse momentum spectra must be due to dynamical effects
occurring after the hard process ('final state effects').

Furthermore, there is strong experimental evidence that the final
state effect responsible for jet quenching is of partonic nature,
i.e., that it occurs prior to hadronization. In particular, correlations of high $p_T$ hadron
  pairs with small angular separation are largely unmodified in
  systems that exhibit strong inclusive suppression \cite{Adams:2003im,Adams:2006yt,Adare:2010ry}. In addition,  there is no evidence that the characteristically different
nuclear 'absorption' cross sections of different hadron species play a
role in understanding the jet quenching effect. 

% Since
% hadron formation times are Lorentz dilated in the lab frame,
% one expects on general grounds that with increasing $p_T$ hadronic
% effects occur at larger distances from the production point, and
% should thus be located outside a finite size medium for sufficiently
% large $p_T$. That the jet quenching suppression occurs with
% approximately equal strength for all $p_T$ thus points to a partonic
% origin of the effect. 

As a consequence of these generic observations, all current efforts to
understand and simulate jet quenching are based on models of parton
energy loss. Further progress then depends on promoting the relation
between jet quenching phenomenology and parton energy loss from a
qualitative to a more and more quantitative one. To this end, we
embark in the following on a critical assessment of current parton
energy loss models.

\subsection{Models of radiative parton energy loss}

%%%%%%%%%%%%%%%%%%%%%%%%%%%%%%
\begin{figure}[ht]
\centering
\includegraphics[width=0.9\columnwidth]{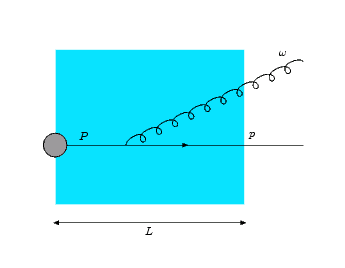}
\vskip -1cm
\caption{Schematic view of a hard parton-parton interactions (grey
blob) producing a highly energetic parton, which subsequently
undergoes parton branching processes.  In a heavy ion collision, this
parton evolution occurs within a dense medium (blue area) that
interferes with the vacuum evolution via {\it a priori} unknown elastic
and inelastic interactions.}
\label{fig1}
\end{figure}
%%%%%%%%%%%%%%%%%%%%%%%%%%%%%%%%%%%%%%%%%%%

The elementary branching process within a medium of finite size $L$ is depicted 
in Fig.~\ref{fig1}. Calculations of radiative parton energy loss aim at determining 
such processes within the framework of perturbative QCD. All approaches studied 
so far involve significant assumptions and approximations about
\begin{itemize}
\item the virtuality and (repeated) branching of the hard parton;
\item the nature of the medium through which the energetic parton propagates; and 
\item kinematical approximations for the interaction between medium and projectile parton.
\end{itemize}
It will turn out that the largest quantitative differences among the
various models arise not so much from differences in their basic
underlying assumptions about the medium and its interaction with the
high energy parton, but rather in the implementation of the
simplifying approximations made to carry out the derivations and
calculations, most importantly, the treatment of energy-momentum
constraints and large angle radiation.

In the remainder of this Section, we will first provide a short general discussion of 
in-medium QCD radiation, before describing the specific implementations.

\subsubsection{Virtuality and parton branching in the medium}

A high-$p_T$ parton, produced by a hard initial collision between
incoming partons, carries initially a high virtuality. Even in the
absence of a medium, the parton will undergo 'vacuum' splitting
processes to reduce its off-shellness. When the
radiations occurs in the medium, the question arises whether this
parton splitting is just the same as in the vacuum, or rather an
additional medium-induced splitting, or a result of the interference
between medium-induced effects and dynamics that would also occur in
the absence of a medium.  None of the existing parton energy loss
calculations treats the entire dynamics depicted in Fig.~\ref{fig1} in
a field-theoretically rigorous fashion. 

The medium effect on parton splitting is brought about by interactions
of the high-energy parton and the radiated gluon with the medium. For
small-angle scattering, the parton and the radiated gluon propagate
along similar paths, leading to significant interference and a finite
formation time of the gluon, which suppresses the gluon radiation
compared to incoherent emission. This effect is often referred to as
the Landau-Pomeranchuk-Migdal effect. A characteristic consequence of
this interference is that the amount of energy loss grows
quadratically with $L$ for in-medium path lengths that are small
compared to the formation time \cite{Baier:1996kr}.

In general, multiple splittings may occur in the medium. A full
calculation of multi-gluon final states would include interference
terms between the different emitted gluons. Recently some progress has
been made on this particular problem, using calculations of the
interference between two emitters (the antenna problem), which show
interesting features as angular ordering or partial decoherence
\cite{MehtarTani:2010ma,MehtarTani:2011tz,MehtarTani:2011jw,CasalderreySolana:2011rz}. The
generalization of these results to multiple gluon emission is not
known at present and all current calculations are based on repeated
application of a single-gluon emission calculation. In the following
we will separately discuss the single-gluon radiation kernel and the
prescription used for multiple gluon emission.

%%%%%%%%%%%%%%%%%%%%%%%%%%%%%%%%%%%%%%%%%%%%%%
\subsubsection{Modeling the medium}
\label{intro:med_model}
The main interest in studying jet quenching is to use it to
characterize the 
medium through which the projectile parton propagates. It is of great interest to implement
different models of the medium in jet quenching calculations to determine how 
the medium model and properties affect jet quenching. In the
current practice, each model description uses a particular set of
simplifications, which can be classified as follows:

\begin{enumerate}
\item {\it The medium is modeled as a collection of static scattering centers}\\
In this approach, the medium is modeled as a set of static colored
scattering centers with a specified density distribution along the trajectory
of the projectile. A decreasing density approximates the effects of
an expanding medium. Because gluon radiation from the scattering
  centers is ignored, calculations in this
set-up lead to gauge-invariant (i.e. physically meaningful) results
only to leading order in a high-energy approximation. By
construction this set-up neglects recoil effects and thus does not
allow for elastic parton energy loss. 

This medium model was pioneered by Baier, Dokshitzer, Mueller,
Peign\'e and Schiff \cite{Baier:1996kr,Baier:1996sk,Baier:1998kq}
(BDMPS) and independently by Zakharov \cite{Zakharov:1996fv,
Zakharov:1997uu}. Gluon radiation is formulated in a path-integral
that resums scatterings on multiple static colored scattering centers.
Wiedemann \cite{Wiedemann:2000za} showed how this path-integral can be
used to include interference effects between vacuum and medium-induced
radiation in such a way that also the $k_\perp$-differential
medium-induced gluon distribution is accounted for. 

In the original BDMPS derivation of the medium-induced gluon
distribution \cite{Baier:1996kr,Baier:1996sk}, the soft gluon
approximation, $x \ll 1$, was used. In later derivations
\cite{Baier:1998kq}, the radiation spectrum is multiplied by an
overall splitting function to take corrections for finite $x$ into
account. In the literature, the term "BDMPS-Z formalism" has been used
for both versions of the formalism. The difference between the
radiation spectrum with and without this splitting function $P_{s\to
g}(x)$ is discussed in Section \ref{sect:AMY_BDMPS_brick} and
Fig. \ref{fig:AMY_BDMPS}.

Most analytical and numerical results from the BDMPS--Z formalism use
a saddle point approximation that amounts effectively to assuming that
the projectile interacts with the medium via \textit{multiple soft
scattering} processes. In the following, the numerical results from the
low-$x$, multiple soft scattering implementation of the BDMPS--Z
formalism are based on the work by Armesto, Salgado and Wiedemann
\cite{Wiedemann:2000za,Salgado:2003gb,Armesto:2003jh}, abbreviated as
ASW--MS.  In the totally coherent limit, in which the entire medium
acts coherently towards gluon production, the multiple-soft scattering formalism results in a
radiation spectrum that is a radiation term for gluon production with
momentum transfer ${\bf q}_\perp$ convoluted with a Gaussian elastic
scattering cross section $\propto \frac{1}{\hat{q}\, L} \exp\left[ -
{\bf q}_\perp^2 / \hat{q}\, L\right]$.  In this limit, the medium is
fully characterized by the transport coefficient $\hat{q}$, the mean
of the squared transverse momentum exchanged per unit path length.

The {\it opacity expansion} was pioneered by Gyulassy, Levai and Vitev
\cite{Gyulassy:1999zd,Gyulassy:2000er} (GLV) and independently by
Wiedemann \cite{Wiedemann:2000za}. It also includes the interference
between vacuum and medium-induced radiation and is based on a
systematic expansion of the calculation in terms of the number of
scatterings. Since the BDMPS--Z path-integral formalism can serve as a
generating functional for the opacity expansion
\cite{Wiedemann:2000za}, the opacity expansion formalism is another
limit for the solution of the BDMPS--Z path integral.

 In most existing calculations, only the leading term ($N=1$) is
included, but the behaviour for larger opacities has been explored in
\cite{Gyulassy:2001nm,Wicks:2008ta}.  The medium is characterized by
two model parameters, the density of scattering centers $n$ or mean
free path $\lambda$, and a Debye screening mass $\mu_D$ used to
regulate the infrared behavior of the single scattering cross section.
In contrast to the multiple soft scattering approximation, this
approach includes the power-law tail of the scattering cross section
expected from QCD, leading to shorter formation times of the radiation
compared to the multiple-soft scattering approximation
\cite{Arnold:2009mr}. There are several model implementations in the
literature that differ significantly in their kinematic
approximations. We will use two implementations (ASW-SH and DGLV
\cite{Djordjevic:2003zk,Wicks:2005gt}) to discuss the
differences in detail in Section \ref{sec:WHDG:ASW}. Extension of the
opacity expansion formalism to a medium with dynamical scattering
centers was recently explored by Djordjevic and Heinz
\cite{Djordjevic:2008iz}.

\item {\it The medium is characterized by matrix elements of gauge
field operators}\\ In principle, multiple gluon exchanges between a
partonic projectile and a spatially extended medium can be formulated
in a field theoretically rigorous fashion by describing the medium in
terms of expectation values of 2-, 4-, 6-, 8-, ... field correlation
functions. Energy loss calculations based on the higher-twist (HT)
approach were pioneered by Guo and Wang
\cite{Wang:2001ifa,Majumder:2009zu}.  The approach includes the
interference between vacuum and medium-induced radiation. Properties
of the medium enter the calculation in terms of higher-twist matrix
elements.  In practice the matrix elements are factorized in the nPDF
and matrix elements describing the interaction between final state
partons and the medium. This factorization is valid at leading order
in the path length $L$ in the medium. As we discuss in some detail in
Section \ref{sect:HT}, the approximations currently employed result in a
formulation of parton energy loss calculations that closely resembles
models starting from a set of static scattering centers.

\item {\it Thermally equilibrated, perturbative medium}\\ The
formulation in rigorous field-theory of parton energy loss in a
weakly-coupled medium in perfect thermal equilibrium was developed by
Arnold, Moore and Yaffe \cite{Arnold:2001ms,Arnold:2002ja} (AMY). The
medium is formulated as a thermal equilibrium state in Hard Thermal
Loop improved finite temperature perturbation theory. As a
consequence, all properties of the medium are specified fully by its
temperature and baryon chemical potential. The calculation does not
incorporate vacuum branching of the projectile parton. In principle,
the perturbative description of the thermal medium applies only at
very high temperature $T \gg T_c$.
\end{enumerate} 

As seen from the list above, different models of parton energy loss characterize the medium in terms
of different primary model parameters. 
In the existing literature, the different approximations used for the
medium in the various approaches have led to different ways to
specify the medium properties. Recently it has become customary to translate 
the primary model parameters into an effective $\hat{q}$, that has the physical 
interpretation of an averaged squared momentum transfer between the
medium and the fast parton per unit path length 
$\hat{q} = \langle q_{\perp}^2 \rangle /\lambda$. This transport
coefficient can be calculated from the differential scattering cross section
$d\sigma/d^2q_{\perp}$ in the medium or the rate of momentum exchange,
$d\Gamma_{el}/ d^{2}q_{\perp}$. In this paper, 
we will use a common approach to calculate $\hat{q}$, using rates from Hard
Thermal Loop (HTL) effective field theory. The basic parameters in the
opacity expansion are the screening length $\mu_D$ and the mean free
path $\lambda$, which can also be calculated in HTL. For details, we refer
the reader to Appendix \ref{sec:commondef}.

% Formally, the Higher Twist (HT) formalism encodes the scattering power
% of the medium into a nonperturbative matrix element of the gauge field
% strength tensor $F^a_{\mu\nu}$.   In practice,
% approximations are made, such as neglecting the scale dependence of
% the matrix element, such that the Higher Twist
% formalism uses directly the transport coefficient
% $\hat{q}$.

\subsubsection{Kinematic approximations in different parton energy loss models}
With regard to the technical approximations employed in parton energy
loss calculations, the four model classes show important
commonalities, but also display characteristic differences. 

The three main assumptions that are made in all the calculations are:
\begin{itemize}
\item Both the parton and the radiated gluon are on eikonal
 trajectories: the parton energy $E$ is much larger than the transverse
 momentum exchanged with the medium $q_{\perp}$, $E \gg q_{\perp}$ and
 the energy of the emitted gluon $\omega$ is also much larger than the
 exchanged momentum: $\omega \gg q_{\perp}$.
\item Small-angle (collinear) radiation: the energy of the gluon is much larger than its transverse momentum $k_T$, $\omega \gg k_{T}$
\item Discrete scattering centers, or some form of localised momentum
 transfer: the mean free path $\lambda$ is much larger than the Debye
 screening length $1/\mu$ $\lambda \gg 1/\mu$.
\end{itemize}

Since the eikonal approximation is used, one often uses the additional
approximation that the gluon energy is much smaller than the parton
energy $\omega \ll E$, or $x=\omega/E \ll 1$, i.e. the soft radiation
approximation. This approximation is not made in the AMY calculation.

Let us first consider the kinematic constraints $E \gg \omega \gg
k_{T}, q_{\perp}$, which is often referred to as the soft eikonal
limit. In the calculations, this limit is used for example to neglect
the changes to the parent trajectory due to multiple scattering, which
simplifies the calculations significantly.  In phenomenological
applications of parton energy loss, however, gluon radiation is
calculated in the entire allowed phase space, up to $\omega = E$ and
$k_T = \omega$, where the approximations are not valid. The
approximations employed in the current calculations lead to finite
radiation probability at the kinematic bounds $\omega \approx E$ and
$k_T \approx \omega$, thus violating energy-momentum
conservation. Most current formalisms remedy this by imposing explicit
cut-offs at the kinematic bounds. The sensitivity of the result to
large-$x$ and large angle radiation can be explored to get an
impression of the accuracy of the result. The two limits warrant
separate discussions.

Consider the large-$x$ regime, where $\omega \approx E$.
For all energy loss models, the gluon energy distribution peaks at a
``typical'' energy (which may be very small). As a result, one can always restrict
the considerations (and in most cases the measurement as well) to a
reasonably high parton energy for which the calculated rate at $x \rightarrow
1$ is small compared to the total rate, thus giving confidence that
the impact of the cut-off at $x=1$ is small. For typical medium
densities at RHIC, the gluon energy $\omega$ peaks at $O(1$ GeV), so
that for parton energies $E \gtrsim 10 $ GeV, the probability density
at the kinematic boundary becomes reasonably small.

In the present calculational framework, the yield in the spectrum for
$\omega > E$, i.e. above the kinematical boundary can be taken as a
probability of total absorption of the parton, `death before
arrival'. We note that simply ignoring the radiation spectrum beyond
the kinematic limit leads to the unphysical result that in some cases
the total radiation probability decreases with increasing density or
path length, when the typical gluon energy is close to $E$. In the AMY
formalism, $\omega> E$ is kinematically allowed because interactions
with dynamic medium allow the propagating parton to absorb energy from
the medium when $E$ is not much larger than $T$, which is then
re-emitted.

The situation is different for large-angle radiation.  In the
presently existing formalisms, the typical transverse momentum of the
radiated gluon $k_T$ depends on the typical transverse momentum
exchanges $q_{\perp}$ and the number of scatterings $L/\lambda$, but not on
the gluon energy $\omega$. As a result, there is always some radiation
with $\omega$ smaller than the typical $k_T$ and thus with a large
probability for radiation at $k_T \rightarrow \omega$. The
quantitative impact of large angle radiation depends on the medium
model (large $q_\perp$ contributions in the medium cross section) and
the choice of parameters. Already in an early publication it was pointed
out that the large-angle regime may be important for phenomenological
applications, for example in \cite{Wiedemann:2000tf}: `` ... the
BDMPS--Z formalism is based on the assumption of small transverse
gluon momentum $\vert k_T \vert \ll \omega$ while we find the main
contribution to radiative energy loss for $\vert k_T \vert =
O(\omega)$. Both features question the validity of the BDMPS--Z
formalism ..."  A detailed discussion of the effect of large angle
radiation in the opacity expansion framework in given in Section
\ref{sec:WHDG:ASW} and in \cite{Horowitz:2009eb}.

%% From Will H:
% For the opacity expansion, AMY and BDMPS calculations, it is assumed
% that the parent parton pathlength is much longer than the mean free
% path of the gluon in medium ($L\gg\lambda$). In GLV and BDMPS this is
% used to neglect poles from propagators multiplied by $\exp(-\mu\Delta
% z)\approx\exp(-\mu\lambda)\ll1$, where $\Delta z$ is the distance
% between successive scattering centers; this approach is probably
% invalid for $L\lesssim\lambda\sim1$~fm.

\subsubsection{Multiple gluon emission}
\label{sect:mult_gluon}
Multiple gluon emission is calculated by repeating the single gluon
emission kernel as needed. 

The simplest procedure for multiple gluon emission is the Poisson
ansatz, where the number of emitted gluons follows a Poisson
distribution, with the mean number given by the integral of the gluon
emission spectrum. The energy distribution of each gluon follows the
single gluon emission kernel. This procedure is used by GLV and
ASW. In most cases relevant to the experimental conditions at RHIC,
the mean number of radiated gluons $\langle N_g \rangle > 1$, so that
the mean total energy loss $\langle\Delta E\rangle$ is larger than the
mean of the energy of a single gluon $\langle \omega \rangle$. In
general, this procedure leads to a distribution of lost energy that
does not conserve energy as the degrading momentum of the parent
parton is not dynamically updated. However, in most cases, the
probability to radiate a larger energy than the incoming parton energy
is limited, so that the uncertainty associated with this effect is
small. For a more detailed study, see \cite{Wicks:2008zz}.
Interference effects between medium-induced radiation and vacuum
radiation are included in the single-gluon emission spectra, but the
parton is assumed to fragment in the vacuum after energy loss.

HT and AMY both use a coupled evolution procedure to calculate
multiple gluon emission. In the case of HT, medium-modified DGLAP
evolution is used, which includes the virtuality evolution in
vacuum. In the AMY approach, rate equations are used and no vacuum
radiation is included. However, recent work \cite{CaronHuot:2010bp} has addressed this issue and showed how
it can be included in the same framework.

The evolution equations used in HT and AMY both include the coupling
between the quark and gluon distributions in the jets and keep track
of the gradual degradation of the jet energy. The emission probability
distribution changes as the jet energy degrades, which is in principle
a significant conceptual improvement over the Poisson convolution
approach.  There is, however, an important point that is not
explicitly addressed in any of the models: as the energy degrades, the
remaining path length through the medium also decreases. So, in
principle, the energy-evolution should be accompanied by an evolution
in coordinate space and the emission probabilities should be
calculated using local information about the parton energy, the
medium, and the remaining path length. Some models do take into account
the evolution of the medium and corresponding change in the local
environment of the propagating parton. However, combining both local
information and the true finite size effect proves to be a difficult
problem in the presence of interference effects, where the gluon
radiation is not a purely local phenomenon, but instead couples to the
parton over an extended area.

\subsection{Schemes of radiative parton energy loss}
\label{sect:intro_sbys}
%%%%%%%%%%%%%%%%%%%%%%%%%%%%%%%%
\begin{figure}[ht]
\centering
\includegraphics[width=0.9\columnwidth]{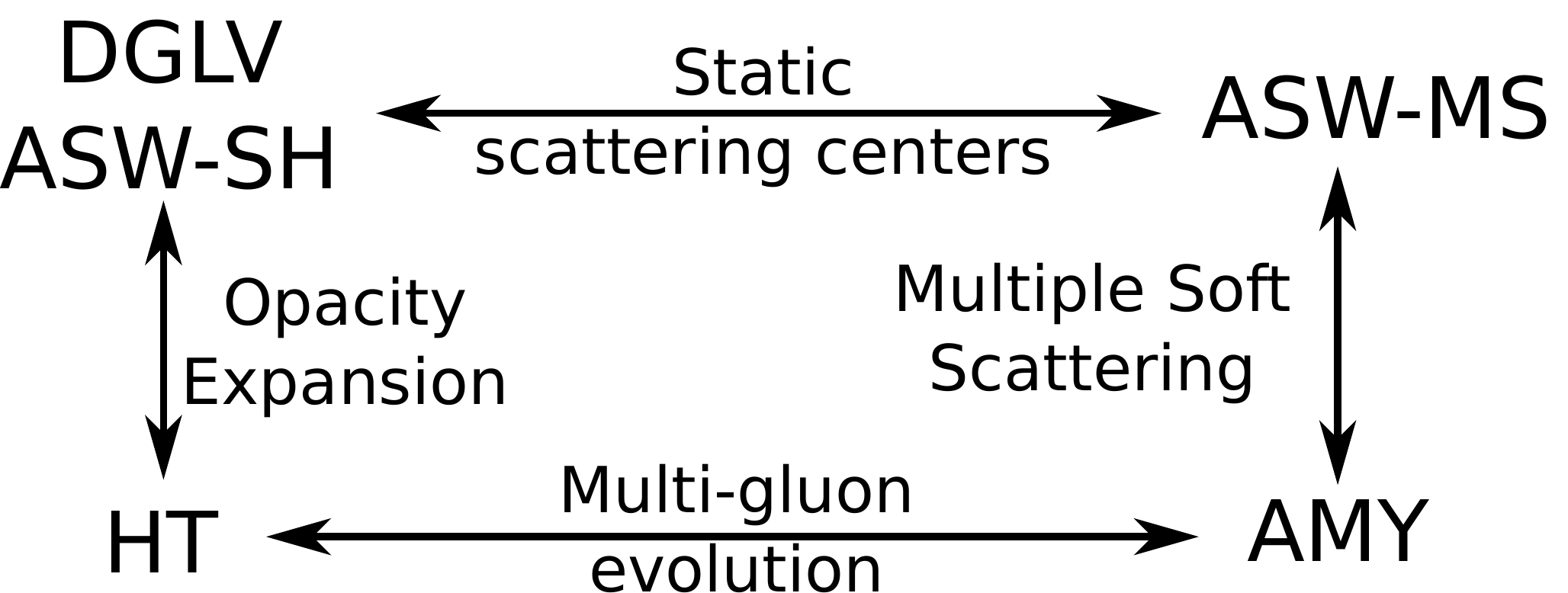}
%{figs/Scenarios.jpg}
\caption{The landscape of pQCD based jet quenching formalisms. Arrows indicate 
common concepts or assumptions between adjacent formalisms.}
%pairs of formalisms for which detailed analytical and numerical comparisons are presented.}
\label{overview}
\end{figure}
%%%%%%%%%%%%%%%%%%%%%%%%%%%%%%%%%

In the previous sections, we outlined both the single-gluon
emission calculations as well as the multiple gluon calculations that
are commonly used in the literature. To gain a better understanding of
the various formalisms, it is useful to compare and contrast closely
related formalisms.
Fig.~\ref{overview} illustrates the close technical and conceptual relations between the different models
of parton energy loss. 
\begin{enumerate}
    \item {\it Multiple soft scattering vs Opacity expansion}\\ The
      multiple soft scattering approximation and the opacity expansion
      can be shown to be different approximations of the same BDMPS--Z
      path integral. In the opacity expansion, the hard scattering
      tail of the scattering potential of the medium is taken into
      account, but the interference between neighboring scattering
      centers is only taken into account order by order in the
      expansion. The most commonly used approximation is the single
      hard scattering ($N=1$) approximation which assumes that there
      is only one dominant hard scattering. The multiple-soft
      scattering approximation resums all the interference terms, at
      the cost of neglecting hard scatterings with the medium.  In
      general, the multiple-soft scattering approximation is expected
      to be valid for thick media, while the single-hard scattering
      approximation is more accurate for thin media.
    \item {\it AMY vs multiple soft scattering approach}\\ Parton
	energy loss within the AMY formalism can be formulated in
	terms of the same path integral entering the BDMPS--Z
	formalism \cite{Salgado:2003gb}. The technical commonalities
	between both approaches are further elaborated in
	\cite{Arnold:2008iy,Arnold:2009mr}.  The main conceptual
	difference between the AMY and the BDMPS-Z calculation is the
	formulation of the medium: an equilibrated high-temperature
	plasma in AMY versus static scattering centers in BDMPS-Z.
        There are also
	important technical differences, however. Most notably: in AMY
	an infinite-length medium is used, while BDMPS-Z includes
	finite-size effects in the emission rates. The effect of these
	differences is discussed in detail in
	\cite{CaronHuot:2010bp}. In addition, the ASW calculation
	include the effect of large angle cut-off on the
	radiation. This effect is not taken into account in the
	standard AMY calculation, but is explored in detail in Section
	\ref{sect:AMY_largeangle} below.
    \item {\it Opacity expansions: DGLV vs ASW-SH}\\ There are
    two implementations of the ($N=1$) opacity expansion in the literature: DGLV
    \cite{Gyulassy:1999zd,Djordjevic:2003zk} and ASW--SH
    \cite{Wiedemann:2000za,Salgado:2003gb}. Both groups calculate the
    same set of multiple scattering Feynman diagrams within the same
    high-energy and collinear approximation. However, in extending
    this result from the kinematic region in which these
    approximations are valid to the entire phase space open for gluon
    production, they take different approximations. The resulting
    numerical differences are discussed in detail in Section
    \ref{sec:WHDG:ASW}. In particular, large angle gluon
    radiation is a source of significant theoretical uncertainty.
    \item {\it Higher Twist vs Opacity Expansion}\\ The Higher Twist formalisms
      and the ($N=1$) Opacity expansion are related in the sense
      that they are formulated for media where the jet scatters a few
      times per emission.  While both formalisms have extended the
      theoretical set-up to include multiple scatterings per emission,
      most comparisons to data still make use of the one scattering
      per emission approximation.  This is referred to as leading
      order in opacity for GLV scheme and as next to leading twist
      in the higher twist scheme. The characterization of the medium
      is also somewhat different as the GLV scheme incorporates the
      full functional form for scattering off a heavy static
      scattering center and the HT scheme takes the leading transverse
      momentum moment of the exchanged transverse momentum
      distribution.  The HT scheme thus assumes a Gaussian
      distribution for the exchanged transverse momentum.

      The formalisms differ significantly in the way multiple-gluon
	emission is treated.  The Higher Twist calculation uses modified DGLAP
	virtuality evolution, while in the Opacity Expansion, multiple
	gluon emission is described using Poisson distributed
	independent emissions (see Section \ref{sect:mult_gluon}). The
	modified DGLAP evolution used in the HT scheme implies that subsequent
	radiated gluons are ordered in transverse momentum $k_{\perp}$
	($k_{\perp}$ of a given radiation is the
	upper bound for the subsequent radiation). There is no such
	restriction in the GLV scheme and each radiation may explore
	the limits set by the kinematic bounds.
\end{enumerate}

\subsection{Summary of the main findings}
\label{sect:conclusion}
A detailed comparison of the different energy loss models touches
on many aspects of the underlying phsyics assumptions, but also the
technical implementation. Most of these aspects are mentioned in the
preceding discussion and are treated in more detail in Section
\ref{sect:detailed_comp}. Here we list the four most important areas
which are fundamental to the problem of radiative energy loss in a QCD
medium.
\begin{enumerate}
\item {\it Effect of kinematic limits:} \\All formalisms make use of
 collinear approximation $\omega \gg k_{T}$. In addition, most of the
 calculations presented in Section \ref{sect:detailed_comp} (ASW-SH,
 ASW-MS, GLV, and HT), except AMY, make use the soft radiation
 approximation $E \gg \omega$. In phenomenological calculations, we
 have to extend the calculations into the large-$x$,
 $\omega\rightarrow E$, and large-angle, $k_T \rightarrow \omega$,
 domain. For sufficiently large parton energies, the uncertainties
 associated with the large-$x$ domain seem to be reasonably small (see
 Sections \ref{sec:WHDG:ASW} and \ref{sec:AMY}).  Large angle
 radiation, however, is expected to be problematic for all
 models. This is explored in detail in Section \ref{sec:WHDG:ASW}, for
 the opacity expansions, where the resulting uncertainty is found to
 be about a factor 2 in the gluon spectra. The resulting uncertainty
 in $R_{AA}$ decreases with increasing $p_{T}$ \cite{Horowitz:2009eb}.
 In Section \ref{sec:AMY} it is shown that part of the difference
 between the AMY and ASW--MS formalism can be attributed to the
 absence of a large angle kinematic bound in the AMY calculation.
\item {\it Effect of treatment of multiple gluon emission:}\\ In the
 existing formalisms, the calculation of the single-gluon emission
 kernel and the calculation of multiple gluon emission are separate
 steps. GLV, ASW-SH and ASW-MS use the Poisson ansatz; AMY uses rate
 equations; and HT uses DGLAP evolution.  We find that differences
 between the energy loss formalisms originate at the single-gluon
 level. It is possible that the differences between the multiple-gluon
 emission treatment further affect the total energy loss, but we have
 not investigated this in detail. Semi-inclusive observables, like
 the fragmentation function or the jet shape, are expected to be more
 sensitive to the details of multiple gluon emission, such as whether
 the gluons are emitted independently (Poisson ansatz) or in a $k_T$
 ordered pattern (DGLAP evolution), than the leading particle energy
 loss or the single hadron suppression $R_{AA}$.
\item {\it Effect of medium model:}\\ In most of the current energy
 loss formalisms, the picture of the medium is intrinsically tied to
 the approach used in the calculation. As a result, it is not possible
 to separate the effect of the medium model from assumptions made in
 the rest of the calculation. For example, the multiple-soft
 scattering calculation ASW-MS is only analytically feasible in the
 limit of gaussian broadening, which corresponds to scattering centers
 with a harmonic oscillator potential. On the other hand, the AMY
 approach uses a medium based on Hard Thermal Loop field theory with
 dynamical scattering centers. Futher work is required to
 systematically compare different medium models.
\item {\it Effect of including vacuum radiation:}\\ All formalisms
include vacuum radiation to some extent, except AMY.  The fact that
AMY shows the largest suppression in the final comparison (Section \ref{sec:syst_comp}) is mostly
due to lack of finite length effects. These effects lead to the $L^2$
dependence of energy loss during the early stage of the evolution. At
later times, or larger $L$, this turns into a linear dependence on
$L$. Because AMY implicitly assumes an infinite medium the $L^2$
dependence is lost.  This behavior is studied in detail in
\cite{CaronHuot:2010bp} using the light cone path integral formalism
including finite length effects. The obtained rates show the expected
$L^2$ dependence of
the energy loss early in the evolution and approach the AMY result at
late times.  

\end{enumerate}

Before we end this Section, we emphasize a common limitation of 
all formalisms of parton energy loss discussed here: They assume
that the hadronization of the leading parton occurs in vacuum. The
formalisms describe the interaction of the leading parton with the
medium strictly at the partonic level, but do not consider any influence
of the medium on the hadronization process or final-state interactions
of the produced leading hadron with the medium. This appears to be
justifiable for leading hadrons with momentum $p_h \gg m_h$, where
$m_h$ is the hadron mass, because the formation time for such a hadron
is Lorentz dilated by a factor $p_h/m_h \gg 1$, and thus the formation
of the hadron may be assumed to occur far outside the medium. Indeed,
as discussed in the introduction, experimental evidence supports this
assumption for light hadrons.
We note however that this assumption is not generally satisfied under present
conditions of heavy ion collisions at RHIC for baryons and mesons
containing heavy quarks.

\subsection{Directions for future work}
It is clear from the preceding discussion that the current energy loss
calculations have a number of specific weaknesses which may have large
quantitative impact on the results. Based on our review of the existing
models, we can formulate a number of questions for future work.

Firstly, we recommend to develop calculations with better control of
the large angle and large energy radiation. In a realistic
calculation, the radiation cross section should naturally go to zero
at the kinematic limits $k_{T}=\omega$ and $\omega = E$. It is most
important to control the large-angle behaviour; the large-energy
regime can to some extent be avoided by concentrating on high-energy
partons/jets. The most natural way to control the large-angle
behaviour of the radiation would be to use the full
Next-to-Leading Order (NLO) matrix elements for $2 \rightarrow 3$
scattering with radiation.

The treatments for multiple gluon radiation should also be further
investigated. So far, the treatments are based on incoherent
superposition and/or $k_{T}$-ordering of the shower. It is worthwhile
to quantitatively investigate deviations from these
assumptions. Semi-inclusive observables such as fragmentation
functions and jet shapes may be more suitable to investigate specific
multi-gluon emission scenarios than inclusive observables, such as the
nuclear modification factor $R_{AA}$.

\section{Detailed Comparisons of Models}
\label{sect:detailed_comp}

In the following sections, we will present more detailed side-by-side
comparisons of the existing energy loss formalisms. The discussion 
follows the structure layed out in Section \ref{sect:intro_sbys}.

%
%    ASW section 
% 

\subsection{The ASW formalism}
\label{sect:ASW}

As discussed in Section \ref{intro:med_model}, the ASW formalism
calculates parton energy loss based on a path-integral formalism
\cite{Wiedemann:2000za,Salgado:2003gb,Armesto:2003jh} which can be
evaluated either in the multiple soft scattering limit or through an
opacity expansion. The GLV $N = 1$ opacity result reproduces the ASW
expression \cite{Wiedemann:2000za} on the level of the Feynman
diagrams and the analytic expression for the $\omega$- and
$k_T$-differential gluon energy distribution.

In this section, we concentrate on the multiple soft scattering
approximation ASW--MS, which has been mostly used in comparisons to
experimental data. A comparison of the single-hard scattering result
ASW--SH to the DGLV opacity expansion \cite{Djordjevic:2003zk} used by
WHDG \cite{Wicks:2005gt} is given in Section \ref{sec:WHDG:ASW}.

\begin{figure*}
\includegraphics[width=0.49\textwidth]{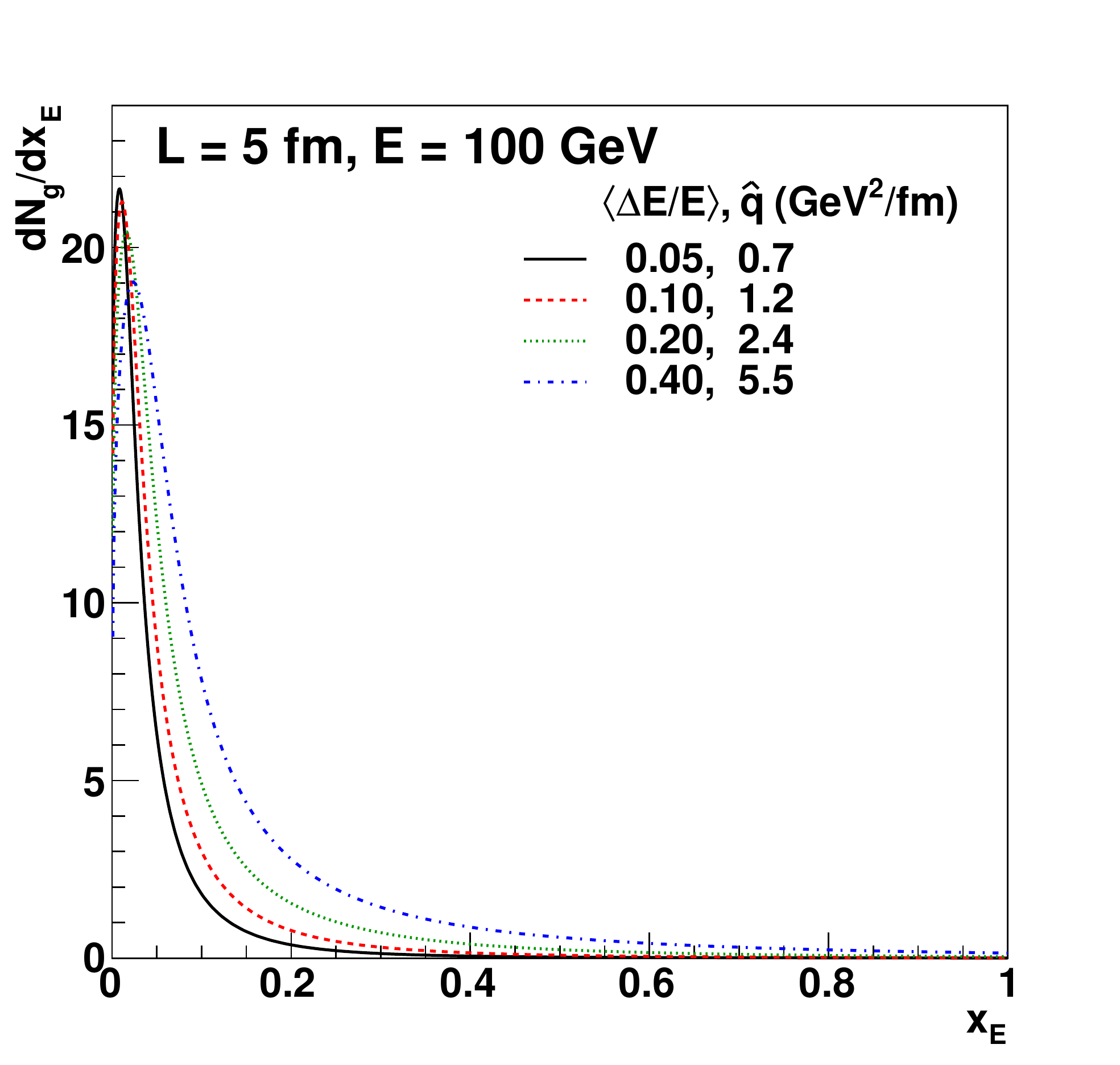}
\hfill
\includegraphics[width=0.49\textwidth]{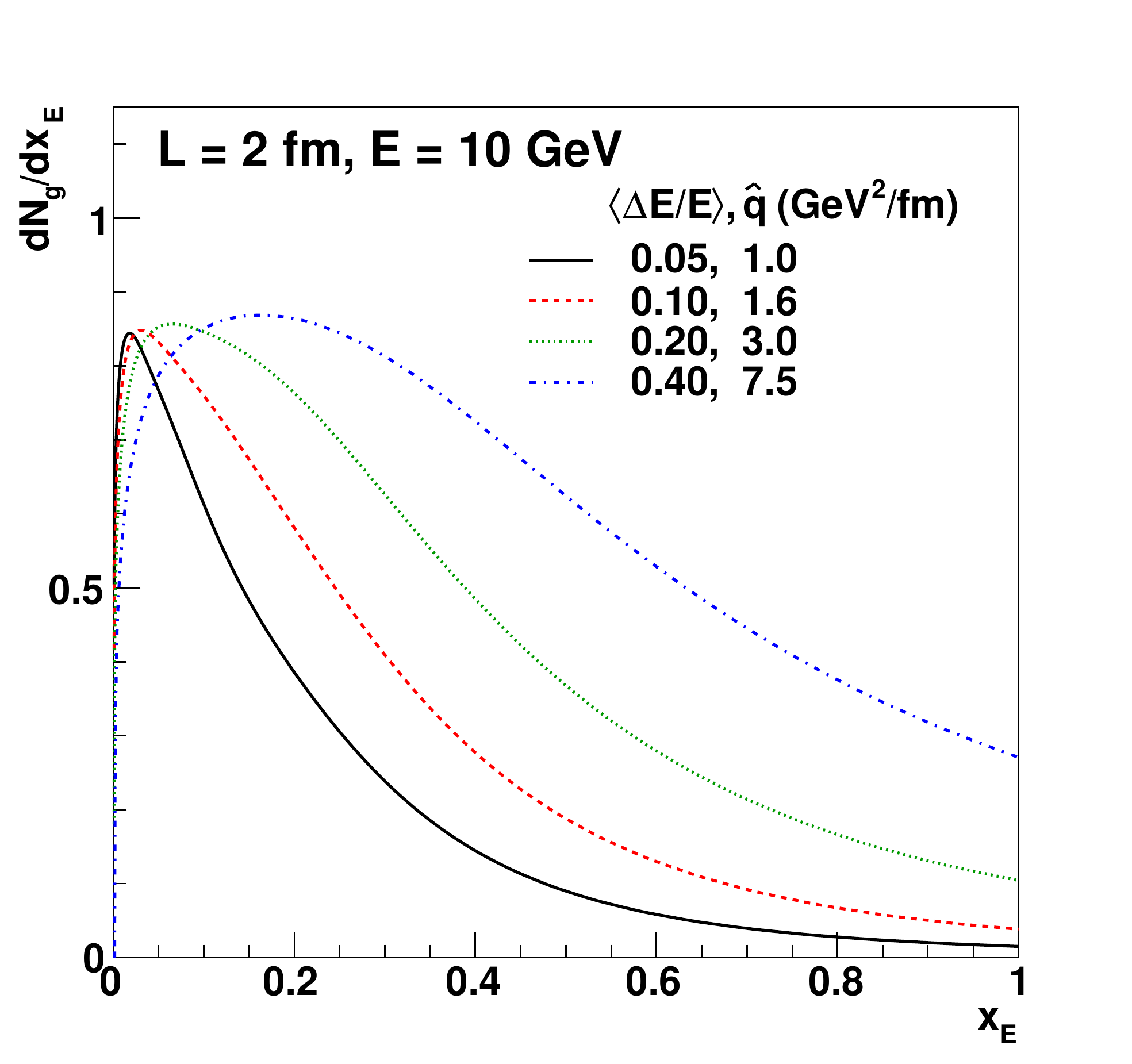}
\caption{Energy spectrum of radiated gluons, for a light quark with energy $E=100$ GeV and path length $L=5$ fm (top panel) and for a light quark with $E=10$ GeV and $L=2$ fm (lower panel). The legends on the plots indicate the average energy loss and the corresponding value of the transport coefficient $\hat q$.}
\label{asw:fig3-4}
\end{figure*}

\subsubsection{Gluon radiation spectrum}

The result of the ASW--MS calculation is publicly available in terms
of the `quenching weights' which represent the energy loss probability
distribution \cite{ASWCode}. However, before discussing the
probability distribution, we will present the single gluon radiation
spectra. The multiple soft scattering limit suppresses the production
of infrared gluons by a destructive interference effect. As a
consequence, all spectra are peaked at finite gluon energies. In
general, the radiated gluons become harder as one increases the
average energy loss (i.e. as one increases $\hat{q}$). If the
projectile energy is sufficiently large and the in-medium pathlength
is sufficiently small, then the radiated gluons carry small fractions
of the projectile energy. This is the case for a projectile quark
energy $E = 100$ GeV shown in the upper panel of
Fig. \ref{asw:fig3-4}.

However, if the projectile energy is too small, see lower panel of
Fig. \ref{asw:fig3-4}, one faces a particular problem: One calculates
the radiated gluon spectrum as if the parton would propagate through a
medium of path-length $L$, although there is a finite probability that
the parton does not have sufficient initial energy to make it through
$L$, and gets stuck before. In the present calculational framework,
finding yield in the spectrum beyond the kinematic boundary $x = 1$
($\omega = E$), signals that one has assumed that the particle
propagates through a length $L$ for which the probability of `death
before arrival' is finite.

\begin{figure*}

\includegraphics[width=0.49\textwidth]{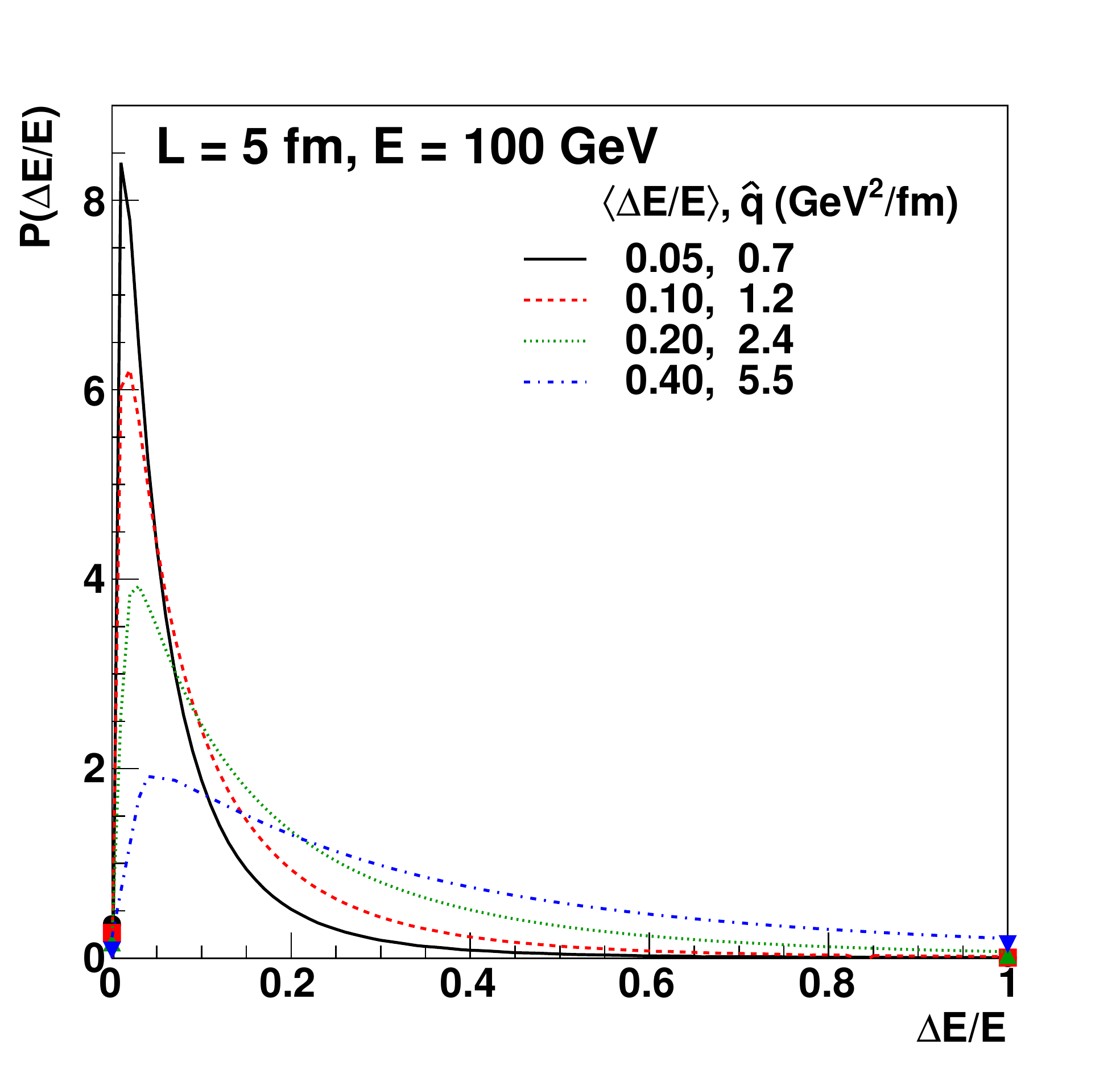} \hfill%
\includegraphics[width=0.49\textwidth]{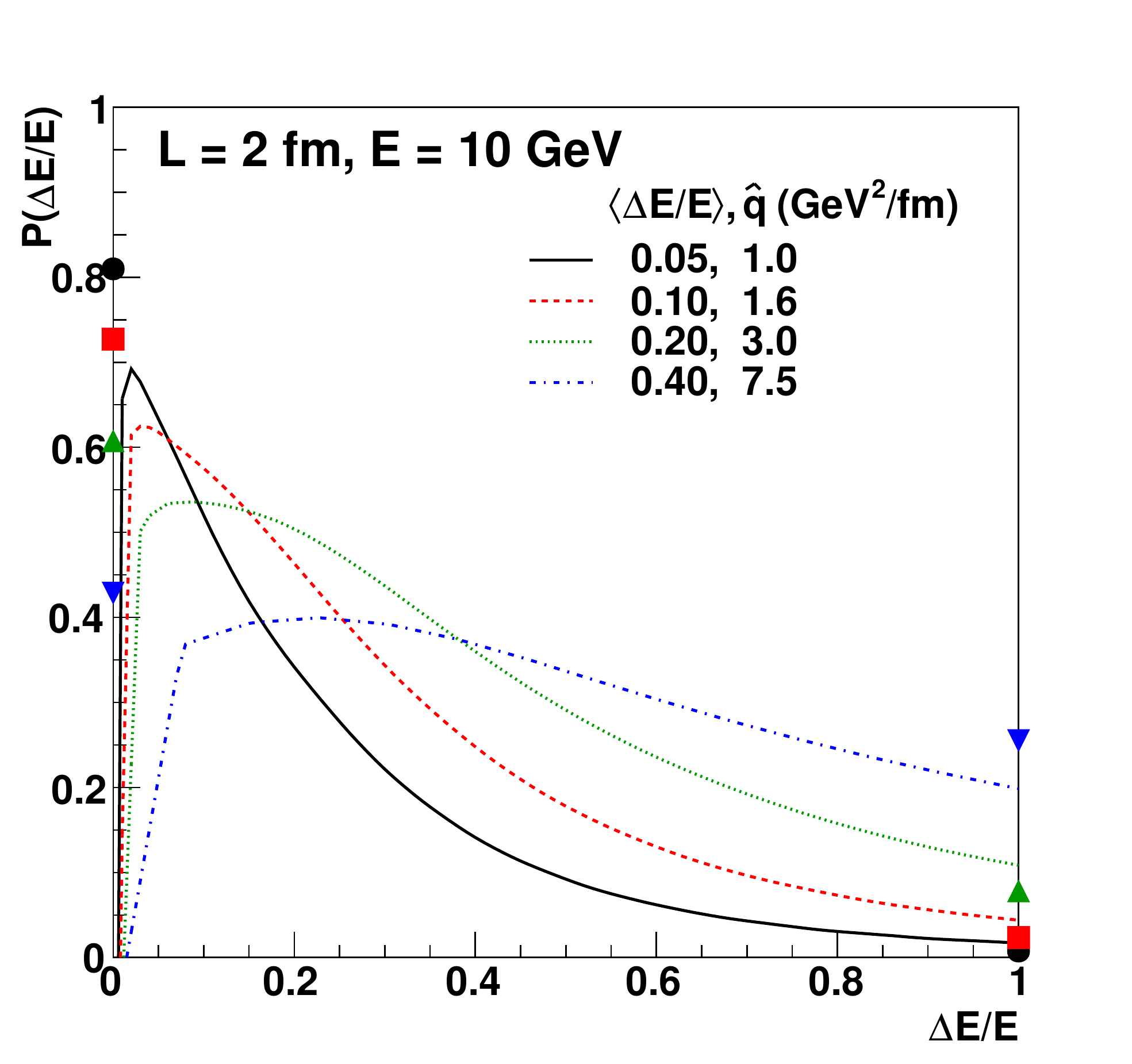}
\caption{Energy loss probability density $P(\Delta E)$ for a light
quark of energy $E=100$ GeV and path length $L=5$ fm (top panel) and $E=10$ GeV and $L=2$ fm. The different
lines are for different average energy loss, as indicated in the
legend, together with the corresponding values of the transport
coefficient $\hat q$.}
\label{asw:fig1-2}
\end{figure*}

\subsubsection{Multiple gluon emission}
In the small $x$ limit, the gluon emission probability is independent
of the incident parton energy. In this case one may convolute with the
Poisson distribution to compute the probability $P(\Delta E)$ for an
incident parton to lose energy $\Delta E$ during its passage through
the medium
\begin{eqnarray}
P(\Delta E) = \sum_{n=0}^{\infty} \frac{e^{-\langle N_{g} \rangle}}{n!} \prod_{i=1}^n \left[\int d\omega
\frac{dI}{d\omega}\right] \delta(\Delta E - \sum_{i=1}^{n} \omega_i)\,,
\label{eq:poisson_mg}
\end{eqnarray}
where $\langle N_{g}\rangle = \int_0^\infty d\omega (dI/d\omega)$ is the mean number
of radiated gluons. This Poisson convolution is used in the ASW
calculations for the multiple soft gluon approximation as well as the
opacity expansion, and also in the GLV-based opacity expansions.

The resulting energy loss probability distribution $P(\Delta E / E)$
are shown in Fig.~\ref{asw:fig1-2} using the ASW multiple soft scattering
limit for an incoming quark with energy $E$ and in-medium path length
$L$. All results have been computed with $\alpha_s = 0.3$.  The
probability distribution has three distinct pieces:

\begin{enumerate}
\item {\it Untouched survival}: There is finite probability that a
  parton does not interact with a medium of length $L$ and that it
  loses no energy. This probability is $P(0)=\exp(-\langle N_g
  \rangle)$ and is represented by a color-coded dot at $\Delta E / E =
  - 0.05$ in Fig. \ref{asw:fig1-2}.
\item {\it Survival with finite energy loss}: This is the continuous probability that a parton makes it through a medium of length $L$ but loses a finite fraction $\Delta E / E$ during its passage. This is denoted by the color-coded curve at finite $0 \le \Delta E/E \le 1$.
\item {\it Death before arrival}: In general, if one shoots a particle into a wall of thickness $L$, it can get stopped on its journey before reaching the length $L$. This probability is denoted by the color-coded dot at $\Delta E / E = 1.05$.
\end{enumerate}

In general, as the average energy loss grows due to an increase in
$\hat{q}$, one observes that:
\begin{itemize}
\setlength{\itemsep}{0pt}
\item[(a)] the probability of untouched survival decreases; 
\item[(b)] the probability of survival with finite energy loss shifts to larger values of $\Delta E/E$; 
\item[(c)] the probability of death before arrival increases. 
\end{itemize}
This is seen clearly in Fig. \ref{asw:fig1-2}, which shows the energy
loss probability distribution $P(\Delta E / E)$ for two different
choices of the parton energy $E$ and path length $L$. The extreme case
is $E = 10$ GeV and $L = 2$ fm (lower panel of Fig.~\ref{asw:fig1-2}),
where requiring an energy loss of $\Delta E = 4$ GeV amounts to a
greater than 40 percent probability of untouched survival and 30
percent probability of death before arrival. This comes close to an
all-or-nothing scenario, where a particle either goes through without
medium modification or gets stuck, but its probability of emergence as
an object with reduced energy is relatively small.

\subsubsection{Uncertainties associated with kinematic limits}
The limitations of the high-energy eikonal approximations used in the
derivation of the path-integral formalism from which both the multiple
soft scattering limit and the opacity expansion stem, were discussed
in the original papers
\cite{Wiedemann:2000za,Salgado:2003gb,Armesto:2003jh}. 
More recently, ASW have used a number different approaches which
are trying to address the shortcomings of the analytical calculations,
such as using DGLAP evolution equations
\cite{Armesto:2007dt} and Monte Carlo algorithms for
final state radiation
\cite{Zapp:2008gi,Zapp:2008af,Armesto:2009fj,Armesto:2009ab}.

%
%         Opacity expansion Section
%

\subsection{Opacity expansion: WHDG and ASW--SH}\label{sec:WHDG:ASW}

This subsection contains a detailed comparison of the radiative part
of the WHDG calculation \cite{Wicks:2005gt}, equivalent to the first
term in the DGLV opacity expansion \cite{Djordjevic:2003zk}, and the
ASW opacity expansion formalism truncated to first order in opacity, here called ASW--SH
\cite{Salgado:2003gb,Armesto:2003jh}. We discuss the importance of the definition of
the splitting variable $x$ and of the kinematic cut-offs made to
enforce the assumptions used in deriving the energy loss formulae.
Quantitative results comparing the radiative components of
WHDG and ASW--SH will also be shown.

\subsubsection{Single emission kernel}

\begin{figure}
\centering
%\vspace{0.5cm}
\includegraphics[width=0.45\textwidth]{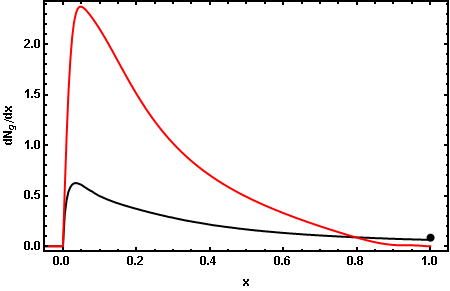}
\caption{\label{WHDG:figone}
Single inclusive gluon radiation distribution, $dN_g/dx$,
from the WHDG implementation of the first order in opacity DGLV
formula, \eq{WHDG:DGLV}, in red, and the ASW--SH implementation of
\eq{WHDG:ASW}, in black, for a 10 GeV up quark traversing a nominal, 2
fm long static brick of QGP held at a constant $T=485$ MeV.  The point
at $x=1$ indicates the integrated weight of $dN_g/dx$ in the ASW--SH
implementation for $x>1$.  }
\end{figure}

One of the major driving forces of the present investigation was the
realization that some of the approximations made to arrive at
analytical results for the medium-induced radiation have a larger
impact on the final result than expected.
An
excellent case in point comes when one attempts a naive comparison
between the single inclusive gluon distribution implemented in a
massless quark and gluon version of the radiative piece of WHDG and
the one found from the ASW--SH code \cite{Salgado:2003gb,ASWCode} for
the opacity expansion. Both purport to compute the single inclusive distribution
of gluon radiation, $dN_g/dx$, to first order in opacity
\cite{Gyulassy:1999zd,Gyulassy:2000er,Wiedemann:2000za} for a medium
of Debeye-screened colored static scattering centers
\cite{Gyulassy:1993hr}.  See \fig{WHDG:figone}, which shows $dN_g^{\rm
DGLV}/dx$ and $dN^{\rm ASW-SH}_g/dx$ for a nominal 10 GeV quark jet in
a static plasma of length 2 fm and temperature $T=485$ MeV \footnote{We
note that the ASW--SH implementation \cite{Salgado:2003gb} is only in
terms of the Poisson convoluted energy loss probability distribution,
which we will discuss in detail further below.  The comparison shown
here was obtained from an independent numerical implementation of the
massless ASW--SH formula that well reproduces the Poisson
convolution results of \cite{Salgado:2003gb}.}.

\begin{widetext}
The DGLV formula for the first order in opacity energy loss is
\cite{Djordjevic:2003zk}:
\be
\label{WHDG:DGLV}
x \frac{dN_g^\mathrm{DGLV}}{dx} = \frac{2C_R \alpha_s}{\pi^3}\frac{L}{\lambda}\int d^2\wv{q}d^2\wv{k}\frac{\mu^2}{\big(\wq^2+\mu^2\big)^2} \frac{\wk\cdot\wq(\wk-\wq)^2-\beta^2\wq\cdot(\wk-\wq)}{\big[(\wk-\wq)^2+\beta^2\big]^2\big(\wk^2+\beta^2\big)} \int dz \left[ 1-\cos\left( \frac{(\wk-\wq)^2+\beta^2}{2xE}z \right)\right] \rho(z) .
\ee
Here $\wv{q}$ is the transverse momentum exchanged with the medium and
$\wv{k}$ the transverse momentum of the radiated gluon, $\mu$ is the
Debeye screening mass, $m_g$ and $M_q$ are the effective thermal mass
of the radiated gluon and the mass of the parent parton, respectively,
$\beta^2 = m_g^2+x^2M_q^2$; $\lambda$ is the mean free path of the
radiated gluon and $\rho(z)$ is the probability distribution for the
distance to the first scattering center \footnote{In general one
needs to consider the distribution in differences in distance between
successive scattering centers.  However at first order in opacity
there is only one scattering center; as one may always set the initial
value of $z$ to 0, $\rho(z)$ is the absolute distance to the first
scattering center.}.  The equivalent expression for ASW--SH, which does
not include the effect of a thermal gluon mass $m_g$, is
\cite{Armesto:2003jh}
\ba
\label{WHDG:ASW}
\omega\frac{dI^\mathrm{ASW-SH}}{d\omega} &=& \frac{2C_R\alpha_s}{\pi^3}\frac{L}{\lambda}\int d^2\wv{q}d^2\wv{k}\frac{\mu^2}{(\wv{q}^2+\mu^2)^2}
\frac{\wk\cdot\wq(\wk-\wq)^2-x^2M_q^2\wq\cdot(\wk-\wq)}{\big[(\wk-\wq)^2+x^2M_q^2\big]^2\big(\wk^2+x^2M_q^2\big)}
\nn \\ 
&& \qquad\qquad
\times\int dz\left[1-\cos\left(\frac{(\wv{k}-\wv{q})^2+x^2M_q^2}{2\omega}\right)\right]\rho(z).
\ea
\end{widetext}

If we set $\omega=xE$ in (\ref{WHDG:ASW}) and $m_g=0$ in $\beta$ in
(\ref{WHDG:DGLV}), then we see that \eqtwo{WHDG:DGLV}{WHDG:ASW} are,
in fact, identical.  Why, then, do the curves obtained by evaluating
these expressions, shown in \fig{WHDG:figone}, differ so drastically?
The answer lies in the details of the implementation of these two
expressions.  It turns out that there are a number of choices that
must be made in going from the differential expressions for the
radiation spectrum to the integrated gluon distribution, $dN_g/dx$.
These are summarized in \tab{WHDG:table}.

\begin{table}[!htb]
\begin{tabular}{ccc}
%\hline
& WHDG & ASW--SH \\
\hline
$k_\mathrm{max}$ & $2x(1-x)E$ & $x E$ \\
%\hline
$x$ & $x_+$ & $x_{\rm E}$ \\
%\hline
$q_\mathrm{max}$ & $\sqrt{3\mu E}$ & $\infty$ \\
%\hline
$\rho(z)$ & $2e^{-2z/L}\theta(z)/L$ & $\theta(L-z)\theta(z)/L$ \\
%\hline
$L/\lambda$ & $L\rho\sigma = L\rho\times9\pi\alpha_s^2/(2\mu^2)$ & $1$ \\
%\hline
$\alpha_s$ & $0.3$ & $1/3$ \\
%\hline
$m_g$; $M_q$ & $\mu/\surd2$; $\mu/2$ & $0$; $0$\\
\hline
\end{tabular}
\caption{\label{WHDG:table} Table of differences between the implementations of \eqtwo{WHDG:DGLV}{WHDG:ASW} in WHDG and in ASW--SH.
}
\end{table}

\begin{figure}[!htb]
\centering
\includegraphics[width=0.9\columnwidth]{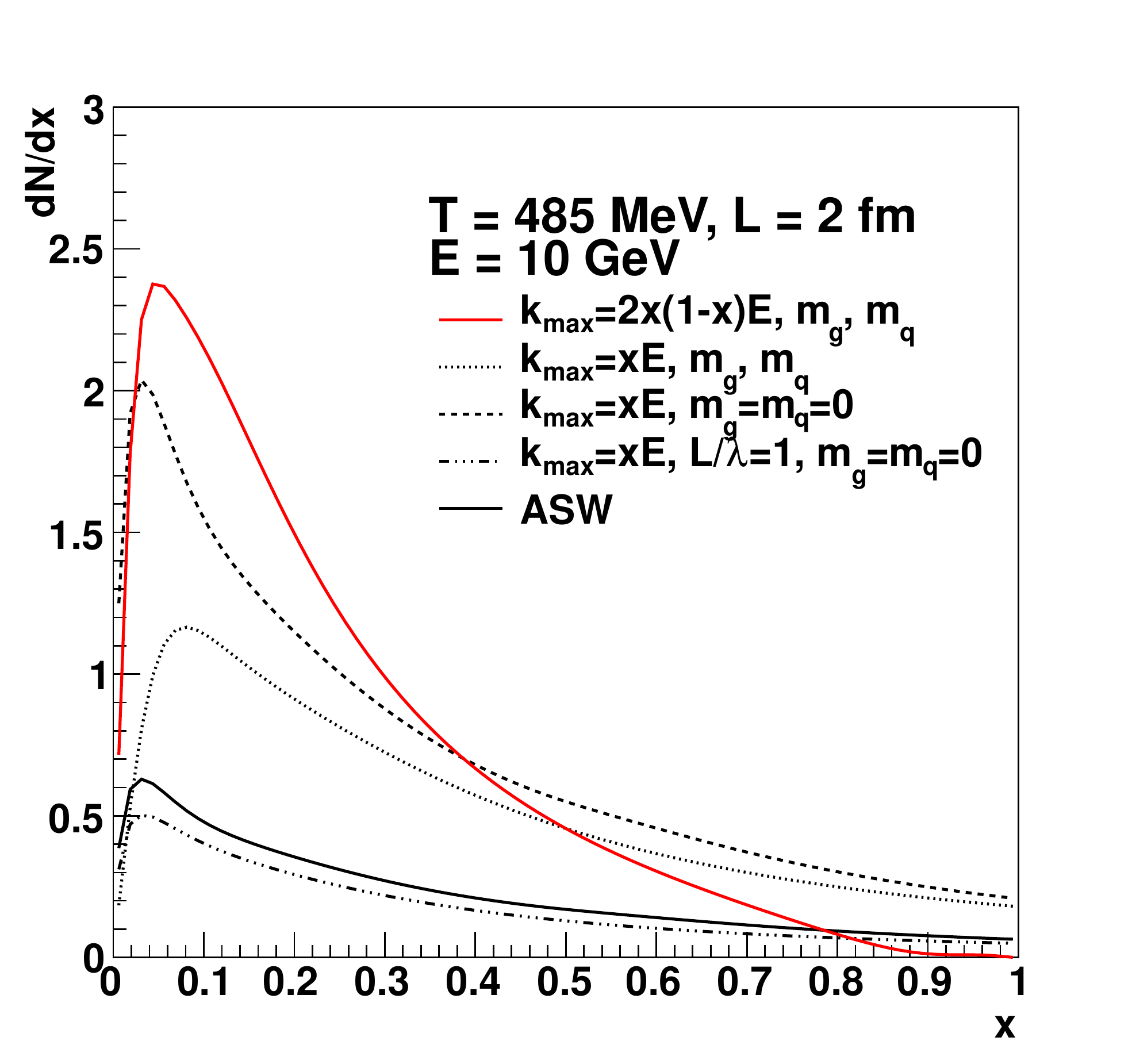}
\caption{\label{WHDG:comp_all}Comparison of the single gluon emission
spectrum for several variants of the
$N=1$ opacity expansion, using a uniform medium with $T=485$ MeV and
$L=2$ fm. All curves use $\alpha_S=0.3$ and
$q_\mathrm{max} = 10 \;\mathrm{ GeV} \gg \sqrt{3\mu E}$. The solid curve
labeled ASW corresponds to what is used in \cite{Armesto:2003jh,Salgado:2003gb}, while the grey
(red) curve corresponds to the radiative WHDG formalism used in \cite{Djordjevic:2003zk,Wicks:2005gt}, except for the
difference in $q_\mathrm{max}$. See text for details.}
\end{figure}

The differences outlined in Table \ref{WHDG:table} have been
individually explored to see which choices have a large effect on the
calculated gluon spectra. The effect of non-zero quark mass $M_q$ is
small, because the mass is small compared to the total energy. The
choices for $\alpha_S$ differ by 10\% which directly translates into a
change in the single gluon spectra by the same amount. The shape of
the distribution also depends on $\alpha_S$ through the value of
$\mu$, but this is a small effect. It also turns
out that the choice of $q_{\mathrm{max}}$ has only a small effect,
because of the small radiation probability for $q > \sqrt{3\mu T}$.
Another
small effect is the choice of $\rho(z)$; the difference between the
exponential profile and a uniform density only leads to characteristic
shapes in the spectrum for small $L \lesssim 2$~fm, but does not effect
the total radiation probability much. The effect of the density
profile $\rho(z)$ on the gluon spectrum can be seen in
Fig. \ref{WHDG:comp_all}, comparing the lower solid line, labeled ASW,
which uses $\rho(z) = \Theta(L-z)/L$ and the dot-dot-dashed line using
$\rho(z) = 2\exp(-2z/L)/L$.

The remaining three items in Table \ref{WHDG:table} have significant
effects on the gluon radiation spectrum. The effect of each of these
on the gluon spectrum is show in Fig. \ref{WHDG:comp_all}. $L/\lambda$
enters as a multiplication factor in the single gluon spectrum
(compare dot-dot-dashed and dashed line). This is a significant factor
at $T = 485$ MeV, $\lambda=0.62$ fm. The choice of $L/\lambda=1$ in
the ASW papers \cite{Armesto:2003jh} was made for convenience in
tabulating the quenching weights $P(\Delta E)$, but this calculation
has never been used to determine the medium density. In the rest of
this paper, we will therefore include $L/\lambda$ different from unity
in the calculation. Introduction of a finite gluon mass $m_g$ leads to
a sizable reduction of the radiation at small $x$ where $\omega\approx
m_g$ (compare dashed and dotted lines in Fig. \ref{WHDG:comp_all}). On
the other hand, due to nontrivial interference effects, its effect on
observables such as $R_{AA}$ is actually not very large
\cite{Horowitz:2009eb}.

Finally, the choice of the cut-off $k_{\mathrm max}$ on the
perpendicular component of the gluon momentum has a large effect,
mostly at small $x$, where the difference between the cut-offs is a
factor 2 and the probability of radiation at finite $k_\perp$ is large
(compare dotted line and solid grey (red) line in
Fig. \ref{WHDG:comp_all}). Because this cut-off is also related to
large angle radiation, it has a fundamental impact on our ability to
calculate the radiation spectrum, which we will discuss in more detail
in the following section.

\subsubsection{Definition of the variable $x$}
The upper bound of the transverse gluon momentum $k_\perp$ is
different in the WHDG calculation than in the ASW--SH calculation. At
small $x$, both formalisms implement a cut to ensure forward emission
of the gluon. The difference in the values of the cut-off arises
because in the derivations descending from the work by
GLV\cite{Gyulassy:1999zd,Gyulassy:2000er,Djordjevic:2003zk,Djordjevic:2009cr},
$x$ is taken to be the positive light-cone momentum fraction
$x=x_+=k^+/P^+$, while the ASW--SH derivation
\cite{Wiedemann:2000za,Armesto:2003jh} uses $x=x_{\rm E}=k^0/P^0$, the
fraction of energy carried away by the radiated gluon.

If we
denote the usual four-momenta with parentheses and light-cone momenta
with brackets then the radiated gluon four-momentum is
\ba
\label{WHDG:k}
k &=& \left(x_{\rm E} E, \wv{k}, \sqrt{(x_{\rm E}E)^2-\wv{k}^2}\right) 
\nn \\
&=& \left[x_+ E^+,\frac{\wv{k}^2}{x_+E^+},\wv{k}\right],
\ea
where $\wv{k}$ is the momentum of the gluon transverse to the
direction of the parent parton and the gluon is assumed to be on mass
shell.  

In light-cone coordinates forward emission implies that
$k^+>k^-$.  This condition implies
\be
\label{WHDG:kmdglv}
k_T<k_\mathrm{max}=x_+P^+=x_+E^+.
\ee
For a massless parent parton, $E^+ = 2E$, so that $k_\mathrm{max} = 2x_{+}E$.
In Minkowski coordinates forward emission implies that $k_z>0$.  This condition implies 
\be
\label{WHDG:kmasw}
k_T<k_\mathrm{max}= \omega = x_{\rm E} E.
\ee
Both values of $k_\mathrm{max}$ restrict emission to angles be less
than 90$^\circ$, which is still a rather wide angle. In general, a
cut-off angle $\theta_\mathrm{max}$ can be defined, leading to 
\be
k_\mathrm{max} = \left\{ \begin{array}{ll}
x_+ E^+ \tan (\theta_\mathrm{max}/2),\quad & x=x_+, \\
x_{\rm E} E \sin (\theta_\mathrm{max}),\quad & x=x_{\rm E}.
\end{array}\right.
\ee

Note that the same physical cut-off at an angle $\theta_\mathrm{max}$
gives different expression for $k_\mathrm{max}$ as a function of
$x_E$ and $x_+$. The exact relations between $x_+$ and $x_{\rm E}$ are
\begin{eqnarray}
\label{WHDG:x}
x_+ & = & \frac{1}{2}\,x_{\rm E}\left(1+\sqrt{1-\left(\frac{k_T}{x_{\rm E}E}\right)^2}\,\right), \\
x_{\rm E} & = & x_+\left(1+\left(\frac{k_T}{x_+E^+}\right)^2\right).
\end{eqnarray}
Note that to lowest order in the expansion parameter is $k_T/xE$,
which controls the extent of collinearity, the two definitions of $x$
are identical. The difference between using $x_{E}$ and $x_{+}$ in the
calculations is only significant at larger angles.

\subsubsection{Transverse momentum cut-off at large momentum fraction}

\eqtwo{WHDG:DGLV}{WHDG:ASW} also have support for all values of $x$.
Nonzero weight in $dN_g/dx$ for $x>1$ of course violates
energy-momentum conservation.  Requiring the continued forward
propagation of the parent parton leads to an additional $k_T$ cut-off
that enforces energy-momentum conservation and also restricts
large-angle radiation at large $x$. 

The momentum of a massless parent parton after energy loss is
\ba
\label{WHDG:p}
p &=& \left((1-x_{\rm E})E,\wv{q}-\wv{k},\sqrt{\big((1-x_{\rm E})E\big)^2-\big(\wv{q}-\wv{k}\big)^2}\right) 
\nn \\
&=& \left[(1-x_+)E^+,\frac{(\wv{q}-\wv{k})^2}{(1-x_+)E^+},\wv{q}-\wv{k}\right] ,
\ea
where $\wv{q}$ is the transverse momentum transfer to the parent
parton from the scattering center.  For completeness the original
parent parton momentum is $P=(E,\wv{0},E)=[E^+,0,\wv{0}]$.  

In light-cone coordinates, forward
propagation of the final state parton implies $p^+>p^-$; in Minkowski coordinates it implies
$k_z>0$.  For the light-cone coordinate case forward propagation leads to
\be
\label{WHDG:kmdglvtwo}
(1-x_+)E^+>|\wv{q}-\wv{k}|\approx k_T,
\ee
where $|\wv{q}|\sim3T<q_\mathrm{max}=\sqrt{6ET}$ is small compared to most values of $|\wv{k}|$; for the Minkowski coordinate case forward emission leads to
\be
\label{WHDG:kmaswtwo}
(1-x_{\rm E})E>|\wv{q}-\wv{k}|\approx k_T.
\ee

For each definition of $x$ there are two cut-offs (e.~g.\
\eqtwo{WHDG:kmdglv}{WHDG:kmdglvtwo} for $x \equiv x_+$ or
\eqtwo{WHDG:kmasw}{WHDG:kmaswtwo} for $x \equiv x_{\rm E}$); this
needs to be taken into account when evaluating the $k_T$ integral over
$dN_g/dxdk_T$.  One possibility would be to take the minimum of the
two definitions; for instance, using light-cone coordinates and taking
$\theta_\mathrm{max}=\pi/2$; this would mean
$k_\mathrm{max}=\min(x_+,1-x_+)E^+$.  The present implementations of
DGLV and WHDG use a smoother function, $k_\mathrm{max}=x_+(1-x_+)E^+$.
Note that the existing ASW--SH implementation
\cite{Salgado:2003gb,ASWCode} does not include a large-$x$ cut-off.

\begin{figure}
\centering
\includegraphics[width=0.9\columnwidth]{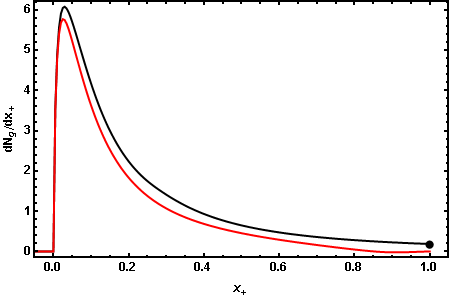}
\caption{\label{WHDG:kmax}
Single gluon spectra using \eq{WHDG:DGLV} with
$k_\mathrm{max}=x_+E^+$ (black) and $k_\mathrm{max}=x_+(1-x_+)E^+$
(red) cut-offs for a 10 GeV up quark traversing a 2 fm static QGP of
$T=485$ MeV and $m_q = m_g = 0$.  The black dot at $x_+=1$ represents the integrated weight
of $dN_g/dx_+$ for $x_+>1$ when $k_\mathrm{max}=x_+E^+$.  }
\end{figure}

Figure \ref{WHDG:kmax} shows the single gluon spectra $dN/dx_g$ with
the two different cut-offs Eq. \ref{WHDG:kmdglv} and
Eq. \ref{WHDG:kmdglvtwo}. It can be seen in the Figure that the effect
of introducing the kinematic cut-off $x \le 1$ is small in regimes
where the $dN_g/dx$ peak is large and at small $x$, for instance for
$L = 2$ fm, with the effective gluon mass set to zero, $m_g = 0$.  In
this sense the $dN_g/dxdk_Tdq_T$ integrand respects the small-$x$
approximation rather well.  In Fig. \ref{WHDG:comp_all} the effect of
lifting the $x \le 1$ kinematic cut-off is more important because the
number of radiated gluons $\langle N_g \rangle$ is smaller when $m_g
\ne 0$; in this case the large-$x$ cut-off causes a 10-20\% change in
$\langle N_g \rangle$. In general, the effect of the large-$x$ cut-off
is expected to increase with increasing path length and to decrease with
increasing parton energy.

\subsubsection{Effect of large angle radiation} 
\begin{figure}
\centering
\includegraphics[width=0.9\columnwidth]{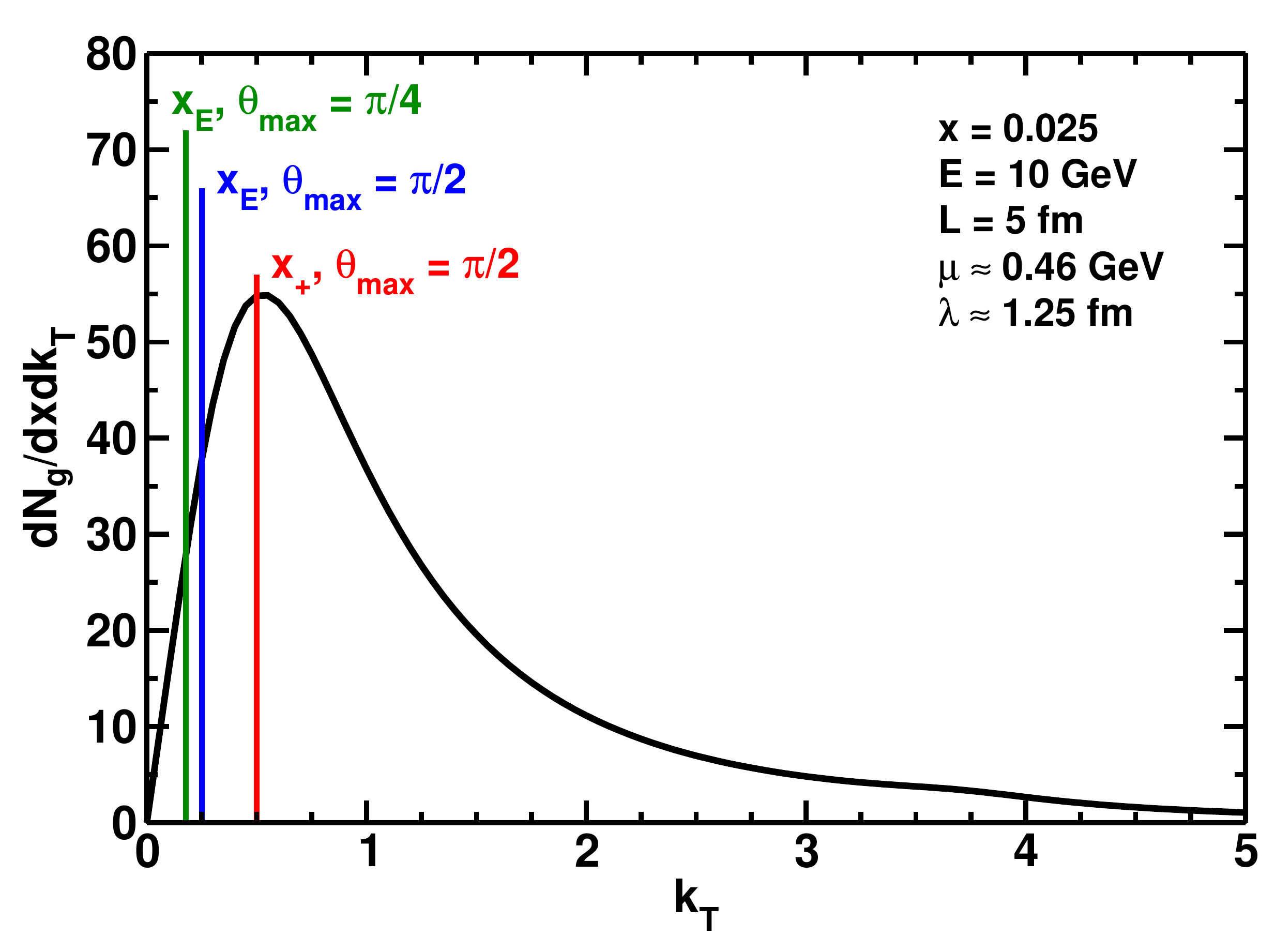}
\caption{\label{WHDG:dndxdk} Transverse momentum spectrum $dN_g/dxdk_T$ of emitted gluons with $x=0.025$, calculated using \eq{WHDG:DGLV} for a light quark with all
masses set to 0, $E=10$~GeV, $L=5$~fm, and representative values of
$\mu\approx 0.46$~GeV and $\lambda\simeq1.25$~fm for a medium density
of $dN_g/dy = 1000$ similar to RHIC conditions \cite{Wicks:2005gt}.
Vertical lines depict the three values of $k_T$ discussed in the text
as possible cut-offs to enforce collinearity in \eq{WHDG:DGLV}.}
\end{figure}

In order to illustrate the sensitivity of the single gluon emission
probability to $k_{\rm max}$, \fig{WHDG:dndxdk} plots $dN_g/dxdk_T$
for $x=0.025$, along with an illustration of three possible cut-offs
for $k_T$:
\begin{itemize}
\setlength{\itemsep}{0pt}
\item[(a)] $x=x_+$ with $\theta_\mathrm{max}=\pi/2$; 
\item[(b)] $x=x_{\rm E}$ with $\theta_\mathrm{max}=\pi/2$;
\item[(c)] $x=x_{\rm E}$ with $\theta_\mathrm{max}=\pi/4$.
\end{itemize}  
Recall that to lowest order in collinearity, the first two cut-offs
are identical; the third is a natural $\mathcal{O}(1)$ variation in
the cut-off that one can use to estimate the systematic uncertainties
deriving from the collinear approximation.  Clearly the assumption of
collinearity is badly violated: for values of $x\sim\mu/E$,
$dN_g/dxdk_T$ reaches its maximum value at $k_T\sim xE$.  For these
values of $x$ the emission spectrum is highly sensitive to the choice
of $k_\mathrm{max}$: $dN_g/dx\sim k_\mathrm{max}^2$.

\begin{figure}
\centering
\includegraphics[width=0.9\columnwidth]{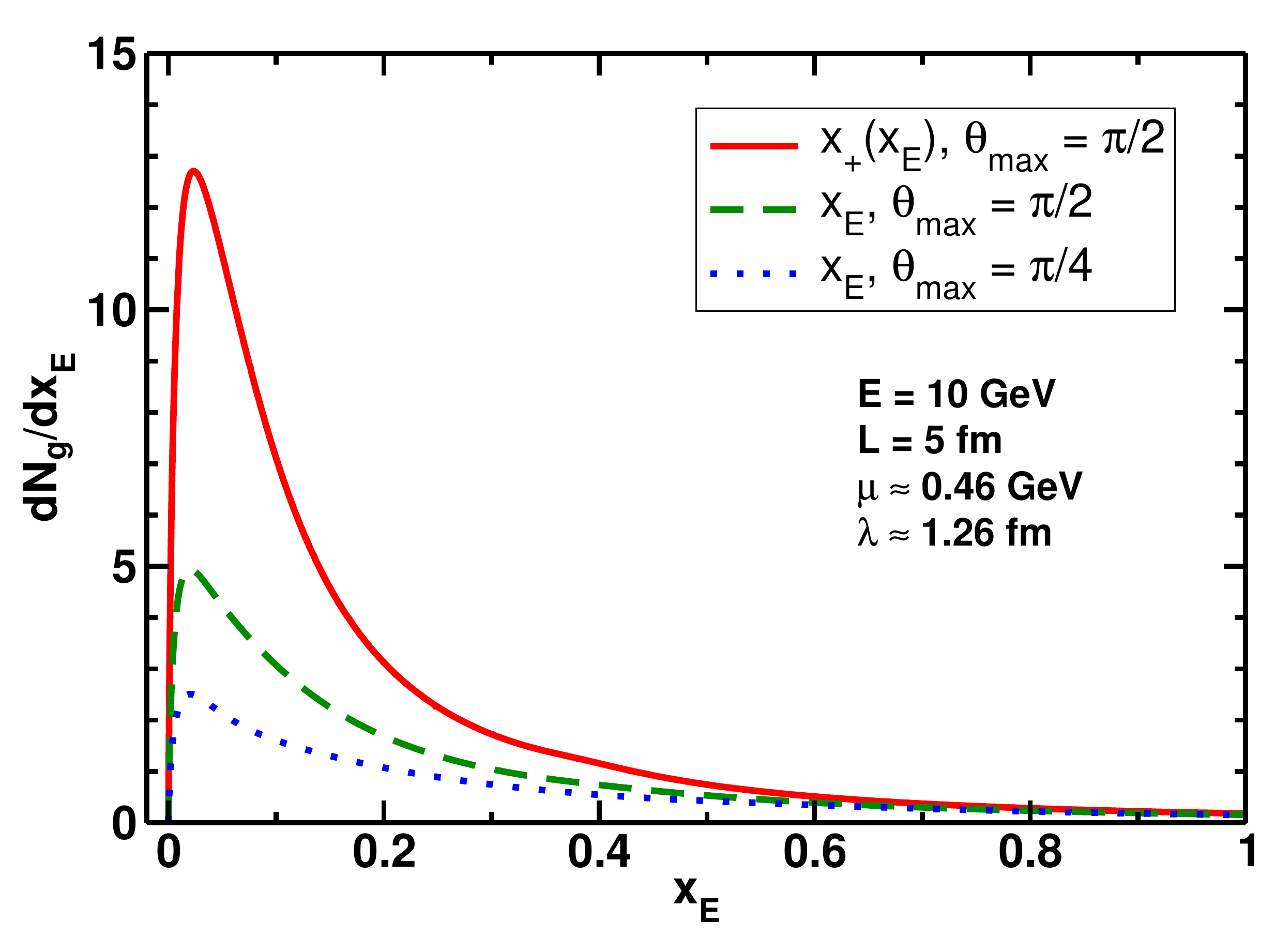}
\caption{\label{WHDG:apples}
Comparison of \eqtwo{WHDG:DGLV}{WHDG:ASW} in the
massless limit and for which the $x_+$ dependence of \eq{WHDG:DGLV}
has been transformed into $x_{\rm E}$; see \eq{WHDG:dndxpxe}.  Also
shown is the result when using the $x_{\rm E}$ interpretation and
reducing $\theta_\mathrm{max}$ to $\pi/4$, a reasonable
$\mathcal{O}(1)$ variation in the $k_T$ cut-off.}
\end{figure}

Since the collinear approximation is so badly broken, it is not a good
approximation to take $x_+\approx x_{\rm E}$.  A meaningful comparison
of results, then, can come only when the emission spectra of
\eqtwo{WHDG:DGLV}{WHDG:ASW} are plotted with respect to the same
variables.  Since one is interested in a differential quantity, a
Jacobian is required.  We choose to transform $x_+$ to $x_{\rm E}$
because, ultimately, one is interested in energy loss, as opposed to
the loss of positive light-cone momentum.  The transformed spectrum is
then given by
\be
\label{WHDG:dndxpxe}
\frac{dN^{\rm J}_g}{dx_{\rm E}}(x_{\rm E}) = \int^{k_{\rm max}} dk_T \frac{dx_+}{dx_{\rm E}}\frac{dN_g}{dx_+dk_T}\big(x_+(x_{\rm E})\big),
\ee
with
\ba
\frac{dx_+}{dx_{\rm E}} & = & \frac{1}{2}\left[1+\left(1-\Big(\frac{k_T}{x_{\rm E}E}\Big)^2\right)^{-1}\right] ,
\label{eq:jacobian}
\\
k_{\rm max} &=& x_{\rm E} E\sin(\theta_\mathrm{max}) .
\label{eq:kmax}
\ea
Note the change in the upper limit (\ref{eq:kmax}) of integration in
\eq{WHDG:dndxpxe}.  The resulting comparison of $dN_g/dx_{\rm E}$ is
shown in \fig{WHDG:apples}.  Note the very large difference in the
results for the two collinearly equivalent definitions of $x$ and that
for the result with a reduced $\theta_\mathrm{max}$.  Of course this
enormous difference implies a large (factor
$2-3$) uncertainty in the extraction of the medium parameters from leading hadron
suppression data \cite{Horowitz:2009eb}.

%
%             AMY BMDPS Section
%
\subsection{AMY, BDMPS--Z and ASW--MS}
\label{sec:AMY}

\subsubsection{AMY transport equations}

The medium-induced radiative energy loss suffered by high energy
partons passing through nuclear matter was first computed in the BDMPS--Z
approach \cite{Baier:1996kr, Baier:1996sk, Baier:1998kq,
Zakharov:1996fv, Zakharov:1997uu}, in which the gluon emission
probability is expressed in terms of the Green's function of a 2-D
Schr\"odinger equation with an imaginary potential proportional to the interaction
cross section with color centers of
a quark-antiquark-gluon system.

In the AMY approach \cite{Arnold:2001ms, Arnold:2001ba, Arnold:2002ja,
Jeon:2003gi}, the gluon emission rates are calculated fully at leading
order in $\alpha_s$ by resumming an infinite number of ladder
diagrams in the context of hard thermal loop resummed QCD. Both
approaches are valid in the multiple soft scattering limit, but differ
in several essential ways: in AMY the medium consists of fully dynamic
thermal quarks and gluons, while in BDMPS--Z the medium is treated as
a collection of static scattering centers. In BDMPS--Z the gluon
emission probability is calculated in configuration space while in AMY
the radiation rate is calculated in momentum space. Salgado and
Wiedemann \cite{Salgado:2003gb} further extended the BDMPS--Z
formalism to include the correct thin plasma limit, which has important
quantitative effects. In addition, different evolution schemes are
used for multiple-gluon emission: the AMY formalism uses rate
equations to obtain the final parton distributions while applications
of the BDMPS--Z calculation convolute the radiation rate with a
Poisson distribution to obtain the quenching weights.

\begin{figure}
\centering
\includegraphics[width=0.45\textwidth]{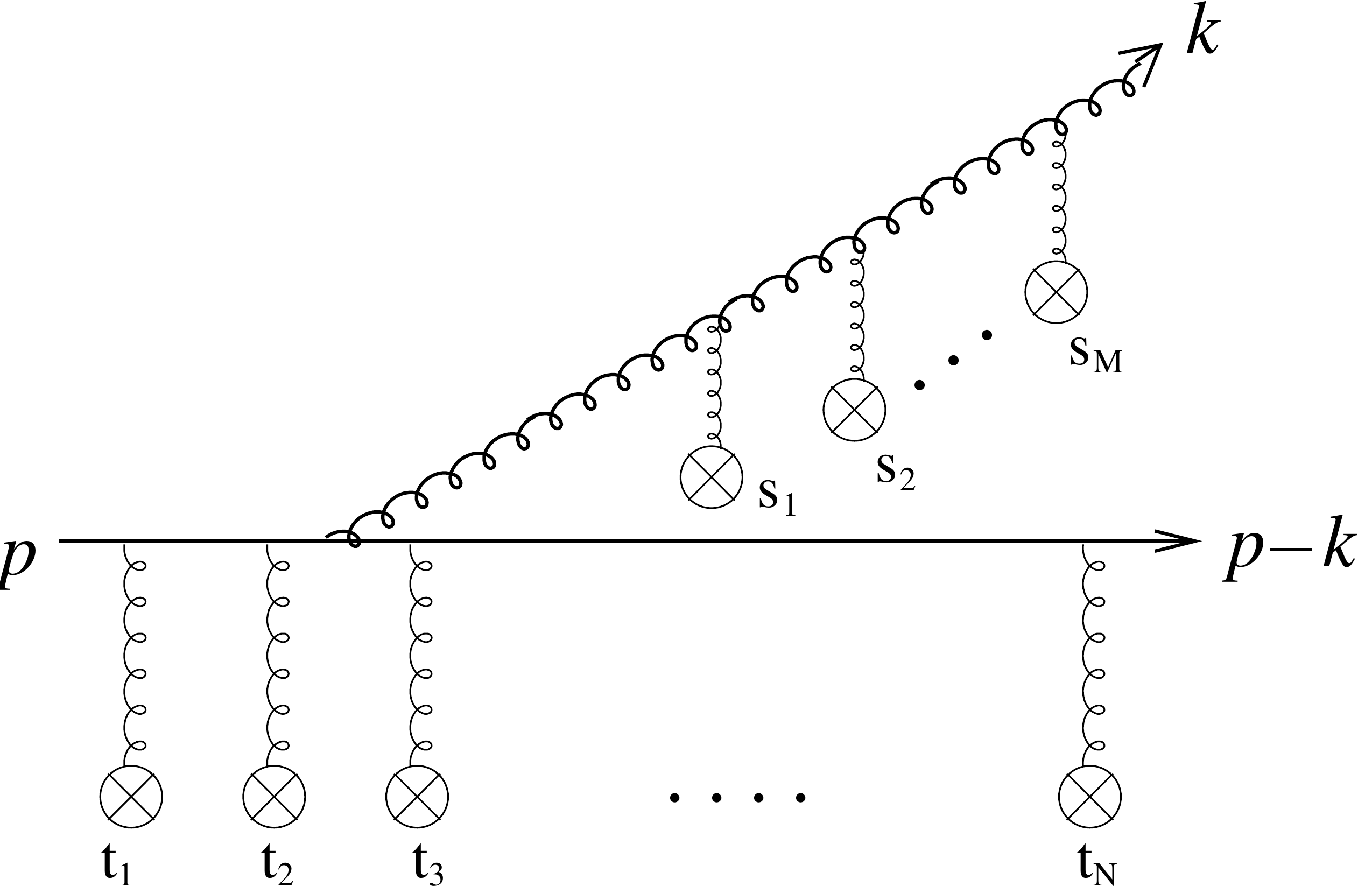}
\caption{ A typical diagram calculated in the AMY and BDMPS--Z approaches. } 
\label{typical_diagram}
\end{figure}

The main assumption in these two formalisms is that the temperature of
the medium is high enough such that the asymptotic freedom of QCD
makes it possible to treat the interactions between a fast parton and
the medium using perturbation theory. In this case, soft exchanges
between the medium and the propagating parton dominate the stimulated
radiation of a hard collinear gluon. At the same time, the effect of
multiple collisions is reduced due to the coherence between multiple
soft scatterings within the formation time of the emitted gluon (LPM
effect). This effect makes it necessary to resum all diagrams as
depicted in Fig. \ref{typical_diagram} to calculate the leading order
gluon emission probability/rate.

In the AMY approach, one considers a hard parton traversing an
extended medium in thermal equilibrium with asymptotically high
temperature $T\rightarrow \infty$.  Due to the small coupling
$g\rightarrow 0$, a hierarchy of parametrically separated scales $T >
gT > g^2T$ makes it possible to construct an effective field theory of
soft modes (modes with momentum $|{\bf k}| \sim gT$) by summing
contributions from hard thermal loops into effective propagators and
vertices \cite{Braaten:1989kk, Braaten:1989mz}. The hard parton
traversing a thermal QGP undergoes a series of soft elastic
scatterings with transverse momentum of order $\sim gT$ off the
thermal particles of the medium. The differential cross section
(interaction rate) at leading order in $\alpha_s$ is
\begin{eqnarray}
\label{diff_rate}
\frac{d\bar{\Gamma}_{\rm el}}{d^2q_\perp} = \frac{1}{(2\pi)^2} \frac{g^2 T m_{\rm D}^2 }{ {\bf q}_\perp^2 ({\bf q}_\perp^2 + m_{\rm D}^2)}\,.
\end{eqnarray}
Note that the rate has been divided by the quadratic color Casimir
$C_R$ of the parent parton, indicated by placing a bar over the rate
$\Gamma_{\rm el}$ (similarly for other quantities).

\begin{figure*}
\centering
\includegraphics[width=0.9\textwidth]{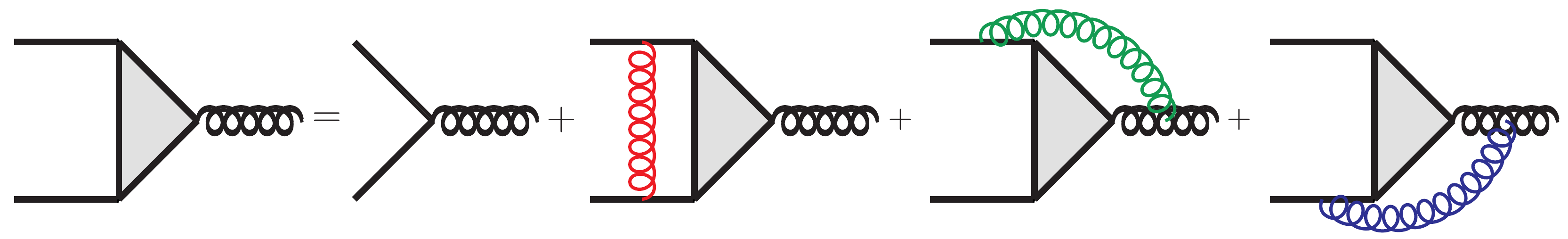}
\caption{ Diagrammatic representation of the Schwinger-Dyson equation for the emission vertices. }
\label{linear_integral_equation_gluon}
\end{figure*}
These soft multiple scatterings induce collinear splitting of
partons. The time scale over which the parton and emitted gluon
overlap is of order $\sqrt{\omega/\hat{q}}$,
which is of greater or equal 
order of magnitude than the mean free time of soft scatterings for $\omega\geq T$, with $\omega$ the energy of the radiated gluon.
To obtain the leading-order gluon
emission rates, one must consistently take into account the multiple
scattering processes. Within the thermal field theory, one essentially
calculates the imaginary parts of an infinite number of gluon
self-energy ladder diagrams. The resummation of these ladder diagrams
can be organized into a Schwinger-Dyson type equation for the
dressed radiation vertex, as depicted in
Fig. \ref{linear_integral_equation_gluon}. The corresponding integral
equation is
\begin{eqnarray}
2{\bf h} &=& i\, \delta E({\bf h},p,k) {\bf F}({\bf h}) + \int {d^2
q_\perp} \frac{d\bar{\Gamma}_{\rm el}}{d^2q_\perp}
\nn \\
& \times &
\Big[ (C_s-C_{\rm A}/2)[{\bf F}({\bf h})-{\bf F}({\bf h}{-}k\,{\bf q}_\perp)]
\nonumber\\ 
&& 
+ (C_{\rm A}/2)[{\bf F}({\bf h})-{\bf F}({\bf h}{+}p\,{\bf q}_\perp)]
\nn \\
&&  
+ (C_{\rm A}/2)[{\bf F}({\bf h})-{\bf F}({\bf h}{-}(p{-}k)\,
 {\bf q}_\perp)] \Big] \,.
\label{linear_int_eq}
\end{eqnarray}
In the above equation, $\delta E$ is the energy difference between the
initial and final states,
\begin{eqnarray}
\label{delta_E}
\delta E({\bf h},p,k)  &=& E_{p-k} + E_k - E_p 
\\
&=& \frac{{\bf h}^2}{2pk(p{-}k)} + \frac{m_k^2}{2k} +
     \frac{m_{p{-}k}^2}{2(p{-}k)} - \frac{m_p^2}{2p} ,
\nn 
\end{eqnarray}
with $p$ the incoming parton momentum and $k$ the momentum of the
radiated hard gluon. $1/\delta E$ is of order of the formation time
for the bremsstrahlung process in the medium. The masses appearing in
the above equations are the medium induced thermal masses. The 2-D
vector $\bf h$ is a measure of non-collinearity of the final states,
defined to be ${\bf h} \equiv ({\bf p} \times {\bf k}) \times {\bf
e}_{||}$, with ${\bf e}_{||}$ the unit vector along the chosen longitudinal direction; here
${\bf h}$ is parametrically of order $g T E (\omega/T)^{1/4}$ and therefore
small compared to ${\bf p}\cdot {\bf k}$, with $\omega\sim k$. For the case of $g\to
q\bar{q}$, the $(C_s-C_A/2)$ term is the one with ${\bf F}({\bf
h}{-}p\,{\bf q}_\perp)$ rather than ${\bf F}({\bf h}{-}k\,{\bf
q}_\perp)$ in the above equation.

The gluon emission rate ${d\Gamma(p,k)}/{dk}$ is obtained by closing
off the dressed vertex with the bare vertex including the appropriate
statistical factors,
\begin{widetext}
\begin{eqnarray}
\label{eq:dGamma}
\frac{d\Gamma(p,k)}{dk} & = & \frac{C_s g^2}{16\pi p^7}
     \frac{1}{1 \pm e^{-k/T}} \frac{1}{1 \pm e^{-(p-k)/T}}
% \nonumber \\ &&
\times \left\{ \begin{array}{cc}
     \frac{1+(1{-}x)^2}{x^3(1{-}x)^2} & q \rightarrow qg \\
     N_{\rm f} \frac{x^2+(1{-}x)^2}{x^2(1{-}x)^2} & g \rightarrow q\bar{q} \\
     \frac{1+x^4+(1{-}x)^4}{x^3(1{-}x)^3} & g \rightarrow gg \\
     \end{array} \right\}
% \nonumber \\ &&
\times \int \frac{d^2 {h}}{(2\pi)^2} 2 {\bf h} \cdot {\rm Re}\: {\bf F}({\bf h},p,k) \, .
\end{eqnarray}
Here $C_s$ is the quadratic Casimir relevant for the process ($C_F =
4/3$ for quarks or $C_A=3$ for gluons), and $x\equiv k/p$ is the
momentum fraction of the gluon (or the quark, for the case $g
\rightarrow q\bar{q}$). 
The subsequent multiple emissions are treated by evolving the momentum distribution $P(p) = dN/dp$ of the traversing hard partons in a set of rate equations,
\begin{eqnarray}
\frac{dP_{q\bar{q}} (p)}{dt} &=& \int_k
     P_{q\bar{q}} (p{+}k) \frac{d\Gamma^q_{\!qg}(p{+}k,k)}{dk}
     -P_{q\bar{q}} (p)\frac{d\Gamma^q_{\!qg}(p,k)}{dk}
     +2 P_{g} (p{+}k)\frac{d\Gamma^g_{\!q \bar q}(p{+}k,k)}{dk} \, , 
\\ 
\frac{dP_{g} (p)}{dt} &=& \int_k 
     P_{q\bar{q}} (p{+}k) \frac{d\Gamma^q_{\!qg}(p{+}k,p)}{dk}
     {+}P_{g} (p{+}k)\frac{d\Gamma^g_{\!\!gg}(p{+}k,k)}{dk}
     -P_{g} (p) \left(\frac{d\Gamma^g_{\!q \bar q}(p,k)}{dk} 
     + \frac{d\Gamma^g_{\!\!gg}(p,k)}{dk} \Theta(k{-}p/2) \!\!\right) .
\label{eq:fokker-planck}
\end{eqnarray}
%\end{widetext}
In the AMY approach, the medium consists of fully dynamic thermal
quarks and gluons; the effect of emitting and absorbing thermal energy
are fully included in the evolution equations. In addition, the
elastic energy loss due to the recoil of the dynamical scattering
centers can also be consistently included in the formalism
\cite{Qin:2007rn,Schenke:2009ik}. However, the transition rates are
calculated in momentum space assuming the thermodynamic limit, i.e.,
the high energy parton experiences a uniform medium on the time scale
of the formation time of the emitted radiation.

\subsubsection{Comparison of BDMPS--Z and AMY}

In the BDMPS--Z formalism, one calculates the amplitude for a quark-antiquark-gluon system evolving in the medium without inducing inelastic reactions (following Zakharov). The general formula for the probability of the gluon emission by a hard parton traversing the medium may be expressed as,
%\begin{widetext}
\begin{eqnarray}
\label{double_t}
k \frac{dI}{dk} = \frac{\alpha  xP_{s\to g}(x)}{[x(1-x)p]^2} \mathrm{Re}\int_0^\infty dt_1 \int_{t_1}^\infty dt_2
\Big(\nabla_{{\bf b}_1} \cdot \nabla_{{\bf b}_2}[G({\bf b}_2, t_2|{\bf b}_1, t_1)
- G_\mathrm{vac}({\bf b}_2, t_2|{\bf b}_1, t_1)]\Big)_{{\bf b}_1={\bf b}_2=0}\,.
\end{eqnarray}
\end{widetext}
Here $I$ is the probability of gluon bremsstrahlung from the high energy particle, $P_{s\to g}(x)$ is the vacuum splitting function for relevant process and $G({\bf b}_2, t_2|{\bf b}_1, t_1)$ is the Green's function of a 2-D Schr\"odinger equation with Hamiltonian
\begin{eqnarray}
H({\bf p}_{\bf b}, {\bf b}, t) = \delta E({\bf p}_{\bf b}) - i \Gamma_3({\bf b}, t)\,.
\end{eqnarray}
The initial condition for the Green's function is
\begin{eqnarray}
G({\bf b}_2, t|{\bf b}_1, t) = \delta^2({\bf b}_2 - {\bf b}_1).
\end{eqnarray}
The kinetic term in the Hamiltonian describes the energy difference between initial and final states, as given by
Eq.(\ref{delta_E}), with ${\bf p}_{\bf b} = {\bf h}/p$, and $\Gamma_3$ in the potential term is the 3-body interaction
rate,
\begin{eqnarray}
\Gamma_3({\bf b},t) &=& \frac{1}{2}C_A \Bar{\Gamma}_2({\bf b},t) + (C_s - \frac{1}{2}C_A)\Bar{\Gamma}_2(x{\bf b},t) 
\nn \\
&& + \frac{1}{2}C_A\Bar{\Gamma}_2((1-x){\bf b},t)\,,
\end{eqnarray}
where $\Gamma_2$ is related to the Fourier transform of the elastic collision rate $d\bar{\Gamma}_{\rm el}/d^2q_\perp$
by
\begin{eqnarray}
\Bar{\Gamma}_2({\bf b},t) = \int d^2q_\perp \frac{d\Bar{\Gamma}_{\rm el}} {d^2q_\perp} \left(1 - e^{i{\bf b}\cdot{\bf
q}_\perp}\right)\,.
\end{eqnarray}
Note that in the original BDMPS--Z formula, the rate $\Bar{\Gamma}_{\rm el}$ for soft scatterings was written as the number density $\rho$ of the static scattering centers in the medium times elastic cross section $\sigma_{\rm el}$ for such scatterings. Here we follow the notation of Arnold \cite{Arnold:2008iy} and write BDMPS--Z formulas more generally in terms of the rate $\Bar{\Gamma}_{\rm el}$ for elastic scatterings off the medium constituents. Since the medium is treated as a collection of static scattering centers, one replaces the denominator of Eq.~(\ref{diff_rate}) with $({\bf q}_\perp^2 + m_D^2)^2$ for the elastic scattering rate. Further assuming that the gluon emission rate is dominated by the region $b \lesssim 1/m_D$, one may approximate the elastic rate by
\begin{eqnarray}
\Bar{\Gamma}_2({\bf b}, t) = \frac{1}{4} \hat{\bar{q}} b^2\,,
\end{eqnarray}
where $C_{R}\hat{\bar{q}}$ is transverse momentum squared transferred to incident parton per unit time ($C_R$ is the appropriate color factor for the parton). Neglecting the effective masses of the particles, the Hamiltonian takes the Harmonic oscillator form,
\begin{eqnarray}\label{hamiltonian}
H({\bf p}_{\bf b}, {\bf b}, t) = \frac{p_b^2}{2M} + \frac{1}{2} M \omega_0^2 b^2\,,
\end{eqnarray}
where
\begin{eqnarray}
\omega_0^2 = -i \frac{(1-x)C_A + x^2 C_s}{2M} \hat{\bar{q}}\,,
\end{eqnarray}
and $M=x(1-x)p$. Making use of the oscillator Green's function and performing the integration over time $t_1$ and $t_2$ for a brick of medium with length $L$, one may obtain the BDMPS--Z formula for multiple-soft scattering,
\begin{eqnarray}
\label{eq:rate-BDMPS}
k\frac{dI}{dk} = \frac{\alpha   x P_{s\to g}(x)}{\pi} \ln|\cos(\omega_0 L)|\,.
\end{eqnarray}

The correspondence between Eq. \ref{double_t}, which is the starting point to
derive the BDMPS--Z formula above, and Eq. \ref{eq:dGamma}, the rates
calculated by AMY, has been explored by Aurenche and Zakharov for photon radiation \cite{Aurenche:2006tg} and by Arnold for gluon radiation
\cite{Arnold:2008iy}. Here we reproduce the main steps. First, note
that the Green's function only depends on the time difference $\Delta t
= t_2 - t_1$ due to time invariance. Performing the integral over
$t_1$ which gives a factor of total time, the resulting non-vacuum
part of Eq.(\ref{double_t}) becomes
\begin{eqnarray}
\label{eq:rate-med}
\frac{d\Gamma}{dk} &=& \frac{\alpha  P_{s\to g}(x)}{x^2(1-x)^2p^3} 
\\
&\times & \mathrm{Re} \int_{0}^\infty d\Delta t\,
\left[\nabla_{{\bf b}_1} \cdot \nabla_{{\bf b}_2} G({\bf b}_2, \Delta t|{\bf b}_1, 0)\right]_{{\bf b}_1={\bf b}_2=0} .
\nn
\end{eqnarray}
One may define the time-integrated amplitude
\begin{eqnarray}
\label{eq:f(b)}
{\bf f}({\bf b}) = 2i \int_0^\infty dt [\nabla_{{\bf b}_1} G({\bf b},t|{\bf b}_1, 0)]_{{\bf b}_1=0}\,,
\end{eqnarray}
which satisfies the following equation,
\begin{eqnarray}
\label{int_fb}
-2 \nabla_{\bf b} \delta^{2}({\bf b}) = H {\bf f}({\bf b})\,.
\end{eqnarray}
In terms of the amplitude ${\bf f}({\bf b})$, the gluon emission rate becomes
\begin{eqnarray}
\label{rate}
\frac{d\Gamma}{dk} = \frac{\alpha  P_{s\to g}(x)}{x^2(1-x)^2p^3} {\rm Re}[ (2i)^{-1} {\nabla}_{\bf b} \cdot {\bf f}({\bf
b})]_{{\bf b}=0}\,.
\end{eqnarray}
Fourier transforming Eq.~(\ref{int_fb}) from impact parameter space to momentum space with the use of
Eq.(\ref{hamiltonian}), one obtains the linear integral equation [Eq.\ref{linear_int_eq}], with ${\bf F}({\bf h}) = p {\bf f}({\bf p}_{\bf b})$. The gluon emission rate becomes
\begin{eqnarray}
\label{eq:rate2}
\frac{d\Gamma}{dk} = \frac{\alpha P_{s\to g}(x)}{4x^2(1-x)^2p^7} \int \frac{d^2h}{(2\pi)^2} {\rm Re}[ 2{\bf h} \cdot {\bf F}({\bf
h})]\,.
\end{eqnarray}
Including the appropriate final state statistical factors, one
reproduces the AMY formula (\ref{eq:dGamma}) for the gluon emission
rate.

%%%%%%%%%%%%%%%%%%%%%%%%%%%

\begin{figure}
 \begin{center}
   \includegraphics[width=8.5cm]{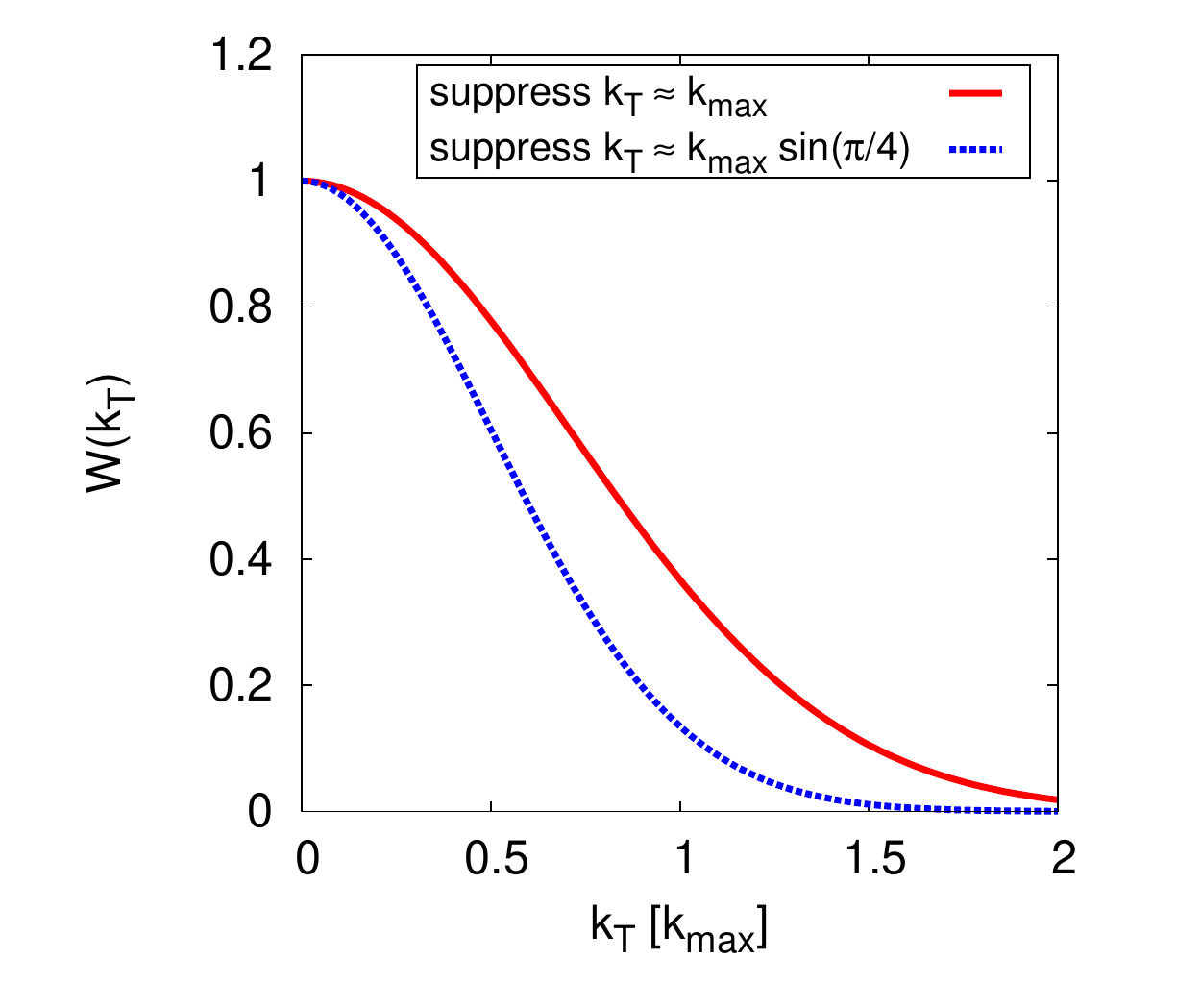}
   \caption{(Color online) Envelope function $W$ for different maximal transverse momenta.}
   \label{fig:envelope}
 \end{center}
\end{figure}

\subsubsection{Large angle radiation in AMY}
\label{sect:AMY_largeangle}
In order to explicitly exclude large angle radiation, which is automatically suppressed for $g\ll 1$ but can contribute for larger $g$, we 
suppress the large angle contributions in the integral over $\mathbf{h}$ in Eq. %(\ref{eq:dGamma}) 
\ref{eq:dGamma}
by introducing an 
envelope function $W(k_\perp,k_{\perp {\rm \max} })$, where $k_\perp=h/p$.

A reasonable choice for $W$ is a Gaussian that cuts off $k_T$ above a value $k_{\perp {\rm \max} }$. We choose a Gaussian because the 
integral equation Eq. \ref{linear_int_eq}
is solved by Fourier transforming into the so called impact parameter space (see \cite{Aurenche:2002pd}), where 
it becomes a differential equation. Hence, we want an envelope function whose Fourier transform is analytic and well behaved.
Explicitly we define
\begin{equation}
 W(k_\perp,k_{\perp {\rm \max} }) = \exp(-k_\perp^2/k_{\perp {\rm \max} }^2)\,,
\end{equation}
where we can vary $k_{\perp {\rm \max} }$, depending on how much we want to restrict the emission angle.
First, we would like to exclude the $k_\perp$ that are kinematically disallowed, so we choose the following $k_{\perp {\rm \max} }$:
\begin{align}\label{eq:amycutoff}
 k_{\perp {\rm \max} }^2 = \left\{ \begin{array}{ll}
       k^2 & ~\text{for}~ |p|>|k| ~\text{and}~ |p-k|>k \\
       (p-k)^2 & ~\text{for}~ |p|>|k| ~\text{and}~ |p-k|<k \\
       (p-k)^2 & ~\text{for}~ |p|<|k| ~\text{and}~ |p|>|p-k| \\
       p^2 & ~\text{for}~ |p|<|k| ~\text{and}~ |p|<|p-k| 
       \end{array} \right.
\end{align}

This suppresses most of the kinematically disallowed $k_\perp$ and makes sure that $k_\perp$ is 
always smaller than the smallest momentum. Note again, that in the limit of $g\ll 1$ those kinematically disallowed
transverse momenta are already suppressed, so that the formalism is perfectly consistent with kinematics in the limit where it is derived.

To also suppress $k_\perp$ which correspond to emission angles larger than a certain $\theta_{\rm \max}$, we
can replace $k_{\perp {\rm \max} }$ by $k_{\perp {\rm \max} } \sin(\theta_{\rm \max})$.
Fig. \ref{fig:envelope} shows the envelope function $W$ for the original $k_{\perp {\rm \max} }$ 
and for a maximal emission angle of $\theta_{\rm \max}=\pi/4$.
Using a Gaussian, not all contributions from $k_\perp>k_{\perp {\rm \max} }$ are removed but they are sufficiently suppressed to give 
a good estimate of the effect of removing these contributions.

%% These figures belong to the next sections. Try to force placement...

\begin{figure*}
   \includegraphics[width=0.49\textwidth]{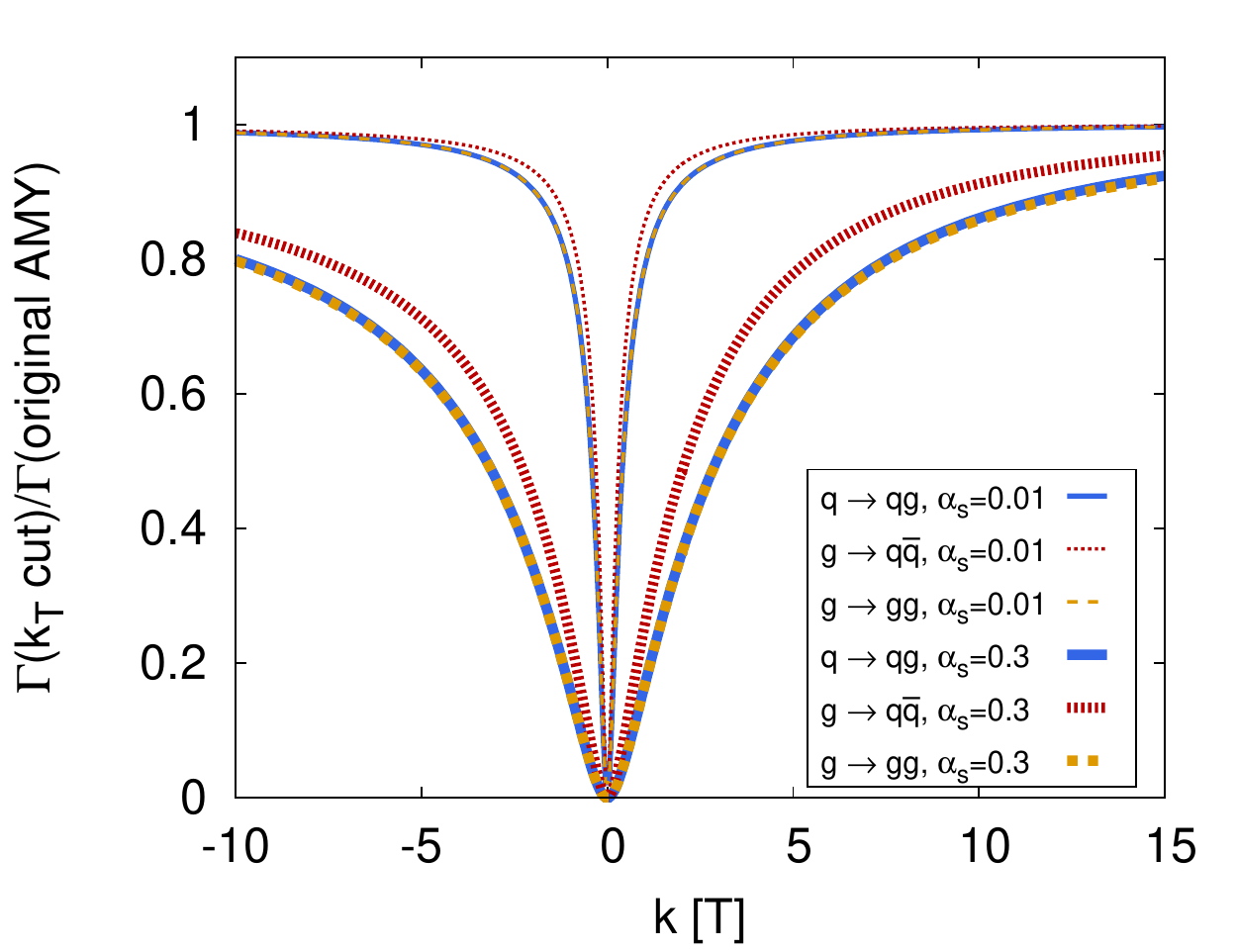}
   \includegraphics[width=0.49\textwidth]{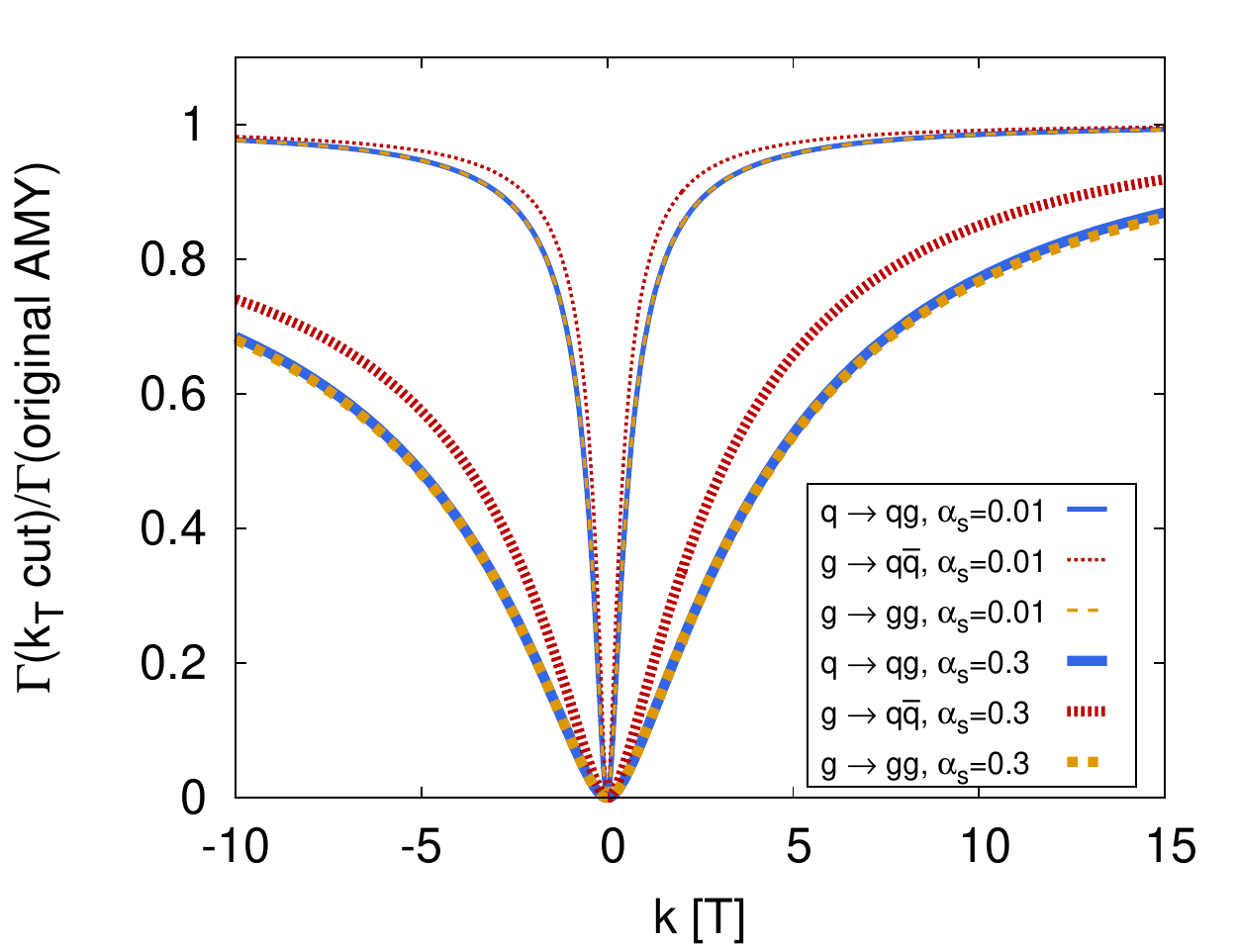}
   \caption{(Color online) Ratio of the rate for the process
   $q\rightarrow qg$ suppressing emission angles larger than $\pi/2$
   (left panel) and $\pi/4$ (right panel) and the original AMY rate
   for two different $\alpha_s$ and $p=40\,T$. The lines for $g \rightarrow gg$ and $q \rightarrow qg$ coincide.}  \label{fig:ratio}
\end{figure*}

%\section{Results}
\label{results}

\subsubsection{Modified rates}

We demonstrate the effect of suppressing large transverse momentum emissions by showing the ratio of the rate including the suppressing
envelope and the original AMY rate in Fig. \ref{fig:ratio} as a function of $k$, the absolute value of
the three-momentum of the emitted gluon. 

For a typical $\alpha_s$ of 0.3, the suppression at small $k$ is significant but decreases for larger $k$.
As expected, the effect is also reduced as $\alpha_s$ decreases, as can be seen in the result for $\alpha_s=0.01$.

\subsubsection{Comparison AMY, BDMPS--Z, ASW--MS in a brick}
\label{sect:AMY_BDMPS_brick}

\begin{figure*}
   \includegraphics[width=\textwidth]{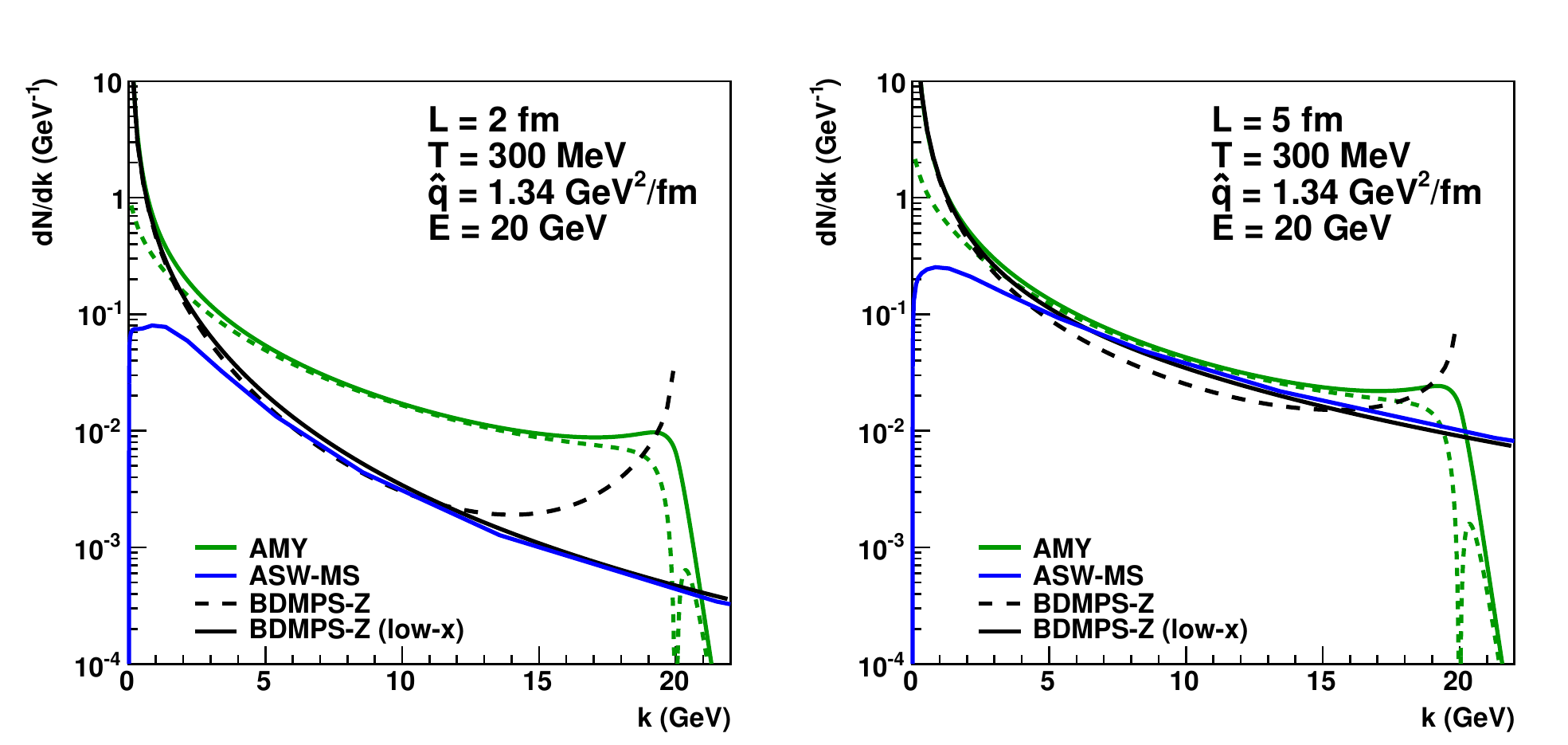}
  \caption{\label{fig:AMY_BDMPS}(Color online) Gluon radiation spectra as a function of
 gluon momentum $k$ for AMY (with and without $k_\perp,\mathrm{max}$
  cut-off), ASW--MS, and the BDMPS--Z multiple soft scattering result Eq. \ref{eq:rate-BDMPS} (with and without small $x$ approximation) for a medium of length $L=2$ fm (left panel) and $L=5$ fm
 (right panel).}
\end{figure*}

Figure \ref{fig:AMY_BDMPS} compares the gluon
emission spectra for AMY, BDMPS--Z and ASW--MS for two path
lengths ($L=2$ and 5 fm) and $T=300$ MeV. 
In the AMY formalism the transport
coefficient $\hat{q}$, which is used to set the medium density in the
BDMPS--Z and ASW--MS calculations, is a derived quantity not a parameter to be set.
For this comparison $\hat{q}$ was calculated using the $q \ll T$ limit of
the HTL scattering rate (Eq. (\ref{diff_rate}), with $q_\mathrm{max}^2=ET$, so that $\hat{q}=6\alpha_s T \mu^2 \ln(\sqrt{ET/\mu^2}))$).

The AMY curve (green line in Fig \ref{fig:AMY_BDMPS}) is obtained by
multiplying the rate $d\Gamma/dk$ Eq. \ref{eq:dGamma} by the medium
length $L$ to obtain a radiation spectrum $dN/dk$. The curve labeled
BDMPS--Z (black dashed) was calculated using Eq. \ref{eq:rate-BDMPS},
while the 'BDMPS--Z (low-x)' curve (solid black) was obtained using in
addition the small-$x$ approximation $x P_{q\rightarrow q}(x) \approx
2$ and $\omega_0 = -i C_A \hat{\bar{q}}/x p$.  Comparing first the AMY
curves with the BDMPS--Z curve in
Fig. \ref{fig:AMY_BDMPS}, one sees that the results are similar for
$L=5$ fm, while for short path lengths ($L=2$ fm), AMY generates more
radiation. This is due to the fact that AMY uses the limit $L \to
\infty$ which ignores the increased effect of finite formation times
at small length \cite{CaronHuot:2008ni}. The ASW--MS calculation (blue
curve in Fig. \ref{fig:AMY_BDMPS}) is based on the BDMPS--Z formalism,
but takes into account finite-length effects. These effects include
vacuuum-medium interference effects, the formation time effect at
small $L$ and the large-angle cut-off $k < k_\perp$, although these
aspects are not easily separated in the calculation
\cite{Wiedemann:2000tf}. The effect of the finite-length corrections
is a pronounced reduction of the radiation at small gluon energies
$k$, which is probably mostly due to the large angle cut-off.

The green dashed curve shows the AMY result with the large-angle
cut-off described in the previous section. This reduces the radiation
at small $k$, as expected, but the reduction is not as pronounced as
in the ASW--MS calculation (compared to BMDPS--Z). The smaller effect
of the cut-off in AMY is likely due to the softer, Gaussian, cut-off
that is used in AMY, compared to a hard cut-off $k_\perp < k$ in
ASW--MS, although it is not excluded that the transverse momentum
spectra are different, which would also affect the cut-off.
The dip in the AMY result around $k=E$ is caused by the cutoff (\ref{eq:amycutoff}) for the case that
a gluon with momentum $k\sim p$ is radiated. In that case the outgoing quark's longitudinal momentum $p-k$ is very small and 
even a small $k_T$ can lead to a large angle for the outgoing quark. Hence, the cutoff becomes relevant and suppresses the rate in this case
as it does for small outgoing gluon momentum.

%
%           Higher Twist (HT) section
%

\subsection{The Higher-Twist (HT) approach}

\label{sect:HT}
In the application of perturbative QCD to hard processes, a
fundamental role is played by the so called factorization
theorems~\cite{Collins:1985ue}. These theorems demonstrate that all
higher order corrections to a particular hard process may be cast in a
form where the hard, short-distance part of a particular process can
be separated order-by-order in coupling from the long-distance, soft
parts of the process as long as one ignores terms which are suppressed
by powers of the hard scale of the process (generically denoted as
$Q^{2}$). In the case of the simplest hard process: The total cross
section in the deep-inelastic scattering (DIS) of an electron on a proton,
these power corrections can be expressed in the simpler language of
the Operator Product Expansion (OPE). In the OPE, the singular product
of operators at closely separated points can be expressed as a series
of terms which contain the product of progressively more complicated
local operators and progressively less singular $c$-functions. The
power suppression (i.e. power of $Q^{2}$ in the denominator) of each
term depends on the ``twist'' of the local operator defined as the
difference of the energy dimensions of the operator and the highest
spin that can be constructed through the product of the various
individual operators in the local product.

So far, factorization has been rigorously proven for the leading twist
part of a set of high $Q^{2}$ processes such as single hadron
inclusive $e^{+} e^{-}$ annihilation, Drell-Yan, DIS, hadron-hadron
collisions etc. Factorization at next-to-leading twist has been proved
for a sub-class of processes, such as the total cross section in
DIS~\cite{Qiu:1990xy}.  As a result, there now exists a formulation of
a class of higher twist corrections to the total cross section in
DIS~\cite{Qiu:1990xxa}.  The higher twist energy loss approach is
based on applying this formalism
%developed in such proofs and in the calculation of higher-twist corrections to the total 
%cross section 
to the problem of calculating the change in the fragmentation function
of a quark produced by the DIS of an electron on a
nucleus~\cite{Guo:2000nz,Wang:2001ifa}. While the initial attempt
focused on computing solely the single scattering correction to the
one gluon emission cross section, this has been subsequently
generalized to include both multiple emissions in a DGLAP like
formalism~\cite{Majumder:2009zu}, as well as multiple
scattering~\cite{Majumder:2009ge}. In the following we will describe
the pertinent details of the HT approach assuming that a quark is
produced in a hard collision at one edge of the QGP brick. We will
first describe the setup where a quark is produced in the DIS of an
electron on a nucleus and then having factorized the initial state and
hard cross section from the final state evolution of the produced
quark, we will consider the change in the evolution of this quark as
it moves through the brick.

\subsubsection{Gluon emission formalism}

In this section, we describe the radiative part of the higher-twist
(HT) calculation as applied to the ``brick problem''. In short, the
higher-twist calculation consists of including a class of medium
corrections to the process of jet evolution in vacuum, brought about
by the multiple scattering of the hard partons in a medium. It is most
straightforwardly derived in the case of single inclusive
deep-inelastic scattering in a large nucleus, with the nucleus playing
the role of the medium. While most scattering corrections are
suppressed by powers of the hard scale $Q^2$, a subset of these are
enhanced by the length of the medium and these are included in the
calculation. Thus, the expansion parameter in the HT approach is
$\alpha_s \hat{q} L /Q^2$ where $\hat{q}$ is the transverse momentum
squared imparted to a single parton per unit length and $L$ is the
length traversed by the parton.

\begin{figure}
%\centering
\includegraphics[width=0.35\textwidth]{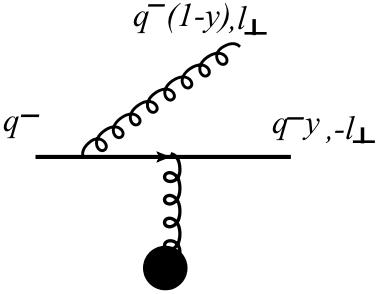}
\caption{A scattering and emission diagram with the corresponding momenta, using the notation convention of the Higher Twist formalism. }
\label{HT-diag}
\end{figure}

Consider the case of deep inelastic scattering (DIS) on a nucleus (in
the Breit frame). The nucleus has a large momentum in the positive
light cone direction $A [p^+,0,0,0]$ with $p^+$ the mean momentum of a
nucleon. The incoming virtual photon has a momentum which may in general
be expressed as $[- Q^2/2q^-,q^-,0,0]$; in the Breit frame $q^- =
Q/\sqrt{2}$ (light-cone coordinates in this section use the convention $p^+=\frac{1}{\sqrt{2}}(p^0+p^3)$ and $p^-=\frac{1}{\sqrt{2}}(p^0-p^3)$). The inclusive cross section to produce a hard hadron,
which carries a momentum fraction $z$ of the initial produced hard
quark may be expressed in a factorized form as
\begin{equation}
\frac{d \sigma }{d z} = \int dx G(x, Q^2) \frac{d \hat{\sigma}}{d Q^2} D(z , Q^2) , \label{diff-cross}
\end{equation}
where $G(x)$ is the parton distribution function (PDF) to obtain a
hard quark in the nucleon with momentum fraction $x$. In the Breit
frame the momentum of the incoming quark is $xp^+ = Q/\sqrt{2}$. Thus
the produced quark has an outgoing momentum $q^- = Q/\sqrt{2}$. The
produced quark is virtual with a virtuality smaller than the hard
scale usually denoted as $\lambda Q$ where $\lambda \ll 1$. The other
two factors are the hard partonic cross section $d \hat{\sigma}/d Q^2$
and the final fragmentation function $D(z, Q^2)$. The scale in the
fragmentation function is the factorization scale and also represents
the maximum possible virtuality of the produced hard jet. The
fragmentation function at the scale $Q$ may be obtained from a
measured fragmentation function at a lower scale using the DGLAP
evolution equations
\begin{equation}
\frac{\partial D_q^h (z, Q^2)}{ \partial \log(Q^2) } = \frac{\alpha_S(Q^2)}{2\pi} \int_z^1 \frac{dy}{y} 
P_{q \to i} (y) D_i^h \left( \frac{z}{y}, Q^2 \right) . 
\label{HT:vac_DGLAP}
\end{equation}
There is an implied sum over flavors $i$ which includes all possible
types of partons that may split off from the hard leading quark
denoted as $q$. The kinematics of the scattering and emission process may be illustrated with the 
diagram in Fig.~\ref{HT-diag}.

In the case of DIS on a large nucleus, the above factorized form may
be assumed to hold with the only change being the replacement of the
vacuum evolved fragmentation function with a medium modified
fragmentation function (as well as a replacement of the nucleon PDF
with a nuclear PDF). The medium modified fragmentation function
contains the usual vacuum evolution piece Eq.~\eqref{HT:vac_DGLAP} and
a medium piece which includes both terms which are interferences
between medium induced radiation and vacuum radiation as well as terms
where both the amplitude and the complex conjugate represent medium
induced radiation.  Once so factorized, the medium modified
fragmentation function can be used to compute the single hadron
inclusive cross section in any process by simply replacing the initial
state parton distribution and hard cross section by those appropriate
for the process in question.

For the Brick Challenge we ignore all the initial state functions and
hard cross sections. We assume that the quark is produced at one edge
of the brick designated as the origin and travels in the negative
light cone direction with a negative light cone momentum $q^-$. We
assign the quark an initial virtuality $Q^2$. Since this is not the
Breit frame in DIS there is no implied relation between $q^-$ and
$Q^2$. The equation for the medium modified fragmentation function
with an initial light-cone momentum $q^-$ and virtuality $Q^2$, which
starts at the location $\zeta_i^-$ and exits at $\zeta_f^-$, is given
as
\begin{widetext}
\be
\frac{\partial {D_q^h}(z,Q^2\!\!,q^-)|_{\zeta_i}^{\zeta_f}}{\partial \log(Q^2)}  = \frac{\alpha_S}{2\pi} \int\limits_z^1 \frac{dy}{y} 
\int\limits_{\zeta_i}^{\zeta_f} d\zeta {P}(y) K_{q^-,Q^2} ( y,\zeta) 
{D_q^h}\left. \left(\frac{z}{y},Q^2\!\!,q^-y\right) \right|_{\zeta}^{\zeta_f}.  \label{HT:in_medium_evol_eqn}
\ee
In the equation above, we have dropped the light-cone ($-$)
superscript on the positions. Note that the medium modified
fragmentation function is now not only a function of $Q^2$ and $z$
but also of $q^-$ and $\zeta$. The calculation of the
evolution equation then requires the evolution of a three dimensional
matrix (in $z,q^-,\zeta$). The medium kernel $K_{q^-, Q^2} ( y, \zeta
)$ for a quark jet is given as~\cite{Majumder:2009ge}
\be
K_{q^-, Q^2} ( y, \zeta ) = \frac{ \left(  \hat{q}(\zeta) -  (1-y) \frac{\hat{q}}{2} + (1-y)^2 \hat{q}_Q  \right)}{Q^{2}}
\left[ 2 - 2 \cos\left( \frac{Q^2 (\zeta - \zeta_i)}{ 2 q^- y (1- y)} \right)  \right] .
\label{eq:HT_kernel}
\ee
\end{widetext}
In the equation above, $\hat{q}(\zeta)$ without any subscripts
represents the position ($\zeta$) dependent transport coefficient of a
gluon, which may be expressed in operator form as
\begin{eqnarray}
\hat{q} = \frac{8 \pi^2 \alpha_s C_A}{N_c^2 - 1} \int d y^- \langle X | Tr \left[  F^{a \mu \nu} (y^-) v_\mu F_{\nu}^{a \sigma}v_\sigma \right] | X \rangle
\end{eqnarray}
where, $F^{a \mu \nu}$ is the position dependent field strength tensor
of the gluon field and $y^{-}$ represents the light-cone separation
between these two field insertions. The state $| X \rangle$ represents
the matter through which the jet propagates.  We note that in the
evaluation of multiple scattering diagrams leading to
Eq. \ref{eq:HT_kernel}, it is assumed that the transverse momentum
exchanged with the medium $\vec{k}_{\perp}$ is soft, so that a Taylor
expansion can be made. The leading non-zero term in this expansion is
the term proportional to the square i.e. $| \vec{k}_{\perp} |^{2}$. In
combination with the gluon vector potential this term yields the
transport coefficient $\hat{q}$. The effect of higher terms in the
Taylor expansion of $\vec{k}_{\perp}$ is ignored, so that distribution
of $\vec{k}_{\perp}$ is approximated as a two-dimensional Gaussian
with a mean $\vec{k}_{\perp} = 0$ and a width proportional to
$\hat{q}$.

The $\hat{q}$ for a quark scattering off the gluon field is
related to the above expression as $\hat{q}_{Q} =
\frac{C_{F}}{C_{A}} \hat{q} $.  In the factorized formalism of the HT
approach, there is no scheme to evaluate these coefficients and thus
their values are external to the formalism. In most cases these are
taken as fit parameters which are dialed to match one data point.

\subsubsection{Single gluon emission}
A gluon emission spectrum $dN/dz$ can be calculated by integrating Eq. \ref{HT:in_medium_evol_eqn} over $Q^2$ using the
gluon radiation kernel Eq. \ref{eq:HT_kernel} and the gluon momentum fraction
$z=(1-y)$, giving
\begin{widetext}

\begin{equation}
\frac{dN}{dz} =  C_F \frac{\alpha_S}{2\pi} \frac{1+(1-z)^2}{z}
 \int_{Q^2_\mathrm{min}}^{Q^2_\mathrm{max}} \frac{dQ^2}{Q^2} \int_0^L d\zeta  \frac{\hat{q}(\zeta)}{Q^{2}}
\left[ 2 - 2 \cos\left( \frac{Q^2 \zeta }{ 2 E z (1- z)} \right)  \right]
%\frac{dN}{dz} =  \int dy \frac{1}{\sigma}\frac{d \sigma }{dy} \delta\left[z-(1-y)\right]
\label{eq:dNdz_HT}
\end{equation}

\end{widetext}
where we have taken only the first term of the emission kernel, and
ignored terms which are suppressed by factors $z=(1-y)$ (soft gluon
approximation). The lower bound for the $Q^2$ integral is given by
the requirement that the formation time of the gluon should be smaller
than the length of the medium: $Q^2_\mathrm{min}=E/L$. The upper
integration limit is given by 
\begin{equation}
Q^2_\mathrm{max}=y(1-y) 2 q^{-}p^{+} = z(1-z)2ME,
\end{equation} 
which is discussed further below (cf. Eq. \ref{eq:xL_lim}).

\begin{figure}
%\centering
\includegraphics[width=0.5\textwidth]{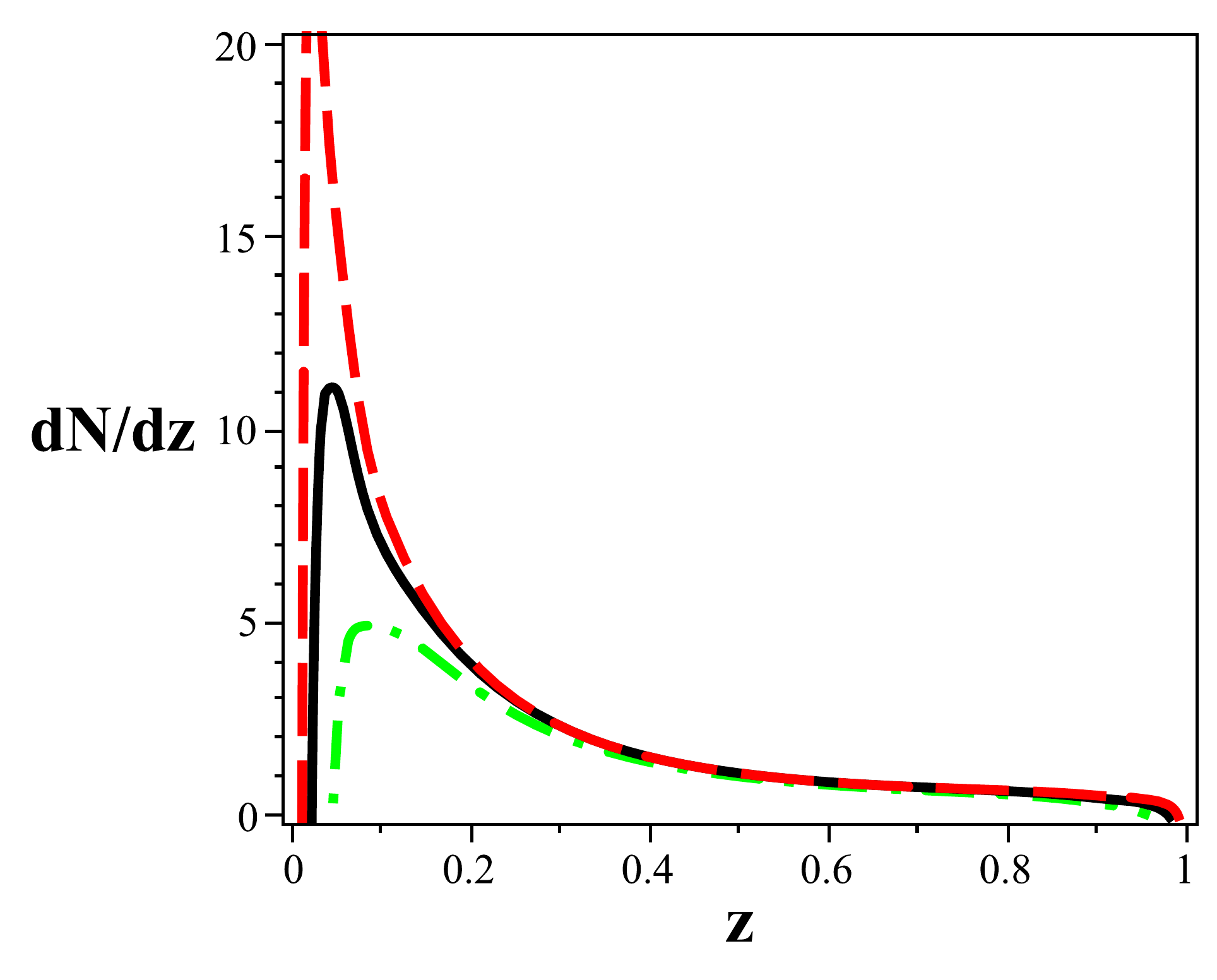}
\caption{The single gluon emission kernel in the HT approach for a
quark with energy $E=20$ GeV in a static medium of length $4$ fm which
has a $\hat{q}_{Q}=1$ GeV$^{2}/$fm with the maximum momentum cutoff
from the medium set at $M=1.2$ GeV (black solid line). The variable $z$
here is equivalent to the variable $x^{+}$ used in the WHDG approach,
earlier in this paper.  Also shown are the single gluon spectra for
$M=2.4$ GeV
(red dashed line) and $M=0.6$ GeV (green dot-dashed line). See text
for details.  }
\label{HT_single_emission}
\end{figure}

The medium induced single gluon emission
distribution in momentum fraction $z$ is plotted in
Fig.~\ref{HT_single_emission}, for a 20 GeV
quark, propagating through a 4 fm long static brick, using $M=3T$.

The medium induced gluon emission kernel has soft divergences at
$y=0,1$ as is the case for the gluon emission kernel in vacuum.
However, given that $\hat{q}$ is very similar to a gluon distribution
function it depends on the momentum fraction of the gluon off which
the hard parton scatters. In the case of DIS, this momentum fraction,
referred to as $x_{L}$, is given as $x_{L} = Q^{2}/(2q^{-}p^{+} y
(1-y))$. Where, $p^{+} = M/\sqrt{2}$, $M$ is the mass of the nucleon
and $q^{-} = \sqrt{2} E$, where $E$ is the energy of the virtual
photon in the nucleus rest frame. One expects $\hat{q} (x_{L})$ to be
dominated by small values of the fraction $x_{L}$ and rapidly
vanish for values of $x_{L}>1$. This introduces a constraint on the
radiated gluon momentum fraction,
\begin{equation}
\frac{1}{2} - \sqrt{\frac{1}{4} - \frac{Q^{2}}{2 q^{-}p^{+} } }  < y < \frac{1}{2} + \sqrt{ \frac{1}{4} - \frac{Q^{2}}{2q^{-}p^{+} } }.
\label{eq:xL_lim}
\end{equation} 
Note that in the above equation $Q^{2} \simeq k_{\perp}^{2}$ the
transverse momentum of the radiated gluon. Since only values of $y$
which lie between 0 and 1 are allowed, the value of $k_{\perp}^{2}$ is
always constrained to be less than $ q^{-} p^{+} $ where $p^{+}$ is
the momentum of the proton.  In a multiple scattering scenario, this
constraint only applies to the largest $k_{\perp}$, i.e. the first
emission.  Successive emissions are ordered in $k_{\perp}$.  While,
this limit is well defined in DIS, it is slightly uncertain in the
case of a quark gluon plasma created in a heavy-ion collision.
However, one argues that the probability of a medium at a temperature
$T$ to radiate a gluon with momentum much larger than the mean
momentum of $3T$ should be very small. Thus one replaces $p^{+} =
M/\sqrt{2} \rightarrow 3T/\sqrt{2}$ in the equation above. Here, $M$
represents the largest momentum that may be exchanged between the
medium and the jet, which is achieved when a thermal parton from the
medium directly impacts with the jet.  The central black solid line in
Fig.~\ref{HT_single_emission} corresponds to the choice of $T=400$ MeV, with $M=3T=1.2$ GeV. The red dashed line corresponds to the replacing the upper limit
with $M=6T = 2.4$ GeV and the dot-dashed green line to replacing the
upper limit with $M=3T/2 = 0.6$ GeV.  We note that the main difference
occurs at at low $z$, i.e. in the soft gluon limit, as would be
expected

\subsubsection{Multiple gluon emission}
We now plot the result of multiple emissions from the single quark
parton produced at one edge of a static brick.  In the soft gluon
limit, the medium modified fragmentation function of a quark is
obtained by replacing the vacuum splitting function with the sum of
the vacuum splitting function and the medium modified splitting
function in Eq.~\eqref{HT:in_medium_evol_eqn}. These equations require
an input fragmentation function at some lower initial scale usually
chosen to be $1$ GeV. The input fragmentation function is usually
chosen to be the vacuum fragmentation function.  Note that even if the
initial input fragmentation function is assumed to have no position
dependence, the evolution equation generates a position dependence in
the evolved fragmentation functions at any higher scale. A more
correct input is to have a position dependent fragmentation function
which is the vacuum fragmentation function at distances beyond the
brick and goes to zero swiftly for fragmentation within the brick. For
the HT plots in this paper we will ignore this sophistication and
ignore the position dependence of the fragmentation function.  Even at
this level of approximation, the equations above are far too
numerically intensive to solve. One usually replaces the position
dependence with the initial position $\zeta \rightarrow \zeta_i$. The
evolution equations now represent the evolution of a two dimensional
matrix and these represent the calculations which will be presented in
this paper. A further approximation which is sometimes used in the
literature is to also drop the energy dependence of the fragmentation
functions. This approximation will not be made and the energy
dependence of the medium modified fragmentation function will be
retained.

\begin{figure}
%\centering
%\includegraphics[width=0.5\textwidth]{figs/delta_func_bounds.jpg}
\includegraphics[width=0.45\textwidth,bb=5 25 350 350,clip=]{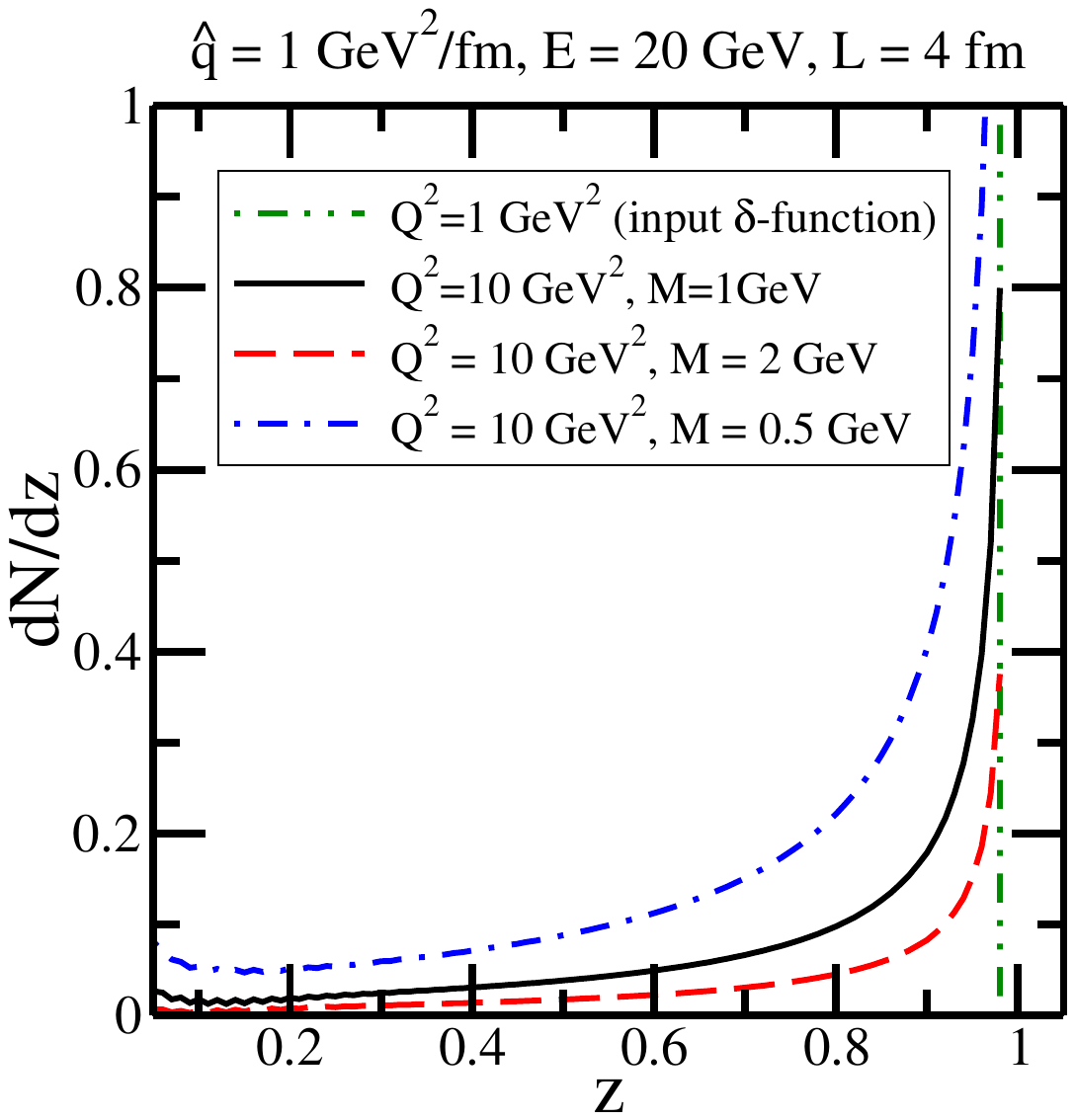}
\caption{The results of evolving an input [$q(\bar{q}) \rightarrow q(\bar{q})$] fragmentation function at $\mu^{2}= 1$ GeV$^{2}$ 
given by a $\delta(1-z)$ (see Eq.~\eqref{HT_delta_input}) up to a
higher scale of $Q^{2}=10$ GeV$^{2}$ in a 4 fm brick. The transport coefficient $\hat{q}_{Q}=1$
GeV$^{2}/$fm with the maximum momentum cutoff from the medium set at 1
GeV (solid black line), 2 GeV
(red dashed line) and 0.5 GeV (blue dash-dotted line). See text
for details. }
\label{HT_delta_evolution}
\end{figure}

In order to compare with other formalisms which present the energy
loss of a single quark, we introduce a quark-to-quark fragmentation
function as the input at the lower scale of $\mu^{2} = 1$
GeV$^{2}$. In actuality our input is a singlet quark fragmentation
function ( $(q+\bar{q})/2 \rightarrow (q+\bar{q})/2$). Our input at $1$ GeV may be expressed as
\begin{eqnarray}
D_{ \frac{u + \bar{u}}{2}  \rightarrow \frac{u + \bar{u} }{2} } ( z, 1 \mbox{GeV}^{2} )&=& \delta(1 - z)  \nonumber \\
D_{ \frac{d + \bar{d}}{2}  \rightarrow \frac{u + \bar{u} }{2} } ( z, 1 \mbox{GeV}^{2} )&=& 0 \nonumber \\
D_{ g  \rightarrow \frac{u + \bar{u} }{2} } ( z, 1 \mbox{GeV}^{2} )&=& 0. \label{HT_delta_input}
\end{eqnarray}
In Fig.~\ref{HT_delta_evolution}, these are represented by the solid
green line [$u(\bar{u}) \rightarrow u(\bar{u})$] along the $z=1$ axis
and the cyan dot-dashed line [$g \rightarrow u (\bar{u})$] along the
$dN/dz =0 $ axis. We do not plot the results of the $d(\bar{d})
\rightarrow u(\bar{u})$ fragmentation. These \emph{distributions} are
then evolved up to the higher scale of $Q^{2} = 10$ GeV$^{2}$. The
results for the three different momentum bounds from the medium $M=$
1.5$T$, 3$T$ and 6$T$ (as discussed above) with $T=1/3$ GeV and a
quark $\hat{q}_{Q} = 1$ GeV$^{2}/$fm are plotted in
Fig.~\ref{HT_delta_evolution}. 
% The solid lines represent the quark
% (anti-quark) to quark (anti-quark) fragmentation function while the
% dot-dashed lines represent the gluon to quark (anti-quark)
% fragmentation function.  The latter arise from the gluon to quark
% (anti-quark) splitting function. 
Note that the full quark and gluon distributions are evolved. This means
that if in one step of the evolution a hard quark loses energy by
splitting into a quark and a gluon, in the next step of the evolution
both the softer quark and gluon will further lose energy by splitting
themselves.

As would be expected, we note that in the case when the bound from the
medium is set at the higher value of $M=2$ GeV, there is more gluon
radiation and thus there is more modification of the input
distribution and vice-versa for the case when the bound is set at a
lower value of $M=0.5$ GeV. The dependence of the evolved distribution
on the bound is enhanced primarily due to the numerical instability of
the vacuum DGLAP evolution equations on an input distribution that is
singular at $z \rightarrow 1$.  Note that the evolution equations
include a pure vacuum part and a medium induced part as expressed in
Eq.~\eqref{HT:in_medium_evol_eqn}.  The pure vacuum splitting
functions are defined using the $(+)$-functions, i.e.,
\begin{eqnarray}
&&\frac{\partial D(z,Q^{2})}{\partial \log Q^{2}}  = \frac{\alpha^{2}}{2 \pi} \int_{z}^{1} \frac{dy}{y} P(y)_{+} D\left( \frac{z}{y}, Q^{2} \right)  \\
&=& \frac{\alpha^{2}}{2 \pi} \int_{z}^{1} \frac{dy}{y} P(y) D\left( \frac{z}{y}, Q^{2} \right)  - D(z,Q^{2} ) \int_{0}^{1} dy P(y). \nonumber
\end{eqnarray}
The above equation assumes that $D(z)$ is well defined at all values
of $z$ over which the evolution is carried out. This condition is not
fulfilled by the strict $\delta$-function. The calculations presented
in Fig.~\ref{HT_delta_evolution} are carried out by using a regulator
to define the $\delta$-function. We should point out that only one
choice of regulator has yielded convergent results as the regulator is
removed from the results, \emph{viz.}
\begin{eqnarray}
\delta(1-z) &=& \frac{1}{\epsilon}, \,\,\,\, \forall \,\,\,\, z > 1-\epsilon, \nonumber \\
		    &=& 0,\,\,\,\, \forall \,\,\,\,  z< 1- \epsilon,
\end{eqnarray}
in the limit $\epsilon \rightarrow 0$. Other choices of regulation,
e.g. a narrow Gaussian with a width given by $\epsilon$ turned out to
yield results which were non-convergent within the limits of numerical
accuracy used.

\subsubsection{Medium modified fragmentation function}

\begin{figure}
%\centering
%\includegraphics[width=0.5\textwidth]{figs/frag_func_bound_high_stat.jpg}
\includegraphics[width=0.45\textwidth,bb=5 25 355 355,clip=]{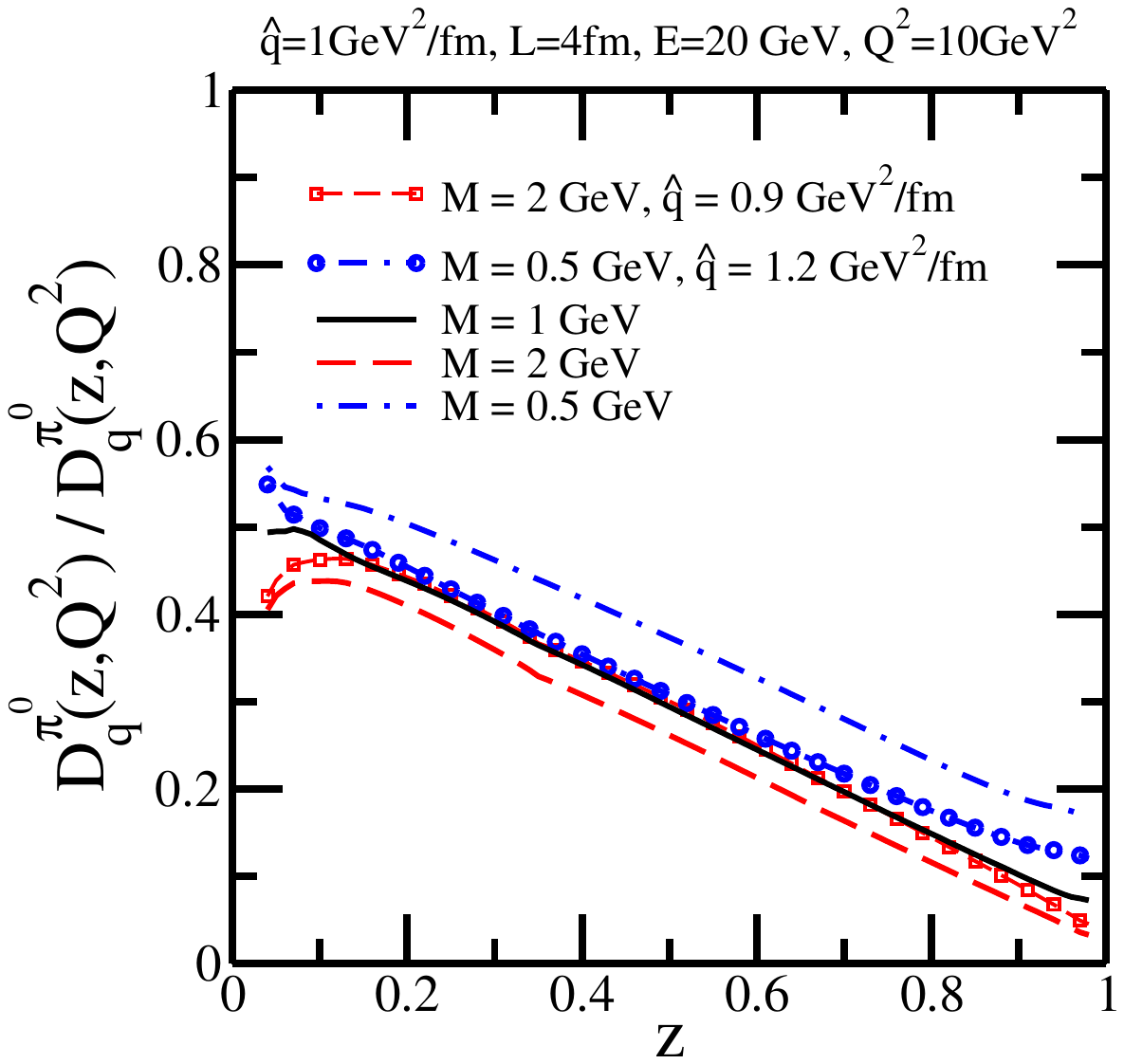}
\caption{The ratio of the medium modified fragmentation function (of a $\pi^{0}$ from a quark with $E=20$ GeV) 
to the vacuum fragmentation function at $Q^{2} = $ 10 GeV$^{2}$.  The
input in both cases is the KKP fragmentation function at
$\mu^{2}~=~1$~GeV$^{2}$.  The medium is a static brick with a length
of 4 fm and a $\hat{q}_{Q}=1$ GeV$^{2}/$fm. Three different values for the maximum momentum cutoff from the medium are used 1~GeV (black solid line), 2~GeV (red dashed line), and
0.5~GeV (blue dot-dashed line). The open red squares and open blue circles indicate the results with $\hat{q}_{Q}$ changed to make the lines with $M=2$ GeV and $M=0.5$ GeV approximately coincide
with the central line with $M=1$ GeV. See text for
details.  }
\label{HT_frag_evolution}
\end{figure}

The DGLAP evolution equations used by the HT scheme require that the
input fragmentation function be well defined at all values of momentum
fraction $z$. To illustrate the degree of control in these
calculations we now repeat the calculation above but using a well
defined input distribution: a $\pi^{0}$ fragmentation function (in
this case taken from the KKP parametrization~\cite{Kniehl:2000fe}).
The input is taken at $\mu^{2} = 1$ GeV$^{2}$ and evolved to $10$
GeV$^{2}$. The results are plotted in Fig.~\ref{HT_frag_evolution} The
brick is again of length 4 fm and has a quark $\hat{q}_{Q} = 1$
GeV$^{2}/$fm. Since fragmentation functions vary over orders of
magnitude as $z$ runs from $0$ to $1$ we plot the ratio of the medium
modified fragmentation function to a vacuum fragmentation function at
the same scale. The denominator is obtained by evolving the
fragmentation function from $1$ GeV$^{2}$ to 10 GeV$^{2}$ with a
$\hat{q}_{Q}= 0$.  The black solid line represents the standard case with
a bound on the momentum from the medium set at $M=1$GeV. The red
dashed and the blue dot-dashed lines represent the cases with $M=2$
GeV and $M=0.5$ GeV. As expected, raising the bound, increases the
phase space of gluon radiation and thus increases the modification.

\subsubsection{Uncertainty due to large angle radiation}
The uncertainty in the bound on the momentum that may emanate from the
medium introduces an uncertainty in the extracted value of $\hat{q}$.
To estimate this uncertainty, we take the case with $M=0.5$ GeV (the
blue dot-dashed) which shows less modification than the case with
$M=1$ GeV and increase the $\hat{q}$ until the blue dot-dashed line
coincides with the black line. The final result is indicated with a
blue dot-dashed line with open circles. This requires a 30\% increase
in $\hat{q}_{Q}$. Similarly we reduce the $\hat{q}_{Q}$ for the case with
$M=2$ GeV (red dashed line) to make it coincide with the black
line. The final result is indicated by the red dashed line with open
squares and requires a $\hat{q}_{Q}$ which is 10\% lower than the central
value of 1 GeV$^{2}/$fm. Thus the uncertainty in the extracted value of
$\hat{q}$ in HT calculations is between 10\% to 30\% when the $\hat{q}_{Q}
\sim 1$ GeV$^{2}/$fm and the length of the medium is of the order of 4
fm.  The reader will note that these values are quite representative
of the respective values at RHIC and thus provide the range of error
in estimates of $\hat{q}$ extracted in RHIC collisions.

%
%        Final comparison
%
\section{Systematic Comparisons}
\label{sec:syst_comp}

In this section, the multiple-soft scattering approximation is
compared with the opacity expansion formalism and the McGill AMY
implementation. Specifically, the following energy loss models are
compared:
\begin{itemize}
\item ASW--MS: The multiple-soft scattering approximation as
formulated by ASW \cite{Salgado:2003gb} and discussed in Sections
\ref{sect:ASW} and \ref{sec:AMY}.
\item ASW-SH: The single hard scattering approximation as described
formulated by ASW \cite{Salgado:2003gb}, and shown in
Eq.~(\ref{WHDG:ASW} and Table \ref{WHDG:table}. The original
formulation uses a fixed value for $L/\lambda=1$ for analytical
convenience. For this work $L/\lambda$ is calculated from the
temperature $T$ in the medium (see Appendix \ref{sec:commondef}).
\item DGLV: The single hard scattering approximation using the same choices for in-medium gluon and quark masses and the kinematic bounds as used by WHDG \cite{Wicks:2005gt}, as given in Eq.~\ref{WHDG:DGLV} and Table \ref{WHDG:table}.
\item AMY radiative: The AMY formalism based on a thermal effective
field theory as described in Section \ref{sec:AMY}. In the interest of
the comparison to the other formalisms, the gluon splitting process $g
\rightarrow g g$ is removed from the calculation.
\end{itemize}

\subsection{Comparison at fixed temperature}

\begin{figure*}[!htb]
\includegraphics[width=0.45\textwidth]{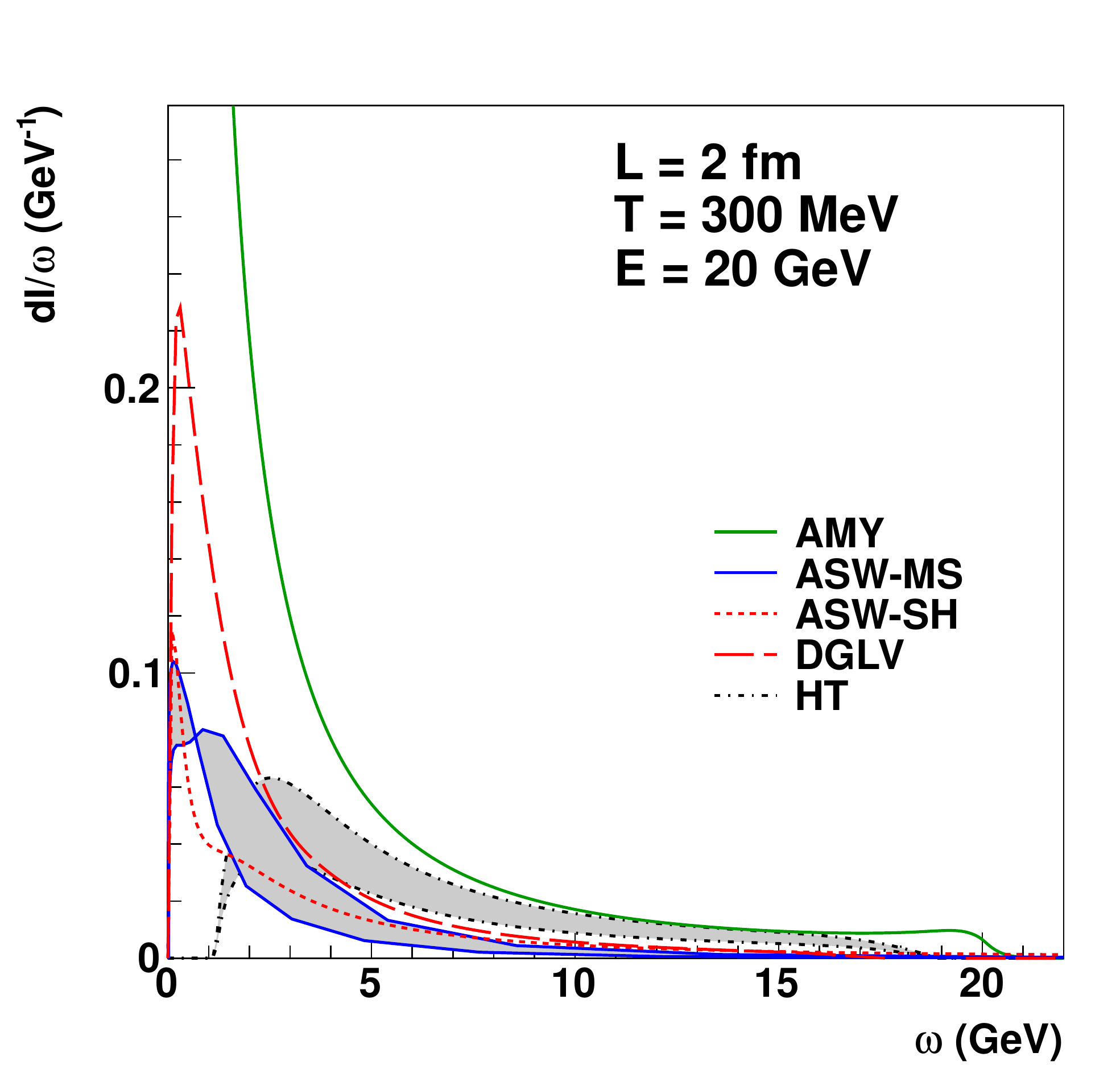}\hfill%
\includegraphics[width=0.45\textwidth]{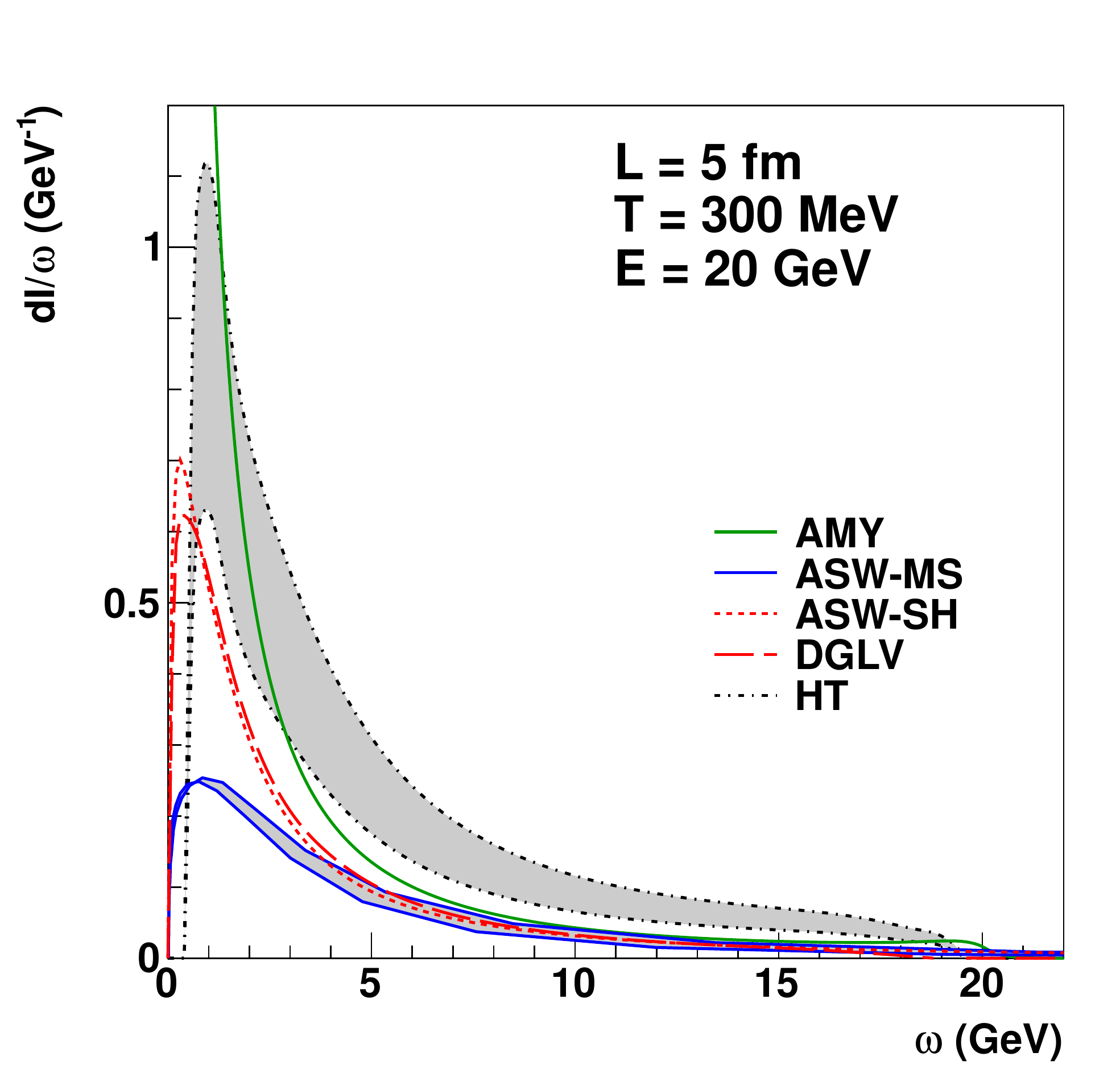}
\caption{(Color online) The single gluon distribution as function of
     gluon energy $\omega$ for a uniform medium with $T = 300$ MeV and
     two different lengths, $L=2$ fm (left panel) and $L=5$ fm (right
     panel). For the AMY calculation, the outgoing gluon spectra are
     plotted, including the evolution via the rate equation
     Eq. \ref{eq:fokker-planck}. The bands for the ASW--MS and HT
     formalisms indicate the uncertainty from evaluating $\hat{q}$
     using Eq. \ref{diff_rate} and \ref{eq:qhat}. }
\label{fig:dIdwSameTemp}
\end{figure*}

\begin{figure*}
\includegraphics[width=0.45\textwidth]{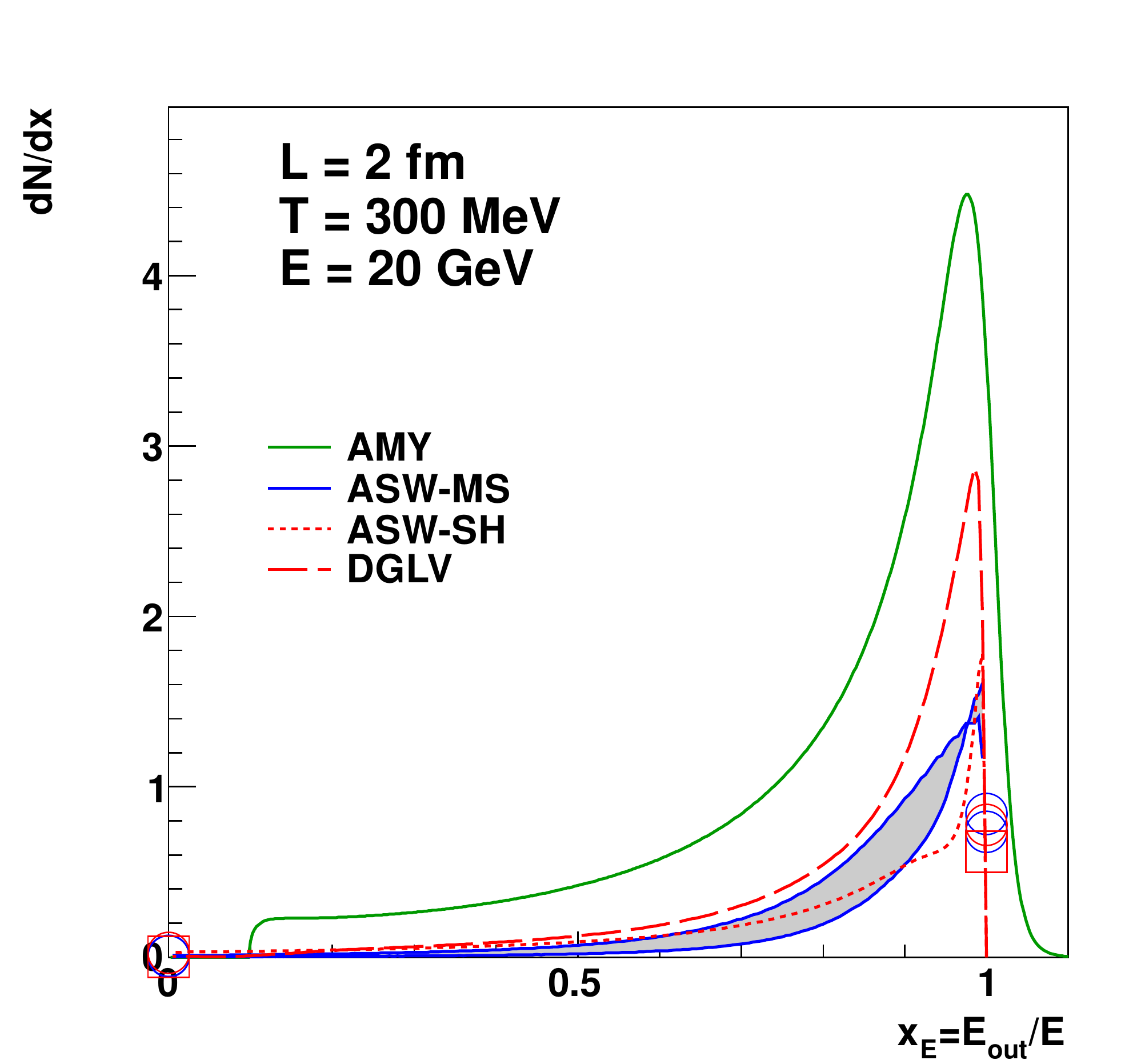}\hfill%
\includegraphics[width=0.45\textwidth]{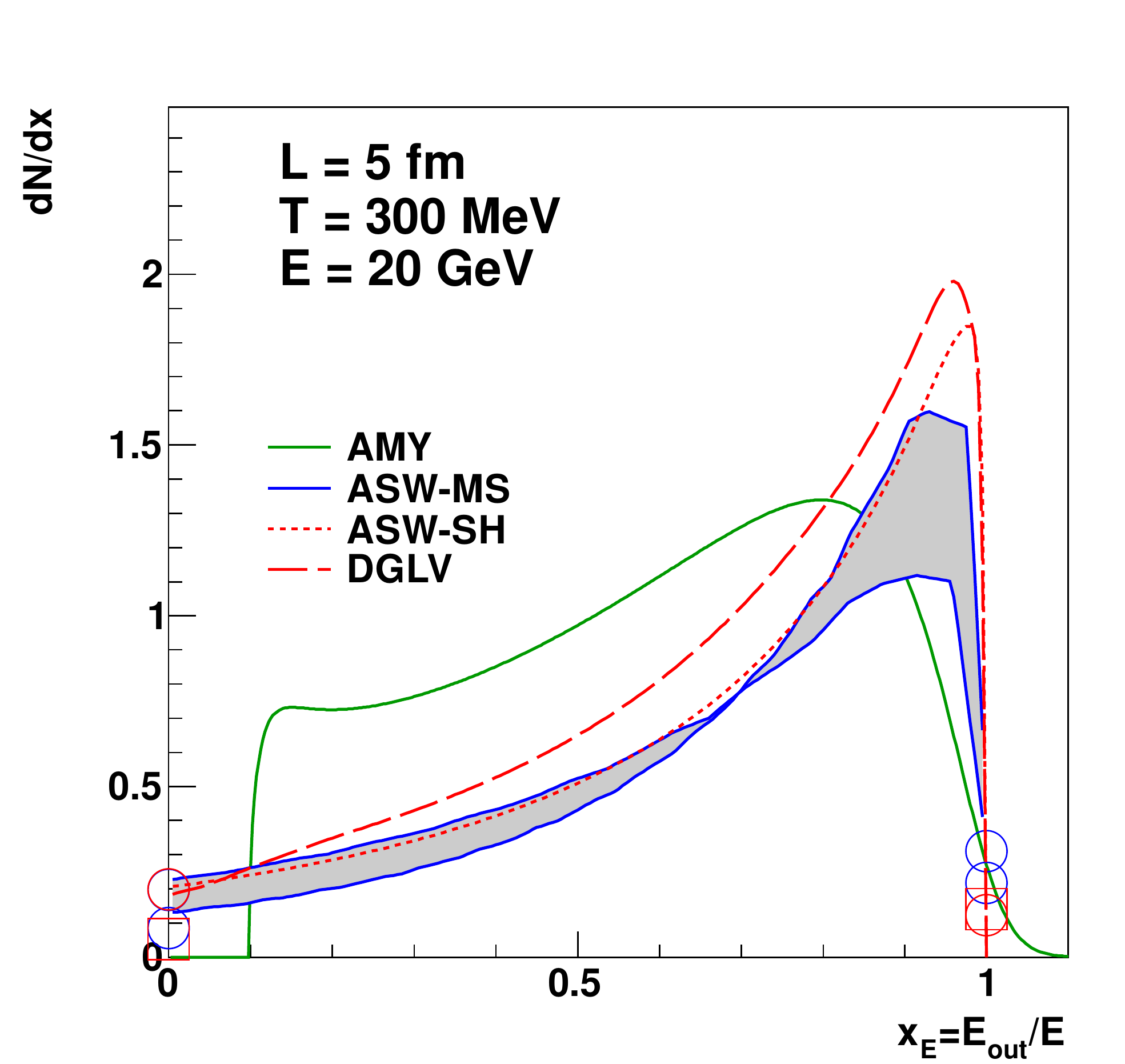}
\caption{(Color online) The final quark energy spectrum as function of
     $x_{E}=E_{out}/E=1-\epsilon$ for a uniform medium with
     temperature $T = 300$ MeV and two path lengths $L=2$ fm (left
     panel) and $L=5$ fm (right panel). The symbols at $x_{E}=0$
     indicate the probability that a quark is absorbed and at
     $x_{E}=1$ the probability that a quark does not interact with the
     medium. Blue circles: ASW--MS. Open red
     squares: DGLV.  red circles: ASW--SH.}
\label{fig:dNdzSameTemp}
\end{figure*}

Figure \ref{fig:dNdzSameTemp} shows the single gluon energy spectra
for medium-induced radiation for a uniform medium with $T=300$ MeV and
two different path lengths $L=2$ fm (left panel) and $L=5$ fm (right
panel) using the four different formalisms discussed above. The
HTL-based AMY formalism (green curve) used here is without the
large-angle cut-off discussed in Section \ref{sect:AMY_largeangle} and
it gives the largest amount of radiation.  The opacity expansions (red
curves, DGLV and ASW--SH) and the Higher Twist formalism (black
curves, HT) are next while the multiple-soft scattering approximation
ASW--MS gives the least radiation.  For ASW--MS and HT two curves are
shown, using different relations between the transport coefficient
$\hat{q}$ and the temperature $T$. One is based on the soft-scattering
limit from HTL field theory, Eq. \ref{diff_rate}, which gives
$\hat{q}=2\,C_{A}\,\alpha_s\,T\,\mu^2\,\log(q_\mathrm{max}^2/\mu^2))$
while the other is Eq. \ref{eq:qhat}, based on Eq. \ref{eq:GammaComb}.
The single gluon spectrum for the HT formalism is evaluated using
Eq. \ref{eq:dNdz_HT} which does not include modified DGLAP-evolution
of the partons. The HT formalism gives more radiation at larger gluon
energies than the opacity expansion and ASW--MS, while the peak at
small $\omega$ is less pronounced for the default
$Q^2_\mathrm{max}=6z(1-z)ET$. See Fig. \ref{HT_single_emission} to see
the effect of changing $Q^2_\mathrm{max}$. We also note that the $L=5$
fm case is beyond the strict validity of the HT formalism,
since the lower virtuality bound $Q^2_\mathrm{min}=E/L\approx 0.4$
GeV$^2$. Despite this, the result is shown for comparison.

A similar ordering between the different formalisms is visible for
$L=2$ fm and $L=5$ fm, although the difference between the AMY result
and the opacity expansions is larger at small $L$. The peak at small
$\omega$ for the HT formalism is also more pronounced for larger
path length. The path-length dependence has been studied in more detail
in \cite{CaronHuot:2010bp}, where is is shown that for small $L$, AMY
overestimates the radiation by ignoring finite-length effects, while
for large path-lengths, the first order ($N=1$) opacity expansion
overestimates the radiation by neglecting interference between
subsequent scatterings. At large path-lengths, the AMY result is close
to the ASW--MS result, but only for large gluon energies. At smaller
gluon energies, the large-angle cut-off strongly reduces radiation in
ASW--MS. This effect is not implemented in AMY (see Section
\ref{sect:AMY_largeangle}).

For ASW--MS and ASW--SH the thermal quark and gluon masses are set
to zero while in DGLV those masses are finite.  Including the thermal
masses reduces soft gluon radiation which can be seen comparing the
ASW--SH and DGLV curves at small $\omega$. Another difference as
discussed in Section \ref{sec:WHDG:ASW} is the cut-off $k_{\rm max}$
on the transverse momentum of the radiated gluon. For ASW--MS and
ASW--SH it is possible that a radiated gluon has larger momentum than
the initial energy of the incoming quark because there is no
kinematical bound on the gluon energy $\omega$.

To calculate the total energy loss, the gluon spectra from Fig.
\ref{fig:dIdwSameTemp} are used as input for a calculation of multiple
gluon emission. For the opacity expansions and the ASW--MS
calculation, a Poisson ansatz is used (Eq. \ref{eq:poisson_mg}), while
AMY use rate equations (Eq \ref{eq:fokker-planck}). The resulting
outgoing quark spectra are shown in Fig. \ref{fig:dNdzSameTemp}. No
results for the HT formalism are shown, because medium-induced and
vacuum radiation cannot be separated in the medium-modified DGLAP
evolution, which makes it difficult to compare to the other formalism on
the same footing. The Poisson ansatz used for the opacity expansions
and ASW--MS gives a finite probability to the parton to lose no energy
(zero gluons radiated) which is indicated by the symbols at $x_{\rm E}
= 1$ in Fig. \ref{fig:dNdzSameTemp}. At the same medium density and $L
= 5$ fm, this probability is approximate twice as large for the
multiple soft scattering approximation as for the opacity expansions
(right panel of Fig. \ref{fig:dNdzSameTemp}). On the other hand, the
gluons that are radiated are softer for the opacity expansions than
for ASW--MS.  For a brick of $L = 2$ fm and a temperature of $T = 300$
MeV the discrete weights at $x_{\rm E} = 1$ are between 0.5 and 1 (see
also Table \ref{tab:charac}), so that the behaviour is dominated by
the discrete weights, although there are large probabilities for small
energy loss $x_{\rm E} \gtrsim 0.8$ as well. The distribution from the
AMY formalism is cut off at $x_{\rm E} < 0.1$, because the formalism
is not valid when then parton energies approach the thermal energy;
partons with energy less than 2 GeV are removed from the evolution.

It is clear from Fig. \ref{fig:dNdzSameTemp} that the different
energy loss formalisms do not result in similar outgoing quark
distributions. 

\begin{figure*}[!htb]
  \centering 
\subfigure[~$L=2$ fm.]{\label{fig:R7vsqhat_L2}
\includegraphics[width=0.49\textwidth]{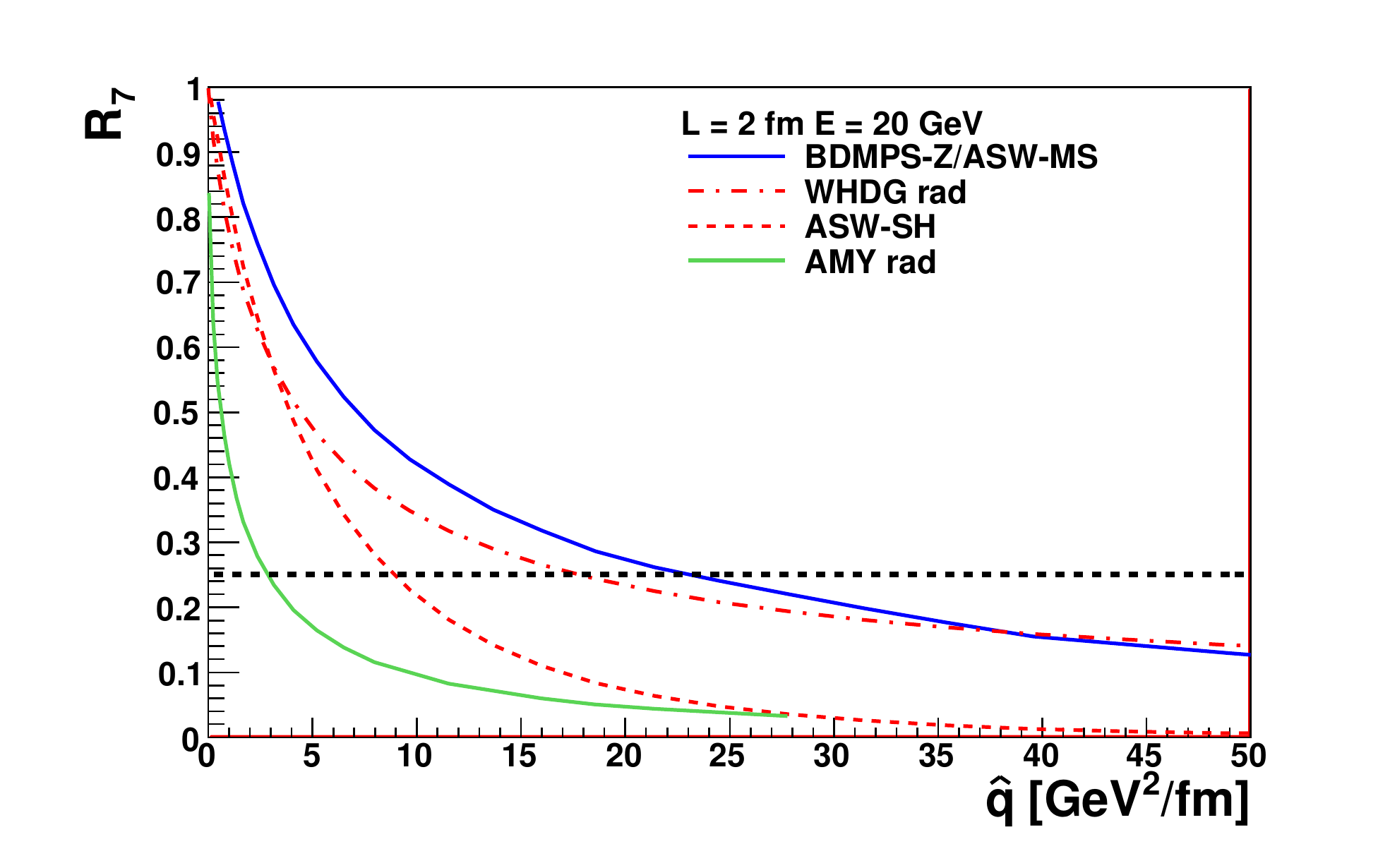}}
\hfill
\subfigure[~$L=5$ fm.]{\label{figs/fig:R7vsqhat_L5}
\includegraphics[width=0.49\textwidth]{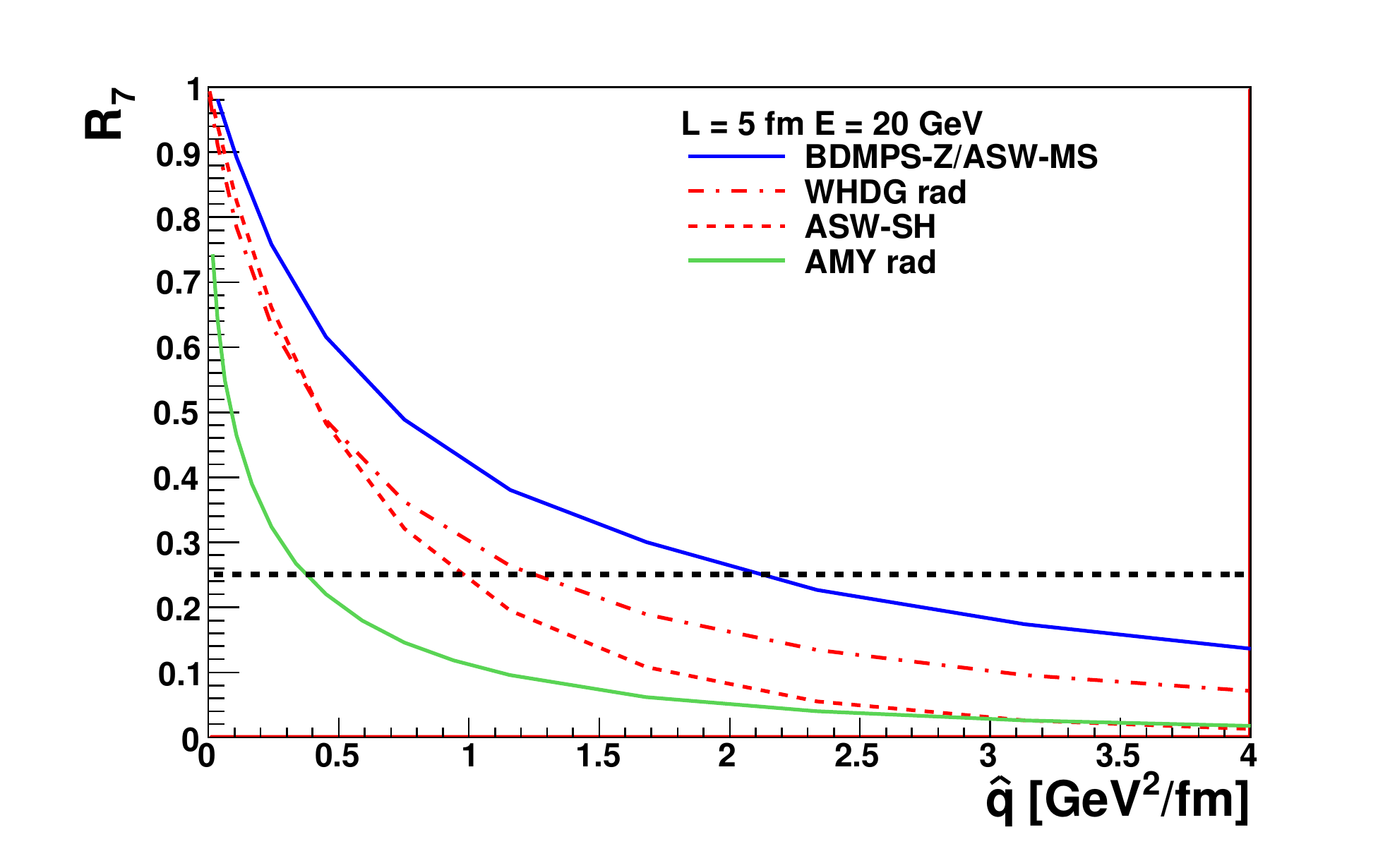}}
\caption{Correlation between $R_{7}$ and $\hat{q}$ for a primary quark with $E=20$ GeV
    for different energy loss formalisms. The horizontal black dashed line indicates $R_{7}=0.25$.}
\label{fig:R7vsqhat}
\end{figure*}

\subsection{Suppression factor in a QGP Brick}

To characterise the energy loss distributions in
Fig. \ref{fig:dNdzSameTemp} in a single number, we calculate an
approximation of the nuclear suppression factor $R_{AA}$ in the
following way.

The measured hadron spectra at RHIC approximately follow a power law:
$dN/dp_{T} = A p_{T}^{-n}$. If the energy of each hadron is reduced
by a fraction $\epsilon$, the hadron momentum after energy loss $p'_{T}=(1-\epsilon)p_T$ and the spectrum after energy loss  
\begin{equation}
\frac{dN}{dp'_{T}} =
A \frac{(1-\epsilon)^n}{p_{T}^{n}}\frac{dp_{T}}{dp'_{T}} =
A \frac{(1-\epsilon)^{n-1}}{p_{T}^{n}}.
\end{equation}
In this case $R_{AA}=(1-\epsilon)^{n-1}$. This can be generalized to
a distribution of energy loss, in which case the nuclear modification
factor $R_{AA}$ can be approximated by the appropriately weighted
average over the energy loss probability distribution $P(\epsilon)$
\begin{equation}\label{eq:Rn}
R_{n} = \int_{0}^{1}d\epsilon (1-\epsilon)^{n-1}P(\epsilon),
\end{equation}
with $\epsilon=\Delta E/E$. In the following, $R_{7}$ will be used as a
proxy for $R_{AA}$, because for RHIC energies the hadron
$p_{T}$ spectrum can be approximated by a power law spectrum with $n=6.5$
for $p_{T}>2.$ GeV/$c$ \cite{Adams:2006nd}.

Figure \ref{fig:R7vsqhat} shows the dependence of the suppression
factor $R_{7}$ on $\hat{q}$ for the different formalisms. The figure
clearly shows that both opacity expansion formalisms generate a larger
suppression at the same medium density (same $\hat{q}$) than the
multiple soft scattering approximation. The AMY formalism generates
the largest suppression at a given density. The values of the
transport coefficient needed to reach a similar suppression as
measured at RHIC \cite{Adams:2003kv,Adler:2003qi,Adler:2003au}, $R_{7}
\approx 0.25$, are listed in Table \ref{tab:variables} and differ by a
factor 5--10 between AMY and ASW--MS.

\begin{table*}
%\begin{tabular}{|c|c|c c c c c c|}
\begin{tabular}{c c c c c c c c}
%\hline
$R_{7}=0.25$ & & $\;\;T$ (MeV) $\;\;$ & $\;\;\hat{q}$ ($\mathrm{GeV^2/fm}$) $\;\;$ &
$\;\;\omega_{c}$ or $\bar{\omega}_c L/\lambda$ (GeV) $\;\;$ & $\;\;R$ or
$\bar{R} L/\lambda \;\;$ & $\;\; L/\lambda \;\;$ & $\;\; m_{\rm D}$ (GeV) $\;\;$ \\
\hline
\multirow{4}{*}{$L = 2$ fm} & ASW--MS & 1030 & 23.2 & 236 & 2393 & --- & --- \\
& WHDG & 936 & 17.8 & 105 & 1063 & 6.25 & 1.82 \\
& ASW--SH & 727 & 8.86 & 49.0 & 500 & 4.85 & 1.41 \\
& AMY & 480 & 2.7 & --- & --- & --- & --- \\
\hline
\multirow{4}{*}{$L = 5$ fm} & ASW--MS & 434 & 2.11 & 134 & 3401 & --- & --- \\
& WHDG & 358 & 1.23 & 36.5 & 925 & 5.97 & 0.69 \\
& ASW--SH & 326 & 0.95 & 27.6 & 702 & 5.44 & 0.63 \\
& AMY & 235 & 0.4 & --- & --- & --- & --- \\
\hline
\end{tabular}
\caption{Values of the model parameters required to reach the typical suppression of $R_{7} = 0.25$.}
\label{tab:variables}
\end{table*}

\subsection{Comparison at fixed suppression $R_{7}$}

\begin{figure*}
 \centering 
\includegraphics[width=0.45\textwidth]{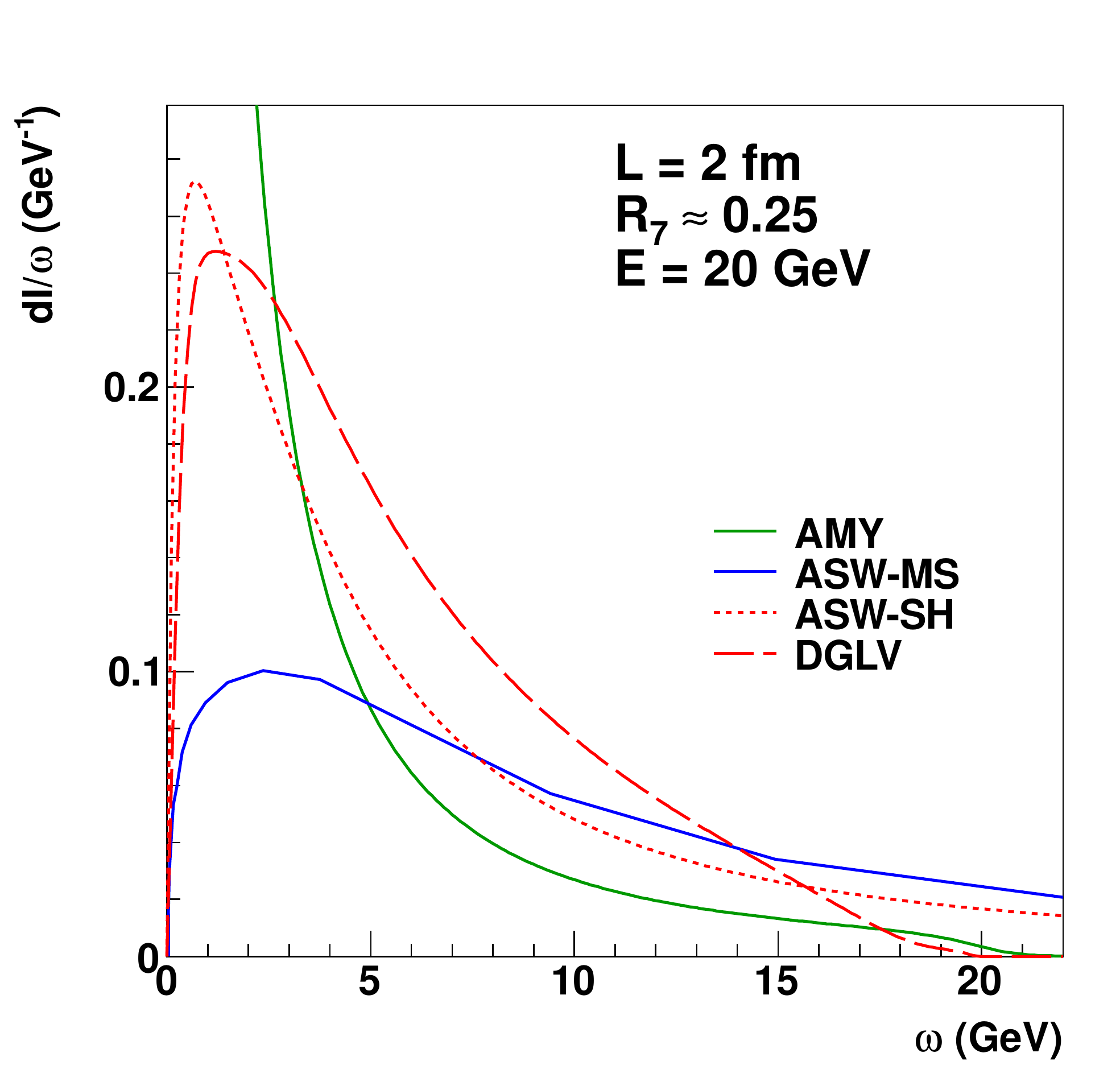}
   \hfill% 
\includegraphics[width=0.45\textwidth]{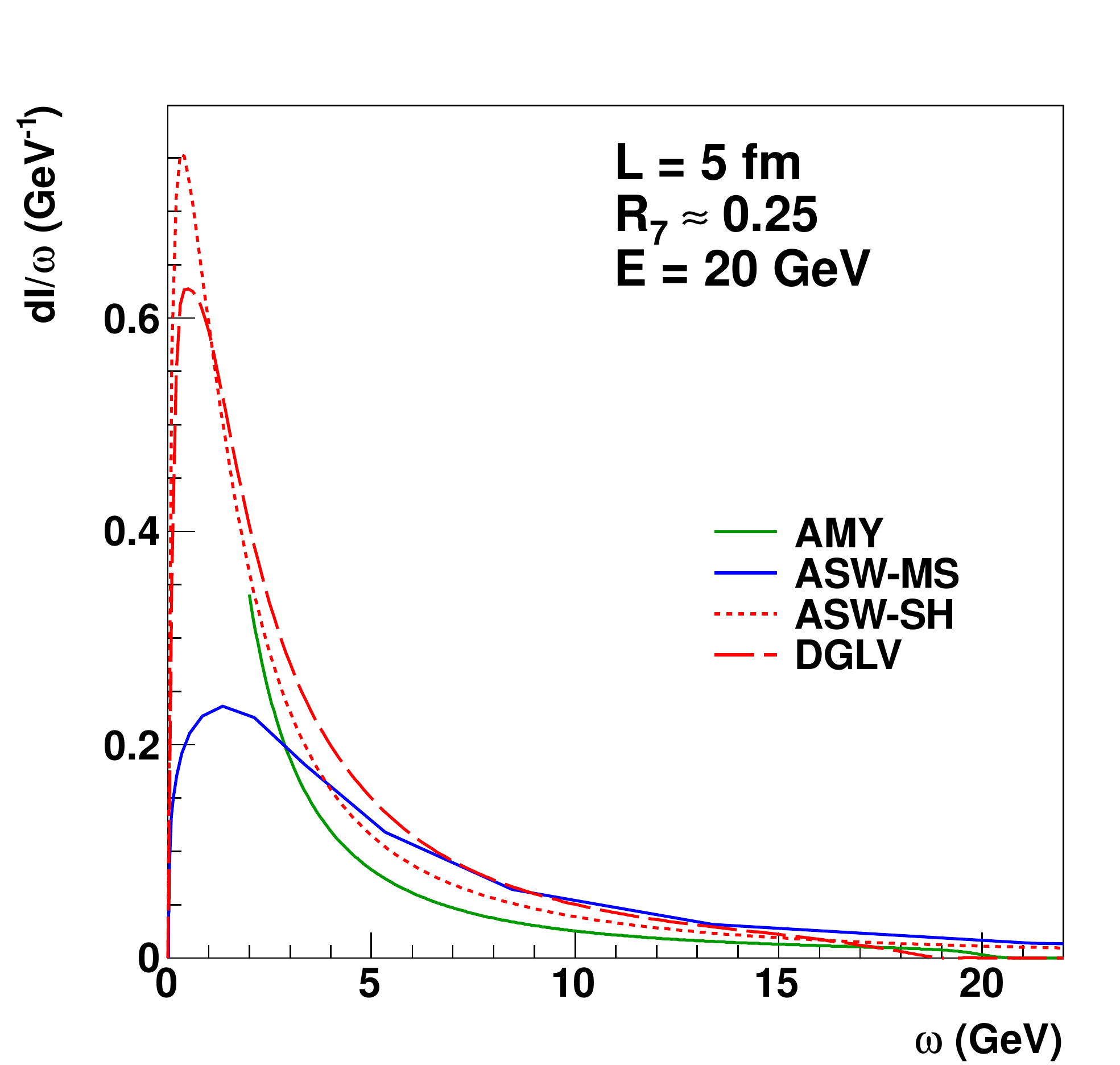}
\caption{\label{fig:dIdw}
 Inclusive gluon radiation spectrum for quarks with $E=20$ GeV with in-medium
 path-lengths $L=2$ fm (left panel) and $L=5$ fm (right
 panel). The medium density has been tuned for each formalism to
 give $R_7 = 0.25$. The AMY result includes the evolution via the rate equations Eq. \ref{eq:fokker-planck}. }
\end{figure*}

\begin{figure*}
 \centering 
\includegraphics[width=0.45\textwidth]{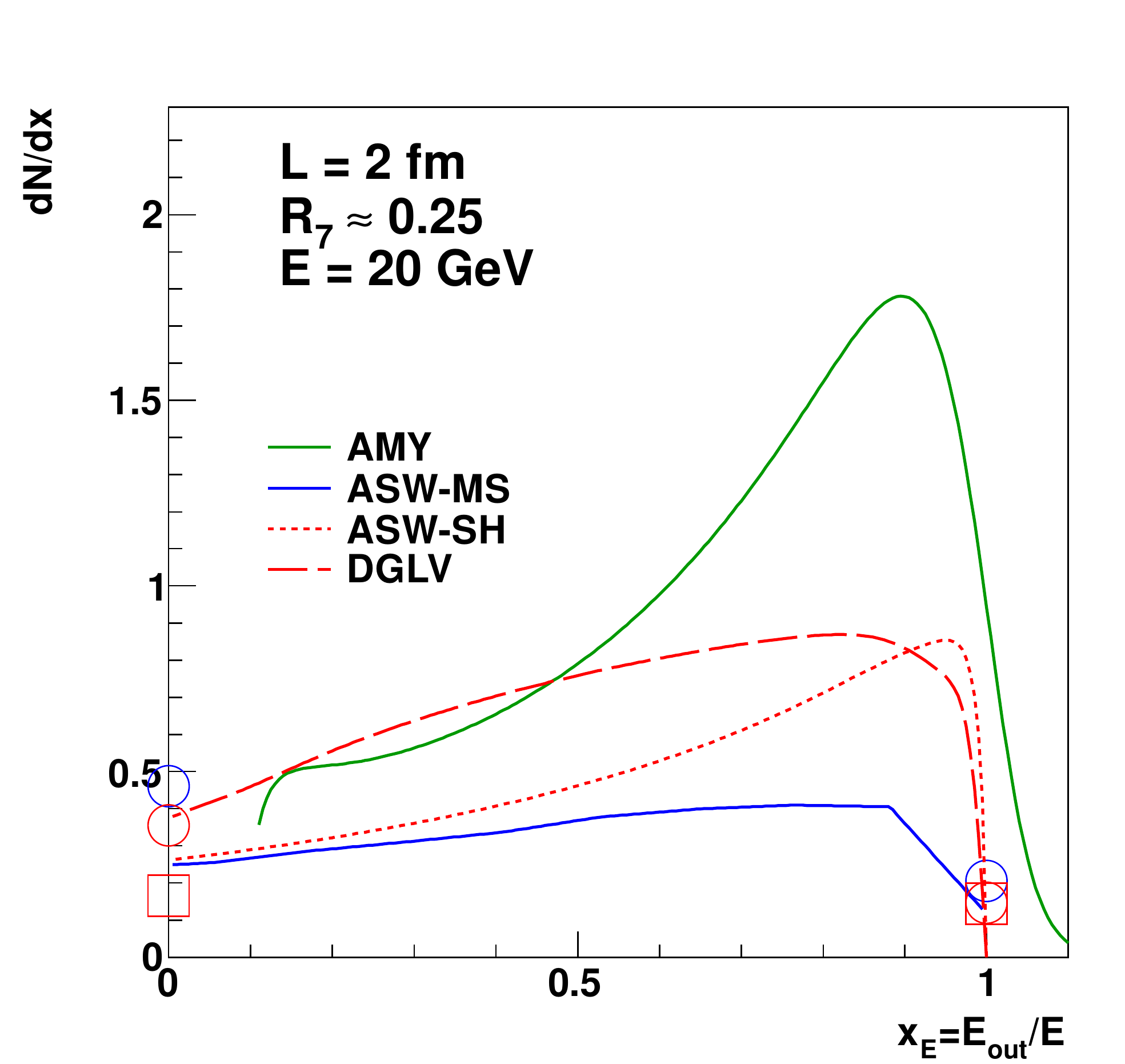}
   \hfill% 
\includegraphics[width=0.45\textwidth]{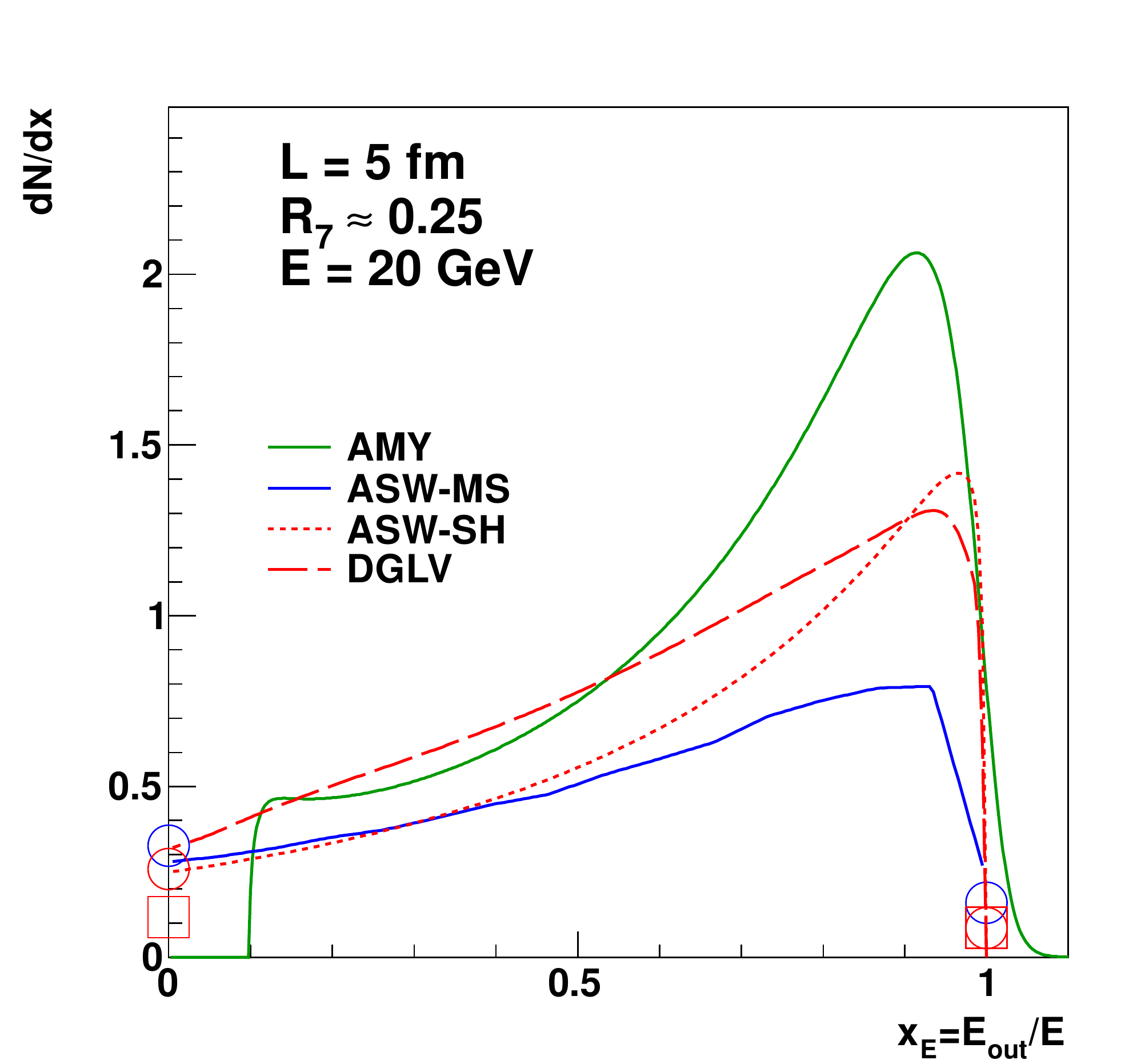}
\caption{The final quark energy spectra as function of $x_{\rm E}=1-\epsilon$ and for
     path-lengths $L=2$ fm (left panel) and $L=5$ fm (right panel). All models are scaled to the same
     suppression $R_{7} = 0.25$. The squares at $x_{\rm E}=0$ indicate the
     probability that a quark is absorbed and at $x_{\rm E}=1$ the
     probability that a quark does not interact with the
     medium. Solid blue squares: BDMPS-Z/ASW--MS. Open red squares:
     DGLV. Solid red squares: ASW--SH.}
\label{fig:dNdz}
\end{figure*}

In Fig.~\ref{fig:dIdw} the inclusive gluon spectra are shown for a
uniform medium of $L=2$ and 5 fm, for a fixed suppression $R_{7}=0.25$
using the medium density values given in Table
\ref{tab:variables}. The single gluon spectrum from DGLV does not
extend beyond $\omega=E$ because that formalism implements a large momentum cut-off to impose forward propagation of the final state quark (cf. Eq. \ref{WHDG:kmdglvtwo}). For the AMY gluon spectrum only $q \to q + g$ splittings
are included. In the AMY formalism it is not possible to distinguish
between thermal and radiated gluons for $\omega < 2$ GeV which is why
in this region for AMY the gluon spectrum is not shown. We note that
the ASW--MS single gluon spectrum at fixed suppression is harder than
that obtained in the opacity expansions.

Figure \ref{fig:dNdz} shows the outgoing quark energy spectrum as
function of $x_{E}=1-\epsilon$ for the two bricks of different lengths
and $R_{7}=0.25$. The probability that a parton is absorbed in the
medium is indicated by the markers at $x_{\rm E}=0$. In this case the
the energies of the multiple radiated gluons add up to a total energy
loss that exceeds the initial energy of the parent parton and the
parton is absorbed in the medium. The large $x$ cut-off on the single
gluon spectra in the DGLV formalisms leads to a smaller probability
for absorption than in ASW--MS and ASW--SH. The corresponding
probability for losing no energy at all is given by the markers at
$x_{\rm E}=1$.

The probabilities for absorption and no interaction of the
multiple-soft scattering approximation are larger than for the opacity
expansions. The continuous part of the energy loss probability
distribution is more relevant in the opacity expansion than for the
multiple soft scattering approximation.

In the AMY formalism, the outgoing quark spectrum peaks at $x_{\rm E}
= 0.9$, which corresponds to an energy loss $\Delta E \sim 2$ GeV. This radiation falls in the the region where we
cannot distinguish the radiated gluons from the thermal gluons in the
AMY formalism, as indicated by the cut in Fig.~\ref{fig:dIdw}.

\begin{table*}
\begin{tabular}{l c @{\hskip 0.2cm} c c c c @{\hskip 0.5cm} c c c c }
%\hline
& & & \hfill $T = 300$ & MeV \hfill & & & \hfill $R_{7} =$ & 0.25 \hfill  & \\
& & $\;\; \langle N_{g} \rangle \;\;$ & $\;\; \langle \Delta E
\rangle \;\;$ & $\;\;\;\; p_{0} \;\;\;\;$ & $\;\;\;\; p_{1} \;\;\;\;$
& $\;\; \langle N_{g} \rangle \;\;$ & $\;\; \langle \Delta E \rangle
\;\;$ & $\;\;\;\; p_{0} \;\;\;\;$ & $\;\;\;\; p_{1} \;\;\;\;$\\
& & & (GeV) & & & & (GeV) & &\\

\hline
\multirow{4}{*}{$L = 2$ fm} & AMY     &      & 3.95 &     &      &      & 6.58 &    & \\
& ASW--MS & 0.17 & 0.3 & 0.84 & 0.01 & 1.58 & 5.9 & 0.2 & 0.46\\
& WHDG & 0.48 & 1.24 & 0.62 & 0.00 & 1.94 & 7.7 & 0.14 & 0.14\\
& ASW--SH & 0.25 & 0.93 & 0.78 & 0.03 & 1.92 & 6.9 & 0.15 & 0.35\\
\hline
\multirow{4}{*}{$L = 5$ fm} & AMY & & 8.30 & & & & 6.47 & & \\
& ASW--MS & 1.17 & 3.95 & 0.31 & 0.08 & 1.83 & 6.4 & 0.16 & 0.33\\
& WHDG & 1.96 & 5.63 & 0.14 & 0.05 & 2.44 & 7.3 & 0.09 & 0.11\\
& ASW--SH & 2.09 & 5.76 & 0.12 & 0.2 & 2.47 & 6.7 & 0.08 & 0.26\\
\hline
\end{tabular}
\caption{\label{tab:charac}Table of different variables for the
 different energy loss models. Left columns: for $T = 300$ MeV for
 all models. Right columns: corresponding medium properties required
 to reach the typical suppression $R_{7} = 0.25$. }
\end{table*}

\subsubsection{Some characteristic quantities}

When all the models are tuned to a fixed amount of suppression
$R_{7}$, the radiated gluon spectra are quite different. To quantify the difference, we calculate a few characteristic quantities for energy loss which are tabulated in Table \ref{tab:charac}. 

The
average number of emitted gluons can be obtained by integrating the
gluon spectrum (e.g. Fig.~\ref{fig:dIdw}) over the gluon energy $\omega$: 
\be \langle
N_{g}\rangle = \int d\omega \frac{dI}{d\omega}.  
\ee 
Note that this
determines the probability for no energy loss:
\be 
p_{0} = e^{-\langle N_{g} \rangle}.  
\ee 
The probability for
complete absorption of the parton is given by
\begin{equation}
p_{1}=\int_E^\infty d(\Delta E) \; P(\Delta E) ,
\end{equation}
where $E$ is the energy with which the parent parton enters the brick. The total average energy loss for the incoming parton is calculated from the energy loss probability distribution $P(\Delta E)$:
\be
\langle \Delta E \rangle = \int_{0}^{E} d(\Delta E) \; P(\Delta E),
\ee
for which only the surviving partons are taken into account. For the AMY formalism the integral starts at negative energies (in principle $-\infty$) because the parton can also absorb energy from the medium (see Figs \ref{fig:dNdzSameTemp} and \ref{fig:dNdz}).

The values collected in Table \ref{tab:charac} show that tuning the
models to the same suppression factor $R_{7}$ does not imply that the
mean energy loss is the same. On the other hand, when the suppression factor $R_{7}$ is
fixed, the number of radiated gluons is similar.

\section{Conclusion}
For a summary of the main findings we refer back to Section \ref{sect:conclusion}.

\section*{Acknowledgements}
The authors would like to acknowledge useful and interesting
discussions with Peter Arnold, Steffen Bass, Ullrich Heinz, and
Bronislav Zakharov.  This research was financially supported by
Ministerio de Ciencia e Innovacion of Spain (grants FPA2008-01177 and
FPA2009-06867-E), Xunta de Galicia (Consellera de Educacion and grant
PGIDIT10PXIB 206017PR), project Consolider-Ingenio 2010 CPAN
(CSD2007-00042), FEDER, the Natural Sciences and Engineering Research
Council of Canada, the Netherlands Organisation for Scientific
Research (NWO), the US Department of Energy under DOE Contracts
DE-FG02-05ER-41367, DE-SC0005396, DE-AC02-05CH11231,
DE-AC02-98CH10886, DEAC-76-00098, DE-SC0004286 and (within the
framework of the JET Collaboration) DE-SC0004104, and by a Lab Directed
Research and Development Grant from Brookhaven Science Associates.

\bibliography{brickreport}

\appendix

\section{Common definition of medium properties $\hat{q}$, $\mu$ and $\lambda$}
\label{sec:commondef}
In this Appendix we review a number of commonly used equations to
relate properties of the medium, such as the transport coeffient
$\hat{q}$, mean free path $\lambda$ and temperature $T$. 
 
\subsection{The basis: a HTL plasma}
The basis for all the equations used are free scattering cross sections, supplemented with elements of Hard Thermal Loop (HTL) field theory to screen the soft divergence in the free cross section.

For a medium with temperature $T$, the Debye screening mass $m_D$ is
\begin{equation}
m_{D}^{2} = (1 + \frac{1}{6}N_{f})g^{2}T^{2},
\end{equation}
where $g$ is the coupling constant ($g^2 = 4\pi \alpha_S$) and $N_f$ is the number of quark flavours (0 for a pure gluon gas).

We also define a number density of the medium $\mathcal{N}$ \cite{Arnold:2008vd}:
\begin{equation}
\mathcal{N} = \frac{\zeta(3)}{\zeta(2)}(1+\frac{1}{4}N_{f})T^{3}.
\end{equation}

The differential
scattering rate for a hard parton traversing the medium is known for two limits of the exchanged momentum $q_\perp$ \cite{Arnold:2008vd}:
\begin{equation}
\label{eq:GammaLowHigh}
%\begin{displaymath}
 \frac{d\Gamma_{el}}{d^{2}q_{\perp}} \simeq \frac{C_{R}}{(2\pi)^{2}}
 \times \left\{ \begin{array}{ll}
   \frac{g^{2}Tm_{D}^{2}}{{\bf q}_{\perp}^{2}({\bf q}_{\perp}^{2}+m_{D}^{2})}
 &\mbox{ if $|{\bf q}_{\perp}| \ll T$}\\
   &\mbox{}\\
   \frac{g^{4}\mathcal{N}}{{\bf q}_{\perp}^{4}}
   &\mbox{ if $|{\bf q}_{\perp}| \gg T$,} \end{array}
\right.
%\end{displaymath}
\end{equation}
the relevant color factor $C_R$ for most energy loss calculations is the color factor of the gluon $C_R = C_A = 3$.

\subsection{Transport coefficient $\hat{q}$}
The transport coefficient $\hat{q}$, which is the only medium parameter that enters in the multiple soft scattering approximation, is defined as the mean momentum kick squared per unit path length
\begin{equation}\label{eq:defqhat}
\hat{q}  = \rho \int d^2q_T\, q_T^2\, \frac{d\sigma}{d^2q_T}  
\equiv \int_0^{q_{\rm max}} d^{2}q_{\perp}\frac{d\Gamma_{el}}{d^{2}q_{\perp}}
{\bf q}_{\perp}^{2},
\end{equation}
in which $\Gamma_{el}$ is the rate for elastic collisions in the
plasma, ${\bf q}_{\perp}$ is the transverse momentum exchanged with
the medium $q_{\rm max}$ is the upper limit for the exchanged
momentum, which represents an ultraviolet cut-off.

\begin{figure}
\centering
\includegraphics[width=0.5\textwidth]{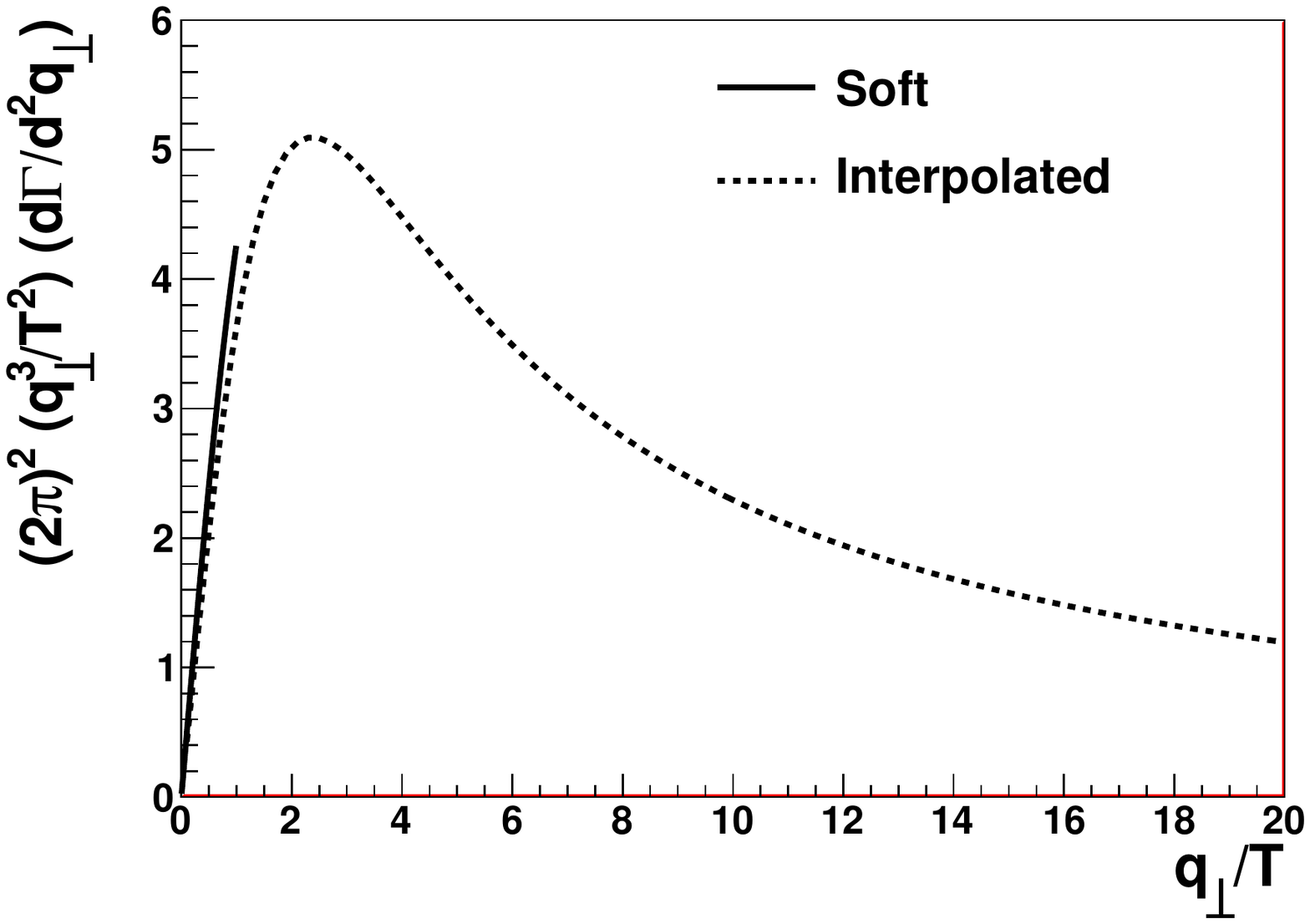}
\caption{Differential elastic cross-section as function of $q_{\perp}$
 for the different energy limits. The curves labeled with {\it Interpolated} correspond to the expression 
 given  in Eq.~(\ref{eq:GammaComb}). Curves labeled with {\it Soft} correspond to the low energy 
 limit in Eq.~(\ref{eq:GammaLowHigh}).}
\label{fig:GammaEl}
\end{figure}

The transport coefficient $\hat{q}$ integrates over all momentum transfers $\hat{q}$ and can therefore in general not be based on only the low or the high $q_{\perp}$ limit given in Eq. \ref{eq:GammaLowHigh}.
The following expression provides a smooth interpolation between the two limits 
\begin{equation} 
\label{eq:GammaComb}
\frac{d\Gamma_{el}}{d^{2}q_{\perp}} \simeq \frac{C_{R}}{(2\pi)^{2}}
 \times \frac{g^{4}\mathcal{N}}{{\bf q}_{\perp}^{2}({\bf q}_{\perp}^{2}+m_{D}^{2})}.
\end{equation}
This interpolation is illustrated in Fig.~\ref{fig:GammaEl} which
 shows low energy limit of Eq.~(\ref{eq:GammaLowHigh}) (curve labeled
 {\it Soft}) together with Eq. \ref{eq:GammaComb} (curve labeled {\it
 Interpolated}). In the soft region the interpolated curve deviates by
 $17\%$ due to the difference in the numerator and approaches the high
 energy limit smoothly at large $q_{\perp}$.

The resulting transport coefficient is 
\begin{equation}
\label{eq:qhat}
\hat{q}(T) =
\frac{C_{R}g^{4}\mathcal{N}(T)}{4\pi}\mathrm{ln}\left(\frac{q_{\rm max}^{2}(T)}{m_{D}^{2}(T)}+1\right).
\end{equation}
In the high energy limit the constant in the argument of the logarithm
vanishes but in the low energy limit its presence is crucial because
otherwise unphysical negative values for $\hat{q}$ are possible. In a
realistic medium created in a heavy-ion collision naturally both cases
will be present.

\begin{figure}
\centering
\includegraphics[width=0.5\textwidth]{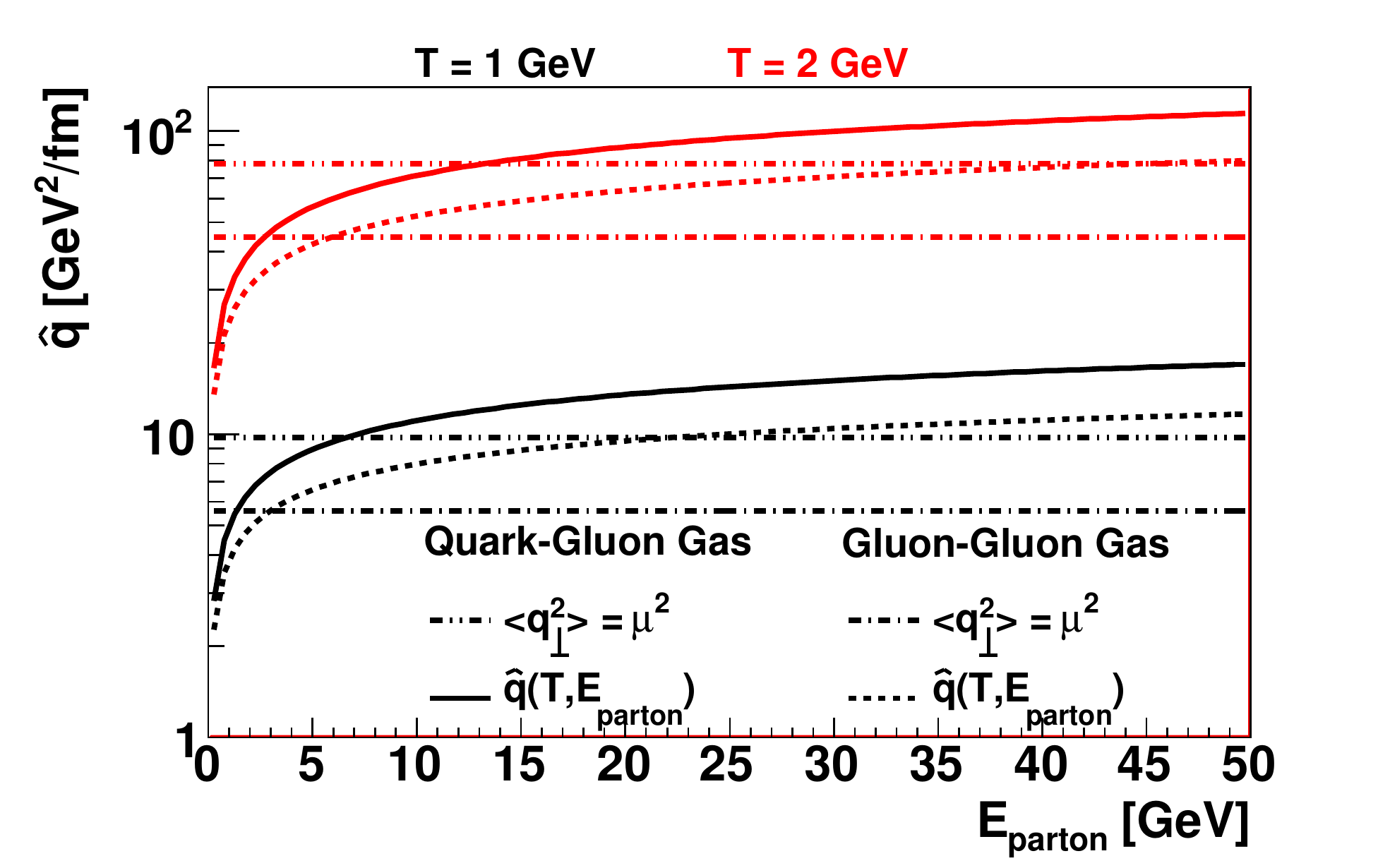}
\caption{The transport coefficient as function of the energy of the
 parton traversing a medium of different tempertures. The transport
 coefficient is calculated in the hard thermal loop formalism as in
 equation \ref{eq:qhat}.\label{fig:qhatvsEParton}}
\end{figure}
The value for $q_{\rm max}$ is expected to fall in the hard scattering
regime, where the scattering centers can be approximated as static,
leading to $q_{\rm max}=g(ET^{3})^{1/4}$  as has been argued in
\cite{CaronHuot:2008ni}. The dependence of $q_\mathrm{max}$ on the
energy of the incoming parton $E$ leads to a logarithmic
dependence of  $\hat{q}$ on $E$. Figure~\ref{fig:qhatvsEParton}
compares the resulting energy-dependent $\hat{q}$ to a commonly used
energy-independent expression for $\hat{q}$
\begin{equation}
\hat{q} = \frac{\langle q^2_\perp \rangle}{\lambda} \approx
\frac{m_D^2}{\lambda},
\end{equation}
where it is assumed that the mean squared momentum exchange per
scattering is equal to $m^2_D$ and the mean free path $\lambda$ is
calculated as given below.

\subsection{Mean free path}
The mean free path $\lambda$ used in the opacity expansion is usually
calculated based on a Quark Gluon Plasma with static scattering centers,
using the scattering rate
\begin{equation}\label{eq:GammaOE}
\frac{d\Gamma_{el}}{d^{2}q_{\perp}} \simeq \frac{C_{R}}{(2\pi)^{2}}
 \times \frac{g^{4}\mathcal{N}}{({\bf q}_{\perp}^{2}+m_{D}^{2})^{2}},
\end{equation}
which is very similar to Eq. \ref{eq:GammaComb}, except for a change
in the denominator which is needed to obtain a finite result for $\lambda$.

This leads to the following expression for the number of scattering
centers 
\begin{equation}
\frac{L}{\lambda} = L\int
d^{2}q_{\perp}\frac{d\Gamma_{el}}{d^{2}q_{\perp}} = 4\pi C_{R}\mathcal{N}\frac{\alpha_s^{2}}{m_{D}^{2}}L.
\end{equation}

\end{document}